%% file: DR16_LRGBAO_FourierSpace.tex
\documentclass[useAMS,usenatbib]{mnras}

\usepackage{graphicx}
\usepackage[dvipsnames]{xcolor}
\usepackage{times}

\usepackage{mn2e-breakabs}

\usepackage{amssymb}
\usepackage{amsmath}
\usepackage{soul} 
\newcommand{\hompc}{\,h\,{\rm Mpc}^{-1}}
\newcommand{\mpcoh}{\,h^{-1}\,{\rm Mpc}}

\usepackage{hyperref}
\hypersetup{
    colorlinks=true,
    linkcolor=blue,
    citecolor=blue,
}

\title[BAO \& FS measurement from eBOSS LRG PS]{The Completed SDSS-IV extended Baryon Oscillation Spectroscopic Survey: measurement of the BAO and growth rate of structure of the  luminous red galaxy sample from the anisotropic power spectrum between redshifts 0.6 and 1.0}

\author[H. Gil-Mar\'in et al.]{\parbox{\textwidth}{
H\'ector Gil-Mar\'in$^{1,2}$\thanks{hectorgil@icc.ub.edu}, 
 Juli\'an E. Bautista$^3$,
Romain Paviot$^4$,
Mariana Vargas-Maga\~na$^5$,
 Sylvain de la Torre$^4$,
Sebastien Fromenteau$^6$, 
Shadab Alam$^7$, 
Santiago \'Avila$^{8}$, 
 Etienne Burtin$^{9}$, 
 Chia-Hsun Chuang$^{10}$, 
 Kyle S. Dawson $^{11}$,  
 Jiamin Hou$^{12}$, 
 Arnaud de Mattia$^{9}$,
Faizan G. Mohammad$^{13,14}$,
Eva-Maria M\"uller$^{15}$,  
Seshadri Nadathur$^{3}$, 
Richard Neveux$^{9}$, 
Will J. Percival$^{13,14,16}$,
Anand Raichoor$^{17}$,  
Mehdi Rezaie$^{18}$, 
Ashley J. Ross$^{18}$, 
Graziano Rossi$^{19}$, 
Vanina Ruhlmann-Kleider$^{9}$,  
Alex Smith$^{9}$, 
Am\'elie Tamone$^{17}$, 
 Jeremy L. Tinker$^{20}$, 
 Rita Tojeiro$^{21}$, 
 Yuting Wang$^{22}$,  
 Gong-Bo Zhao$^{22,\,3}$,  
 Cheng Zhao$^{17}$, 
 Jonathan Brinkmann$^{23}$,  
 Joel R. Brownstein$^{11}$,   
 Peter D. Choi$^{19}$,
 Stephanie Escoffier$^{24}$, 
 Axel de la Macorra$^5$, 
 Jeongin Moon$^{19}$,
 Jeffrey A. Newman$^{25}$,  
 Donald P. Schneider$^{26}$,
 Hee-Jong Seo$^{18}$, 
 Mariappan Vivek$^{26,27}$
} \vspace*{10pt} \\ 
$^{1}${
Institut de Ci\`encies del Cosmos,  Universitat  de  Barcelona,  ICCUB,  Mart\'i  i  Franqu\`es  1,  E08028  Barcelona,  Spain
}\\
$^{2}${
Institut  d’Estudis  Espacials  de  Catalunya  (IEEC),  E08034  Barcelona,  Spain 
}\\
$^{3}${
Institute of Cosmology \& Gravitation, University of Portsmouth, Dennis Sciama Building, Portsmouth, PO1 3FX, United Kingdom
}\\
$^{4}${ 
Aix Marseille Univ, CNRS, CNES, LAM, Marseille, France.
}\\
$^{5}${
Instituto de F\'isica, Universidad Nacional Aut\'onoma de M\'exico, Apdo. Postal 20-364, Ciudad de M\'exico, M\'exico
}\\
$^{6}${
Instituto de Ciencias F\'isicas, Universidad Nacional Aut\'onoma de M\'exico, Av. Universidad s/n, 62210 Cuernavaca, Mor., Mexico
}\\
$^{7}${
Institute for Astronomy, University of Edinburgh, Royal Observatory, Edinburgh, EH9 3HJ, United Kingdom 
}\\
$^{8}${
Instituto de F\'isica Te\'orica UAM/CSIC, Universidad Autonoma de Madrid, 28049 Madrid, Spain
}\\
$^{9}${
IRFU/CEA Universit\'e Paris-Saclay, F91191 Gif-sur-Yvette, France
}\\
$^{10}${
 Kavli Institute for Particle Astrophysics and Cosmology, Stanford University, 452 Lomita Mall, Stanford, CA 94305, USA
}\\
$^{11}${
Department Physics and Astronomy, University of Utah, 115 S 1400 E, Salt Lake City, UT 84112, USA
}\\
$^{12}${
Max-Planck-Institut f\"ur Extraterrestrische Physik, Postfach 1312, Giessenbachstr., 85748 Garching bei M\"unchen, Germany
}\\
$^{13}${ 
Waterloo Centre for Astrophysics, University of Waterloo, Waterloo, ON N2L 3G1, Canada
}\\
$^{14}${
Department of Physics and Astronomy, University of Waterloo, Waterloo, ON N2L 3G1, Canada
}\\
$^{15}${
University of Oxford, Oxford OX1~3RH, United Kingdom 
}\\
$^{16}${
Perimeter Institute for Theoretical Physics, 31 Caroline St. North, Waterloo, ON N2L 2Y5, Canada
}\\
$^{17}${
Institute of Physics, Laboratory of Astrophysics, Ecole Polytechnique Federale de Lausanne (EPFL), Observatoire de Sauverny, 1290 Versoix, Switzerland 
}\\
$^{18}${
Department of Physics and Astronomy, Ohio University, Clippinger Labs, Athens, OH 45701, USA
}\\
$^{19}${
Department of Physics and Astronomy, Sejong University, Seoul, 143-747, Korea
}\\
$^{20}${
Center for Cosmology and Particle Physics, Department of Physics, New York University, New York, NY 10003, USA
}\\
$^{21}${
School of Physics and Astronomy, University of St Andrews, St Andrews, KY16 9SS, United Kingdom
}\\
$^{22}${
National Astronomy Observatories, Chinese Academy of Science, Beijing, 100012, P.R. China
}\\
$^{23}${
Apache Point Observatory, P.O. Box 59, Sunspot, NM 88349
}\\
$^{24}${
Aix Marseille Univ, CNRS/IN2P3, CPPM, Marseille, France
}\\
$^{25}${
PITT PACC, Department of Physics and Astronomy, University of Pittsburgh, Pittsburgh, PA 15260, USA
}\\
$^{26}${
Institute for Gravitation and the Cosmos, Pennsylvania State University, University Park, PA 16802, USA
}\\
$^{27}${
 Indian Institute of Astrophysics, Koramangala, Bangalore 560034, India
 }\\
}
\date{Accepted XXX. Received YYY; in original form \today}

\pubyear{2020}

\begin{document}
\label{firstpage}
\pagerange{\pageref{firstpage}--\pageref{lastpage}}
\maketitle
\begin{abstract}
We analyse the clustering of the Sloan Digital Sky Survey IV extended Baryon Oscillation Spectroscopic Survey Data Release 16 luminous red galaxy sample (DR16 eBOSS LRG) in combination with the high redshift tail of the Sloan Digital Sky Survey III Baryon Oscillation Spectroscopic Survey Data Release 12 (DR12 BOSS CMASS). We measure the redshift space distortions (RSD) and also extract the longitudinal and transverse  baryonic acoustic oscillation (BAO) scale from the anisotropic power spectrum signal inferred from  377,458 galaxies between redshifts 0.6 and 1.0, with effective redshift of $z_{\rm eff}=0.698$ and effective comoving volume of $2.72\,{\rm Gpc}^3$. After applying reconstruction we measure the BAO scale and infer $D_H(z_{\rm eff})/r_{\rm drag} = 19.30\pm 0.56$ and $D_M(z_{\rm eff})/r_{\rm drag} =17.86 \pm 0.37$. When we perform a redshift space distortions analysis on the pre-reconstructed catalogue on the monopole, quadrupole and hexadecapole we find,  $D_H(z_{\rm eff})/r_{\rm drag} = 20.18\pm 0.78$, $D_M(z_{\rm eff})/r_{\rm drag} =17.49 \pm 0.52$ and $f\sigma_8(z_{\rm eff})=0.454\pm0.046$. 
We combine both sets of results along with the measurements in configuration space  and  report  the following  consensus  values: $D_H(z_{\rm eff})/r_{\rm drag} = 19.77\pm 0.47$, $D_M(z_{\rm eff})/r_{\rm drag} = 17.65\pm 0.30$ and $f\sigma_8(z_{\rm eff})=0.473\pm 0.044$, which are in full agreement with the standard $\Lambda$CDM and GR predictions. These results represent the most precise measurements within the redshift range $0.6\leq z \leq 1.0$ and are the culmination of more than 8 years of SDSS observations.

\end{abstract}

\quad

\section{Introduction}

The large-scale structure of the Universe (LSS) contains valuable information of how the Universe has been evolving in the last $\sim7\times10^9$ years, when the Dark Energy domination era started. The current state-of-the-art spectroscopic LSS observations allow to utilise the standard ruler baryon acoustic oscillations (BAO), first detected in \cite{eisenstein_detection_2005} on the Sloan Digital Sky Survey dataset (SDSS) and \cite{cole_2df_2005} on the two-degree Field Survey (2dF, \citealt{colless_2df_2003}),  to determine with precision the background expansion history of the Universe at late-time. During the last decade the BAO technique has evolved in both precision and accuracy becoming mature. Consequently a plethora of measurements have been performed on spectroscopic galaxy surveys at different epochs:  6-degree Field Survey (6dF; \citealt{jones_6df_2009,beutler_6df_2011}) at $z=0.106$, WiggleZ \citep{drinkwater_wigglez_2010,blake_wigglez_2011-1, kazin_wigglez_2014} at $z=0.44,\, 0.6,\, 0.73$, and Baryon Oscillation Spectroscopic Survey (BOSS) galaxies \citep{dawson_baryon_2013,anderson_clustering_2012,anderson_clustering_2014-1,anderson_clustering_2014,alam_clustering_2017} at $z=0.38,\, 0.51,\,0.61$, and BOSS Lyman-$\alpha$ forests \citep{bautista_measurement_2017, du_mas_des_bourboux_baryon_2017} at $z=2.40$. Additionally, if we want to obtain a direct measurement of the growth of structures from these same spectroscopic surveys we need to measure the effect of redshift space distortions (RSD; \citealt{kaiser87}). Consequently, we obtain both an expansion history and a growth of structure measurement from the same dataset. Parallel to the BAO technique development, RSD analyses have also matured both in modelling and observational systematics treatment during the last decade: RSD in 2dF \citep{percival_2df_2004}, in 6dF \citep{beutler_6df_2012}, in WiggleZ \citep{blake_wigglez_2011}, in VIPERS  \citep{guzzo_vimos_2014,de_la_torre_vimos_2013, pezzotta_vimos_2017}, in FastSound  \citep{okumura_subaru_2016}, as well in BOSS galaxies \citep{alam_clustering_2017}. 

Anisotropic BAO studies provide a direct measurement of the background expansion at the epoch of the observed galaxies, $z$, through the absolute and relative BAO peak position in the anisotropic multipoles of the power spectrum or correlation function.  Under the assumption of a functional form of the background expansion, $H(z;\Omega_m)$ one can obtain a direct measurement of the density of matter in the Universe $\Omega_m$. Note that the BAO peak position is not directly sensitive to $H(z)$, but to $H(z)r_{\rm drag}$, and to the comoving angular diameter distance over the comoving sound horizon at the epoch where the baryon-drag optical depth equals unity, $D_M(z)/r_{\rm drag}$. From these measurements one can either infer $\Omega_m$ from the product of the two (which  is independent of $r_{\rm drag}$), or assume an extra prior on $r_{\rm drag}$, which can either come from cosmic microwave background (CMB) measurements, or from a functional form of $r_{\rm drag}$ given by priors on the baryon, $\Omega_b$ and radiation density $\boldsymbol{\Theta}_{\rm rad}$ (which are typically not measured by LSS), and infer $H_0$ (see for e.g. \citealt{Addisson17}). Within the SDSS collaboration we opt to analyse these results under the less restrictive set of priors, and thus only assume a functional form for $H(z)$, but  no restriction on $r_{\rm drag}$ as a function of cosmology. The motivation for proceeding this way is the robustness of the cosmological interpretation under potential changes of the cosmological paradigm if, for example,  the state-of-the-art value of $r_{\rm drag}$ changes significantly in the future or $\Lambda$CDM is ruled out, as one would just only need to re-interpret the quantities $H(z)r_{\rm drag}$ and $D_M(z)/r_{\rm drag}$ rather than reanalysing the data. In this paper we choose to work with the `Hubble distance', $D_H$, defined as $D_H(z)\equiv c/H(z)$, where $c$ is the speed of light. The parameter $D_H(z)/r_{\rm drag}$ has the advantage of being dimensionless, of order unity and directly proportional to the scale factor which is actually measured. 

Redshift space distortions are a measurement of the peculiar velocity field of the galaxies along the line-of-sight (LOS). As this velocity field is only detected along the LOS, it generates an anisotropic signal in the power spectrum expansion as a function of the cosine of the LOS with the vector separation of the galaxy pair. This velocity field is generated by over-densities of matter, and therefore is coherent with the growth of these density perturbations. Thus, by measuring the redshift space distortion effect on the power spectrum of galaxies one can set constraints on the logarithmic growth of structure parameter, $f$. For the 2-point statistics this parameter is degenerate with the parameter $\sigma_8$, the amplitude of dark matter fluctuations at the scale of $8\, \mpcoh$. For this reason power spectrum or correlation function redshift space distortion analyses are sensitive to the combination, $f$ times $\sigma_8$, which we just refer as $f\sigma_8$. 

In this paper we perform two complementary analyses, BAO and full shape analyses in order to extract $D_M(z)/r_{\rm drag}$, $D_H(z)/r_{\rm drag}$ and $f\sigma_8$ from the power spectrum of the final Data Release 16 (DR16) SDSS-IV eBOSS LRG catalogue in combination with the high redshift tail of the Data Release 12 (DR12) SDSS-III BOSS LRG catalogue (for simplicity we refer to this combined catalogue as the DR16 CMASS+eBOSS LRG catalogue). The catalogue consists of 377,458 galaxies between redshifts 0.6 and 1.0, with effective redshift of $z_{\rm eff}=0.698$ and effective comoving volume of $2.72\,{\rm Gpc}^3$. The BAO analysis is focused exclusively on identifying the position of the BAO features in the power spectrum,  whereas the full shape analysis models the anisotropic power spectrum shape to extract information. In order to enhance the BAO detection, we utilise the standard reconstruction algorithm \citep{Eisrecon07,Burden14}. Thanks to reconstruction we are able to remove most of the non-linear bulk flow effect and enhance the significance of the BAO features. For the BAO analysis, we therefore perform the standard analysis on the reconstructed catalogues, whereas the full shape analysis is performed on the original, pre-reconstructed catalogues. The results extracted from the analysis of the same sample in configuration space are presented in the companion paper \citep{LRG_corr}. Since these two results are expected to be highly correlated (as they are both extracted from the exact same catalogue) we perform a consensus results which is presented at the end of both papers. 

The cosmological implication is presented instead in the companion paper \citep{eBOSS_Cosmology} along with the measurements of the rest of the galaxy and Lyman-$\alpha$ samples of BOSS and eBOSS. These samples correspond to \footnote{A summary of all SDSS BAO and RSD measurements with accompanying legacy figures can be found here: \href{https://www.sdss.org/science/final-bao-and-rsd-measurements/}{sdss.org/science/final-bao-and-rsd-measurements/} .  The full cosmological interpretation of these measurements can be found here: \href{https://www.sdss.org/science/cosmology-results-from-eboss/}{sdss.org/science/cosmology-results-from-eboss/}},
\begin{itemize}
\item Luminous Red Galaxy sample (LRG), $0.6<z<1.0$, power spectrum analysis (this paper) and correlation function analysis \citep{LRG_corr}
\item Emission Line Galaxy sample (ELG), $0.6<z<1.1$, power spectrum analysis \citep{demattia20a}, correlation function analysis \citep{tamone20a} and catalogue description \citep{raichoor20a}
\item Quasar clustering sample (QSO), $0.8<z<2.2$, power spectrum analysis \citep{neveux20a} and correlation function analysis \citep{hou20a} 
\item Lyman-$\alpha$ cross- and auto-correlation analysis \citep{2020duMasdesBourbouxH} with quasars $z>2.1$ . 
\end{itemize}
In addition, \cite{eBOSS_Cosmology} includes as well the results from the two low- and middle-redshift overlapping bins from SDSS-III BOSS \citep{alam_clustering_2017}, as they do not overlap with any of the eBOSS samples. 
An essential component of these studies is the generation of data catalogs \citep{ross20a,lyke20a}, mock catalogs \citep{lin20a,zhao20a}, and {\it N}-body simulations for assessing systematic errors on the LRG \citep{rossi20a,smith20} and ELG samples \citep{Avila20,Alam20}. Additionally in \cite{Wang20,gongbo} the cross-correlation signal between LRG and ELG samples is presented and studied. 

Previous to the final DR16 analysis these samples were already studied for the two-year observation catalogues Data Release 14 (DR14): DR14 eBOSS LRG BAO \citep{bautista_sdss-iv_2018}, DR14 eBOSS LRG RSD \citep{icaza-lizaola_clustering_2019}, DR14 eBOSS quasar BAO \citep{ata_clustering_2018}, DR14 eBOSS quasar RSD \citep{hou_clustering_2018,zarrouk_clustering_2018,gil-marin_clustering_2018} and DR14 Lyman-$\alpha$ \citep{de_sainte_agathe_baryon_2019,blomqvist_baryon_2019}. Other studies which included redshift-weighting techniques of the DR14 quasar sample were also presented by \cite{Ruggerietal19,Wangetal18,Zhaoetal19,Zhuetal18}. 

This paper is organised as follows. In \S\ref{sec:data1} we briefly present the actual and synthetic  galaxy catalogues used in this paper. In \S\ref{sec:methodology} we describe the methodology followed for performing the power spectrum estimation and the models used for both BAO and full shape analysis. In \S\ref{sec:results} we present the results of this paper as well as the consensus along with the complementary configuration space analysis. In \S\ref{sec:systematics} we perform an exhaustive systematic study to quantify the potential systematic effect that could affect the inferred cosmological parameters. In \S\ref{sec:consensus} we present the Fourier and configuration space consensus results and in \S\ref{sec:cosmo} we compare our findings with the standard $\Lambda$CDM model predictions. Finally in \S\ref{sec:conclusions} we present the conclusions of this work.

\section{Dataset}\label{sec:data1}

We briefly describe the DR16 LRG dataset along with the synthetic mock catalogues we use. A detailed description of the DR16 dataset is presented in \cite{ross20a}; the synthetic fast \textsc{EZmocks} used for estimating the covariance are fully described in \cite{zhao20a}; and the mocks based on \textsc{OuterRim} {\it N}-body simulation used for validating the pipeline are described in \cite{rossi20a}. Additionally, we make use of a series of {\it N}-body simulations used for previous BOSS analyses \citep{alam_clustering_2017}, which we refer as \textsc{Nseries} mocks.

\subsection{LRG galaxy sample}\label{sec:data}

The Sloan Digital Sky Survey fourth generation spectroscopic observations (SDSS-IV, \citealt{Blanton17}) employ two multi-object spectrographs \citep{Smeeetal13} installed on the Apache Point Observatory 2.5-meter telescope located in New Mexico, USA \citep{SDSSTelescope}, to carry out spectroscopic measurements from a photometrically selected eBOSS LRGs sample \citep{eBOSSoverview}. Such LRGs were previously selected from the optical SDSS photometry DR13 \citep{Albaretietal13} with the supplementary infrared photometry from the WISE satellite \citep{Langetal14}. The same instrument was already used for the previous BOSS program. 

A description of the final targeting algorithm is presented in \cite{Prakashetal16}, which produced 60 LRG targets per square-degree over a sky footprint of 7500 deg$^2$, of which $\sim50/{\rm deg}^2$ were spectroscopically observed. Such observations returned mainly objects between $0.6\leq z \leq 1.0$ as tested by The Sloan Extended Quasar, ELG and LRG Survey (SEQUELS, \citealt{eBOSSoverview}).

The estimation of the redshift of each LRG spectrum was performed using the publicly available \textsc{RedRock} algorithm,\footnote{\textsc{RedRock} is available at \href{https://www.sdss.org/dr16/software/products}{ sdss.org/dr16/software/products}} which improved the redshift efficiency of its predecessor, \textsc{RedMonster} \citep{Hutchinsonetal16}, from 90\% up to 96.5\% in terms of objects with a confident redshift estimate, with less than 1\% catastrophic redshift errors. 

A description of the catalogue creation is presented in detail in \cite{ross20a}. In short, a synthetic catalogue of randomly generated objects is created over the same footprint of the eBOSS targeted objects matching its angular and radial geometry. We refer to this as the {\it random catalogue} of the data, as it does not contain any intrinsic clustering structure, other than that spuriously generated by the selection function.  Both data and random catalogue are filtered through a series of masking processes to remove  regions with bad photometry, target collisions with quasar spectra (quasar objects had priority in being spectroscopically observed over LRGs when a fibre collision occurred) and centre-post regions, among other effects. This series of masking processes removed 17\% of the initial LRG eBOSS footprint. In addition to these effects, 3.4\% of the LRG targets were not observed because of {\it fibre collisions} with another LRG target. For BOSS and eBOSS this occurs when two photometrically selected targets are closer than $62^{\prime\prime}$. Some of these close objects could be spectroscopically observed when the same group of objects of the sky was observed by more than one plate. In this catalogue we treat these collided groups by up-weighting all group objects by the same weight value, $w_{\rm cp}=N_{\rm targ}/N_{\rm spec}$, where $N_{\rm targ}$ is the number of targeted objects and $N_{\rm spec}$ the number of objects with actual spectroscopic observation. Note that this differs from the treatment previously applied to the DR14 eBOSS and DR12 BOSS analyses. A similar procedure is followed for those galaxies with no reliable redshift information, due to catastrophic redshift failures. These types of failures represent $2.1\%$ of the LRG targets. In this case a redshift failure weight, $w_{\rm noz}$ is assigned to such galaxies as a function of the location of its spectrum on the CCD camera and the overall signal-to-noise ratio of the spectrograph in which it was observed. By multiplying the redshift-failure and close-pair weight, we obtain the total eBOSS collision weight, 
\begin{equation}
w_{\rm col}^{\rm eBOSS}=w_{\rm cp}\cdot w_{\rm noz}.
\end{equation}
Note that a galaxy that does not suffer from any of these effects would have a collision weight of unity. 

The density of objects with spectroscopic information per sky-area in the galaxy catalogues is not constant over the eBOSS sky footprint, due to both observational systematics (varying observational features across the imaging survey) and geometrical effects (for example whether a region has been simultaneously observed by more than one plate). We refer to this whole effect as {\it completeness}, without separating the observational and geometrical contributions.  Qualitatively, the completeness  generates spurious signals we need to filter out in order to measure the intrinsic clustering. Within the eBOSS collaboration we define the completeness as the ratio of the number of weighted spectra (including also objects classified as stars and quasars) to the number of targets, which is computed per sky sector, this is, a connected region of the sky observed by a unique set of plates. In order to account for the effect of completeness  we downsample each object of the random catalogue by the completeness of its corresponding sky sector. In this way, the definition of completeness includes the systematic weight, $w_{\rm sys}$, as well as other effects, including the variation of the mean density as a function of stellar density and galactic extinction. For further details on the catalogue creation we refer the reader to \cite{ross20a}. 

Additionally, a minimum variance weight is also applied, the FKP weight \citep{feldman_power-spectrum_1994}. This accounts for the radial mean density dependence, $w_{\rm FKP}(z)=1/[1+n(z)P_0]$, where $P_0$ is chosen to be the amplitude of the power spectrum $P(k)$ at the scales of BAO, $k\sim0.1\,\hompc$, $P_0=10,000\,(\mpcoh)^3$. 

The objects contained by the LRG galaxy catalogue have the following total weight which accounts for the 4 effects described above,
\begin{equation}
\label{eq:wtot1}w_{\rm tot}=w_{\rm FKP}\times w_{\rm sys}\times w^{(i)}_{\rm col}.
\end{equation}
In this paper we merge the eBOSS LRG galaxy catalogue with the BOSS CMASS SDSS-III catalogue above redshift $0.6$ \citep{Reid2016} into a single LRG catalogue, over which we perform our analysis.
Note that for those galaxies observed by BOSS the weighting scheme is different that the one described above. We refer the reader to the BOSS catalogue paper for details \citep{Reid2016}. In short the total collision weight for BOSS galaxies reads, 
\begin{equation}
w_{\rm col}^{\rm BOSS}=w_{\rm cp}+w_{\rm noz}-1,
\end{equation}
where the collision and failure weights have been obtained using the traditional nearest neighbour approach. 

\begin{figure}
\centering
\includegraphics[scale=0.3]{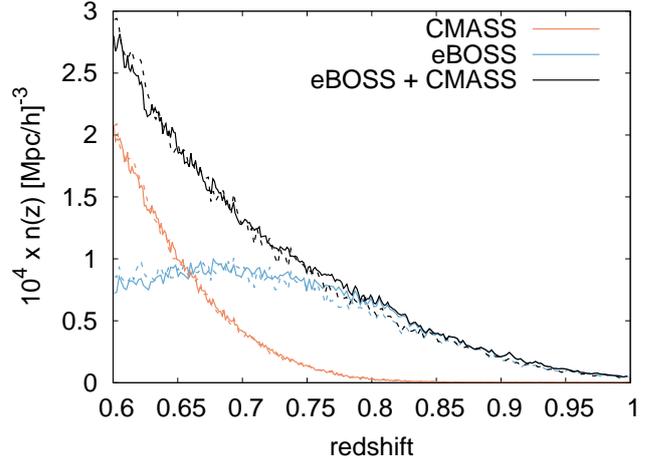}

\caption{ Number density of objects with spectroscopic observations for DR12 BOSS CMASS LRGs (in blue) and DR16 eBOSS LRGs (in orange), for the NGC (solid lines) and SGC (dashed lines). In black is shown the addition of CMASS and eBOSS densities. Note that such additions only correspond to those regions with overlapping area between eBOSS and BOSS CMASS galaxies, which approximately correspond to the whole eBOSS LRG area. The effective redshift of the combined sample corresponds to $z_{\rm eff}=0.698$ according to the definition of Eq. \ref{eq:zeff}.   }

\label{Fig:densities}
\end{figure}

Fig.~\ref{Fig:densities} displays the mean density of objects as a function of redshift for the eBOSS-only LRG (blue) and CMASS (red) galaxies, and the combined CMASS+eBOSS LRG catalogue (black). The solid lines stand for the density of the north galactic cap (NGC) and the dashed lines for the south galactic cap (SGC). 

We have quantified the difference between the NGC and SGC using the mocks to infer the errors and covariance among redshift bins. Unlike the CMASS sample, we find that CMASS+eBOSS LRG $n(z)$ distribution between NGC and SGC is significantly different, which we have imprinted in the \textsc{EZmocks}.  

\begin{figure}
\centering
\includegraphics[scale=0.31]{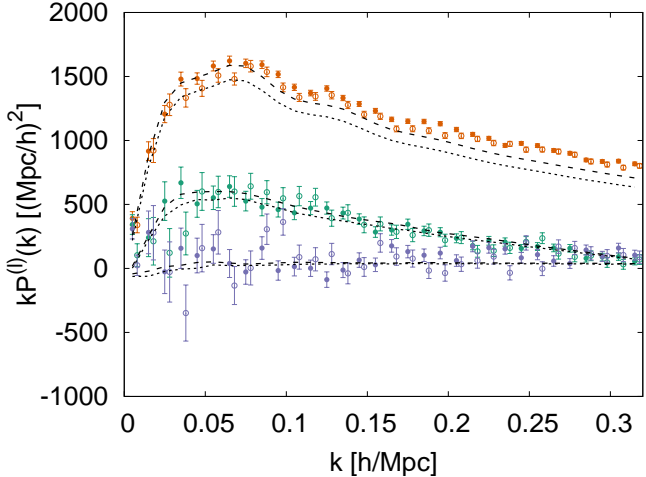}
\caption{Power spectrum multipoles measured from the DR16 CMASS+eBOSS LRG sample, monopole (orange symbols), quadrupole (green symbols) and hexadecapole (purple symbols). The filled and empty symbols correspond to measurements from the NGC and SGC, respectively. The empty symbols are displaced horizontally for visibility. The black dashed and dotted lines correspond to the clustering of the mean of the 1000 realisations of the \textsc{EZmocks} with all the systematics applied, for NGC and SGC, respectively. The amplitude mismatch, more evident for the monopole, is due to the effect of completeness on the normalisation factor of the power for data and mocks.
}
\label{fig:data_mocks}
\end{figure}
\subsection{Synthetic Catalogues}

In this paper we employ several type of mocks in order to estimate the covariance, quantify the impact of systematic errors and to validate the pipeline and methods employed on the data. 

\subsubsection{  \textsc{EZmocks}}

The \textsc{EZmocks} consist of a set of 1000 independent realisations using the fast approximative method based on Zeldovich approximation \citep{ezmocks} with the main purpose of estimating the covariance of the data. Such mocks consist of light-cones with the radial and angular geometry of the CMASS+eBOSS LRG dataset, with observational effects, such as fibre collision, redshift failures and completeness. These light-cones are drawn from 4 and 5 snapshots at different redshifts, for CMASS and eBOSS galaxies, respectively. A full description of these mocks is presented in \cite{zhao20a}. These mocks are generated using fast-techniques, which are a good approximation of an actual {\it N}-body simulation at large scales, but which eventually fail to reproduce the complex gravity interaction and peculiar motions at small scales. Because of this, we use them to estimate the covariance matrix of the data, but their performance for reproducing physical effects such as BAO and RSD is not guaranteed at sub-percent precision level. Thus, we do not estimate the potential modelling systematics based on these mocks, but on full {\it N}-body mocks. However these mocks are useful to estimate the {\it relative} change on cosmological parameters when applying each of these observational features. We use them to quantify the potential impact of observational systematics in the final data results. In order to analyse these mocks we use the covariance drawn from themselves.

\subsubsection{\textsc{Nseries} mocks}

The \textsc{Nseries} mocks are full {\it N}-body mocks populated with a fixed Halo Occupation Distribution (HOD) model similar to the one corresponding to the DR12 BOSS NGC CMASS LRGs.  Their effective redshift, $z_{\rm eff}=0.56$ is slightly smaller compared to the effective redshift of the DR16 CMASS+eBOSS LRG sample, $z_{\rm eff}=0.698$, as they were initially designed to test the potential systematics on the modelling used for the BOSS CMASS sample. They were generated out of 7 independent periodic boxes of $2.6\, h^{-1}{\rm Gpc}$ side, projected through 12 different orientations and cuts, per box. In total, after these projections and cuts 84 pseudo-independent realisations were produced. The mass resolution of these boxes is $1.5\times10^{11}\, M_\odot/h$ and with $2048^3$ particles per box. The large effective volume, $84\times 3.67\,[{\rm Gpc}]^3$ makes them ideal to test potential BAO and RSD systematics generated by the analysis pipeline, as to test the response of the arbitrary choice of reference cosmology on the BAO and full shape model templates, in the galaxy catalogues when converting redshifts into distances, and its impact on the inferred cosmological parameters. We use the NGC \textsc{MD-Patchy} mocks \citep{kitaura} to describe the covariance of these mocks. We rescale the covariance terms by 10\% based on the ratio of particles, as the  \textsc{MD-Patchy} mocks have fewer particles than the  \textsc{Nseries} mocks due to veto effects on DR12 CMASS data, which was also imprinted into the \textsc{MD-Patchy} mocks but not into \textsc{Nseries} mocks. When we run reconstruction on the \textsc{Nseries} mocks, we consistently also use the covariance from reconstructed \textsc{MD-Patchy} mocks.

\subsubsection{ \textsc{OuterRim-HOD} mocks}

The \textsc{OuterRim-HOD} mocks are drawn from the \textsc{OuterRim} {\it N}-body simulation \citep{ORsim} and populated with different types of HOD models (see \citealt{rossi20a} for a full description), some of them similar to the LRG sample, but also others having different properties. The original simulation corresponds to a single cubic box realisation with periodic boundary conditions whose size is $3\,h^{-1}{\rm Gpc}$. This box is divided into 27 cubic sub-boxes of $1\,h^{-1}{\rm Gpc}$ per side, without the periodicity of cubic-boxes. For those galaxy catalogues whose HOD models are close to the actual data sample studied here (those labelled `Hearin-Threshold-2', `Leauthaud-Threshold-2' and `Tinker-Threshold-2', see \citealt{rossi20a} for a description of all models), we place the galaxies in a larger box of $3\,h^{-1}{\rm Gpc}$ per side with empty space between the galaxies and the box edges, and generate a random catalogue with the same distribution but with no clustering. In this way when performing the discrete Fourier transform the non-periodicity conditions do not impact the results. We refer to this process as  {\it padding}. Additionally, we also apply reconstruction on these {\it padded} catalogues. 

The effective volume of each sub-box of the `Hearin-Threshold-2', `Leauthaud-Threshold-2' and `Tinker-Threshold-2', corresponds to $\sim1.1\,{\rm Gpc}^3$. For the rest of the HOD-models, the effective volume varies between $2.1$ and $2.7\, {\rm Gpc}^3$, as the number density of objects, and consequently $\bar{n}P$,  is much higher.  

In order to deal with the covariance of these mocks we have used the covariance derived from the \textsc{EZmocks} and re-scaled by the difference in particle number. These re-scalings correspond to the factors 1.0, 0.64, and 9 for `Standard', `Threshold-1' and `Threshold-2', respectively, for Hearin, Leauthaud and Tinker HOD-types. For Zheng HOD-type we use 0.60, 2.37 and 0.60, for `Standard', `Threshold1' and `Threshold2', respectively.

\subsection{Reference Cosmology}
 In this paper we choose a set of cosmological parameters within the flat $\Lambda$CDM model to define a reference cosmology, which is used to {\it i}) transform the redshifts of galaxies into comoving distances; and {\it ii)} produce a linear template used to build a fitting model. We use as our main baseline analysis the fiducial set of parameters, $\boldsymbol{\Theta}_{\rm fid}$, listed in the first row of Table~\ref{tab:cosmo} as a reference cosmology. In addition, we also analyse the mocks and data using other sets of reference cosmologies to check the impact of this arbitrary choice. Among these cosmologies we choose to use as reference cosmology the underlying cosmology of the \textsc{Nseries} mocks, $\boldsymbol{\Theta}_{\rm Nseries}$, the \textsc{OuterRim} derived mocks, $\boldsymbol{\Theta}_{\rm OR}$ and 3 high-$\Omega_m$ cosmologies, $\boldsymbol{\Theta}_X$, $\boldsymbol{\Theta}_Y$ and $\boldsymbol{\Theta}_Z$, whose properties are listed in Table~\ref{tab:cosmo}. In particular, $\boldsymbol{\Theta}_{Y}$ and  $\boldsymbol{\Theta}_{Z}$ have a very different $r_{\rm drag}$ value compared to the one inferred from the usual CMB-anisotropy experiments \citep{planck_collaboration_planck_2018,WMAP}. In case of $\boldsymbol{\Theta}_Y$ this is driven by a large value of the total number of neutrino species, and for $\boldsymbol{\Theta}_Z$ by a high value of the baryon density. The $\boldsymbol{\Theta}_Y$ and $\boldsymbol{\Theta}_Z$ correspond to a very disfavoured cosmologies compared to the state-of-the art CMB observations. However, our LSS results are presented in a compressed set of variables which do not depend on these CMB priors. Consequently the results inferred from LSS observations by assuming any of the tested cosmologies as `reference-cosmology' are valid, as we will demonstrate in \S\ref{sec:systematics}. 
\begin{table*}
\caption{List of reference cosmology models used along the paper. For our baseline analysis of mocks and data we use the fiducial set of cosmology parameters, $\boldsymbol{\Theta}_{\rm fid}$, as a reference cosmology. For all cosmologies, $\Omega_k=0$.}
\begin{center}
\begin{tabular}{|c|c|c|c|c|c|c|c|c|c|c|c}
Model & $\Omega_m$ & $\Omega_mh^2$ & $\Omega_b$ & $\Omega_bh^2$ & $10^3\times\Omega_\nu$ & $h$ & $n_s$ & $A_s\times10^9$ & $\sigma_8^0$ & $r_{\rm drag}\,[{\rm Mpc}]$ & $N_{\rm eff}$ \\
\hline
\hline
$\boldsymbol{\Theta}_{\rm fid}$ & 0.310 & $0.1417$ &  $0.0481$ & $0.0220$ & 1.400 & 0.676 & 0.97 & 2.040 & 0.8 & 147.78 & $3.046$ \\
$\boldsymbol{\Theta}_{\rm EZ}$ & 0.307 & $0.1411$ & $0.0482$ & $0.0220$ & 0 & 0.678 & $0.96$  & 2.115 & 0.8225 & 147.66 & $3.046$\\
$\boldsymbol{\Theta}_{\rm Nseries}$ & 0.286 & $0.1401$ & $0.0470$ & $0.0230$ & 0 & 0.700 & 0.96 & 2.146 & 0.82 &147.15 & $3.046$\\
$\boldsymbol{\Theta}_{\rm OR}$ & 0.265 & $0.1335$ & 0.0448 & $0.0226$ & 0 &0.710 &0.96 & 2.159 & 0.8 &149.35 & $3.046$ \\
$\boldsymbol{\Theta}_{\rm X}$ & 0.350 & $0.1599$ & 0.0481 & $0.0220$ & $1.313$ &0.676 &0.97 & 1.767 & $0.814$ &143.17 & 3.046 \\
$\boldsymbol{\Theta}_{\rm Y}$ & 0.350 & $0.1599$ & 0.0481 & $0.0220$ & $1.313$ &0.676 &0.97 & 2.040 & $0.814$ &138.77 & 4.046 \\
$\boldsymbol{\Theta}_{\rm Z}$ & 0.365 & $0.2053$ & 0.0658 & $0.0370$ & 0 &0.750 & 0.96 & 2.146 & $0.9484$ &123.97 & $3.046$\\
\end{tabular}
\end{center}
\label{tab:cosmo}
\end{table*}%

In order to determine the effective redshift of the sample, we perform the following weighted pair-count,
\begin{equation}
z_{\rm eff}=\left(\sum_{i>j}w_iw_j(z_i+z_j)/2\right)/\left(\sum_{i>j}w_i w_j  \right),
\label{eq:zeff}
\end{equation}
where $w_i$ is the total weight of the $i^{\rm th}$ galaxy. When we run the above formula over all the pairs separated by distances between $25$ and $130$ $\mpcoh$ we obtain
$z_{\rm eff}=0.698$.\footnote{For the NGC sample we find $z_{\rm eff}=0.695$ and for SGC we find  $z_{\rm eff}=0.704$. For the combined NGC-SGC sample we simply approximate $z_{\rm eff}=0.70$ in the power spectrum linear templates.} Such limits correspond to those used by \cite{LRG_corr} in their FS analysis. Relaxing these limits and accounting for pairs with separations $0<s\,[\mpcoh]<200$ does not modify the effective redshift at 3 significant figures. We therefore take this value of $z_{\rm eff}$ for the analysis performed here, although the correspondence to the Fourier space $k$-ranges to the configuration space ranges is not exact.

\subsection{Reconstruction}
The BAO peak detection significance can be enhanced by applying the reconstruction technique \citep{Eisrecon07}. We use the algorithm described by \cite{Burden14,Burden15} in which the underlying dark matter density field is inferred from the actual galaxy field by assuming a value of the growth of structure and bias, which can be estimated from a full shape-analysis on the pre-reconstruction catalogue, and used to remove both the non-linear motions and the redshift-space distortions of galaxies. 

We make use of the publicly available code\footnote{Reconstruction code available at \href{https://github.com/julianbautista/eboss_clustering}{github.com/julianbautista/eboss$\_$clustering}} employed for performing reconstruction of the DR14 LRG sample \citep{bautista_sdss-iv_2018}. In this paper we apply this code to the combined CMASS+eBOSS sample, by assuming a bias value of $b=2.3$ and a growth rate consistent with $f(z)=\Omega_m^\gamma(z)$, which in this case is $f=0.82$, and using a smoothing scale of $15\,\mpcoh$. Recently \cite{Carter19} showed how the inferred cosmological parameters were not sensitive to these arbitrary choices. Potential systematics arising from reconstruction are checked in \S\ref{sec:systematics}.

\subsection{Power Spectrum estimator}\label{sec:psestimator}
In order to measure the power spectrum multipoles we start by defining the function \citep{feldman_power-spectrum_1994}, 
\begin{equation}
F( {\bf r}_i)=w_{\rm tot}({\bf r}_i)[n_{\rm gal}( {\bf r}_i)-\alpha_{\rm ran} n_{\rm ran}( {\bf r}_i)]/I_{2}^{1/2},
\end{equation}
where $w_{\rm tot}$ is the total weight applied to the galaxy sample described by Eq. \ref{eq:wtot1},
 $n_{\rm gal}$ and $n_{\rm ran}$ are the number density of galaxy and random objects with spectroscopic data, respectively, at position $ {\bf r}_i$, and $\alpha_{\rm ran}$ is the ratio between the weighted number of data-galaxies and randoms. The $w_{\rm tot}$ quantity at each cell position, ${\bf r}_i$, is inferred using the mass interpolation scheme chosen to assign individual objects into a grid. In this fashion, we compute the weighted galaxy density per cell by assigning individual galaxies to a grid weighted by its own individual total weight.  
 
 In this work we use 50 times more density for the random catalogue of the actual LRG dataset and 20 times more for the randoms of the \textsc{EZmocks}. The difference in the estimated power spectrum using the $\times20$ and $\times50$ random catalogue is smaller than 0.5\% per $k$-bin in the power spectrum monopole with no systematic offset.
 As described previously, both data and mocks catalogues total weight $w_{\rm tot}$ is made by the product of the systematic weight, $w_{\rm sys}$ which contains both completeness and imaging weight, the collision weight $w_{\rm col}$ which contains both failures and close pairs collisions, and the FKP-weight. Further details of how these weights were constructed are given in \cite{ross20a}.
The normalisation factor $I_2^{1/2}$ normalises the amplitude of the observed power spectrum and is defined as, $I_2\equiv \int  d{\bf r}\, [n_{\rm gal}w_{\rm tot}({\bf r})]^2$. Later in this section we will comment on how this parameter is inferred and its impact on the final results.

In order to measure the power spectrum multipoles of the galaxy distribution we follow the same procedure described in previous works \citep{gil-marin_clustering_2017}. Briefly, we assign the objects of the data and random catalogues to a regular Cartesian grid, which allows the use of Fourier Transform (FT) based algorithms. We embed the full survey volume into a cubic box of side $L_b=5000\, \mpcoh$, and subdivide it into $N_g^3=512^3$ cubic cells, whose resolution and Nyqvist frequency are $9.8\,\mpcoh$ and $k_{\rm Ny}=0.322\,\hompc$, respectively. We assign the particles to the cubic grid cells using a $3^{\rm rd}$-order B-spline mass interpolation scheme, usually referred to as Piecewise cubic shape (PCS), where each data or random particle is distributed among $5^3$  grid-cells. Additionally, we interlace two identical grid-cells schemes displaced by 1/2 of the size of the grid-cell; this allows us to reduce the aliasing effect below 0.1\% at scales below the Nyqvist frequency (\citealt{HockneyEastwood81}, \citealt{Sefusattietal:2016}).

We estimate the power spectrum using \textsc{Rustico}\footnote{Rapid foUrier STatIstics COde \href{https://github.com/hectorgil/rustico}{github.com/hectorgil/rustico}.} which relies on the Yamamoto estimator approach \citep{Yamamotoetal:2006}, and in particular the implementation presented by \cite{Bianchietal:2015} and \cite{Soccimarro:2015}, to measure the power spectrum multipoles accounting for the effect of the varying LOS,

\begin{eqnarray}
\nonumber P^{(\ell)}(k)&=&(2\ell+1) \int \frac{d\Omega_k}{4\pi} \int d{\bf r}_1 F({\bf r}_1) e^{-i {\bf k}\cdot{\bf r}_1}\\
&\times&   \int d{\bf r}_2 F({\bf r}_2) e^{+i {\bf k}\cdot{\bf r}_2} \mathcal{L}_\ell(\hat{\bf k}\cdot\hat{\bf r}_h),
\label{eq:Pyama}
\end{eqnarray}
where, ${\bf r}_h=({\bf r}_1+{\bf r}_2)/2$, and $\mathcal{L}_\ell$ is the Legendre polynomial of order $\ell$. We approximate $r_h=r_1$, which allows us to perform the two integrals separately using fast FT methods. This approximation introduces wide-angle effects in the power spectrum multipoles as well as the associated window function. However, these effects have been shown to not impact current FS and BAO studies significantly \citep{wideangle}. 
 The $\ell=0$ corresponds to the power spectrum monopole and can be trivially measured using FT without any approximation as $\mathcal{L}_0(x)=1$. The quadrupole and hexadecapole need to expand $\mathcal{L}_\ell$ in powers of its argument. Note that how one distributes these powers of $({\bf k}\cdot{\bf r})$ among the galaxies of the pair is \emph{a priori} arbitrary. For the quadrupole one could expand  $\mathcal{L}_2(x)=(3x^2-1)/2$ as,
 \begin{equation}
  \mathcal{L}_2(\hat{\bf k}\cdot\hat{\bf r}_h)\simeq\frac{1}{2}(3 (\hat{\bf k}\cdot\hat{\bf r}_1)^m(\hat{\bf k}\cdot\hat{\bf r}_2)^{2-m}-1) 
 \end{equation}
 which is equivalent to writing,
  \begin{equation}
  \mathcal{L}_2(\hat{\bf k}\cdot\hat{\bf r}_h)\propto \mathcal{L}_1^m(\hat{\bf k}\cdot\hat{\bf r}_1)\mathcal{L}_1^{2-m}(\hat{\bf k}\cdot\hat{\bf r}_2) 
 \end{equation}
 for $0\leq m \leq 1$, where $\mathcal{L}_1(x)=x$. The obvious option would be to pick either $m=0$ or $m=1$, but note that under this approximation all range of possibilities are equally valid. Note that the option $m=0$ corresponds to $\mathcal{L}_2(\hat{\bf k}\cdot\hat{\bf r}_h) \rightarrow \mathcal{L}_2(\hat{\bf k}\cdot\hat{\bf r}_1)$, whereas option $m=1$ corresponds to $\mathcal{L}_2(\hat{\bf k}\cdot\hat{\bf r}_h)\rightarrow \mathcal{L}_1(\hat{\bf k}\cdot\hat{\bf r}_1) \mathcal{L}_1(\hat{\bf k}\cdot\hat{\bf r}_2)$.  In this work we opt for $m=0$ as it involves FT with Legendre polynomials of even order. For the hexadecapole the number of options increases as it involves a polynomial of 4th order. Among the possible expansions are $\mathcal{L}_4(\hat{\bf k}\cdot\hat{\bf r}_h)\rightarrow \mathcal{L}_4(\hat{\bf k}\cdot\hat{\bf r}_1)$, as used in \cite{Bianchietal:2015}, or $\mathcal{L}_4(\hat{\bf k}\cdot\hat{\bf r}_h) \rightarrow \mathcal{L}_2(\hat{\bf k}\cdot\hat{\bf r}_1)\mathcal{L}_2(\hat{\bf k}\cdot\hat{\bf r}_2)$, as used in \cite{Soccimarro:2015}. Note also the possibility involving polynomials of odd orders, $\mathcal{L}_4(\hat{\bf k}\cdot\hat{\bf r}_h)\rightarrow \mathcal{L}_3(\hat{\bf k}\cdot\hat{\bf r}_1) \mathcal{L}_1(\hat{\bf k}\cdot\hat{\bf r}_2)$. We do not intend to perform a detailed study  of the difference in signals and variances of these different expansions. In this paper for simplicity we choose, $\mathcal{L}_4(\hat{\bf k}\cdot\hat{\bf r}_h)\rightarrow \mathcal{L}_2(\hat{\bf k}\cdot\hat{\bf r}_1)\mathcal{L}_2(\hat{\bf k}\cdot\hat{\bf r}_2)$, as it involves the same type of FT as for the quadrupole, saving a significant amount of computational time. In this fashion the multipole estimators reads, 
\begin{eqnarray}
\label{eq:P0}P^{(0)}( k) &=&\int\frac{d\Omega_k}{4\pi}  |A_0( { k})|^2-P_{\rm noise}, \\
\label{eq:P2}P^{(2)}( k)&=& \frac{5}{2}\int\frac{d\Omega_k}{4\pi}  A_0( { k})\left[ 3A^*_2( { k})-A_0^*( { k}) \right], \\
\label{eq:P4}P^{(4)}( k)&=&\frac{9}{8}\int\frac{d\Omega_k}{4\pi} \{ 35A_2 [ A_2^*-2A_0^*]+3|A_0|^2 \}.
\end{eqnarray}
where,
\begin{equation}
A_n( { k})=\int d { r}\,(\hat { k}\cdot\hat { r})^n F( { r})e^{i { k}\cdot  { r}}.
\end{equation}
Under this approach, measuring the monopole, quadrupole, and hexadecapole requires to consider those cases with $n=0,\,2$. The case $n=0$ can be trivially computed using FT based algorithms, such as \textsc{fftw}.\footnote{Fastest Fourier Transform in the West: \href{http://fftw.org}{fftw.org}} The $n=2$ case can also be decomposed into 6 Fourier Transforms (FT) by expanding the scalar product between $ {\bf k}$ and $ {\bf r}$ and pulling the $k$-components outside the integral, as shown in eq. 10 of \cite{Bianchietal:2015}.  $P_{\rm noise}$ is the shot noise component, which under the Poisson assumption reads as the expression of Eq. \ref{eq:Pnoise}.

Unless stated otherwise, we perform the measurement of the power spectrum linearly binning $k$ in bins of $\Delta k=0.01\,\hompc$ up to $k_{\rm max}=0.32\,\hompc$, although not all the $k$-elements will be necessarily used in the final analysis. The resulting power spectrum multipoles for the combined CMASS+eBOSS LRG sample are displayed in  Fig.~\ref{fig:data_mocks}. We observe a significant mismatch between the amplitude of the mocks and data. This difference is caused by an early version of the mocks (with no completeness) being fitted to reproduce an early version of the data (with completeness). The normalisation of the data was initially set in such a way that the overall amplitude depended on the value of the overall completeness. As a consequence, when the completeness was applied in the final version of the mocks, mocks and data did not match. This mismatching only appears to be evident in Fourier space, but not in configuration space (see for example fig. 2 of \citealt{LRG_corr}). 
Therefore, this effect must correspond to a mismatch at scales of around $s\sim1-5\, \mathrm{Mpc}/h$ in configuration space. We conclude that this effect has no impact on the final covariance of the data. On the other hand, the overall normalisation of the data has no impact on the cosmological signal extracted, as it is appropriately modelled by the window function as we describe below. 

We account for the selection function due to the survey geometry and radial $n(z)$ dependence using the formalism described in previous works \citep{wilson_rapid_2017, beutler_clustering_2017}. We define the window selection function as the random pair-counts weighted by a $\ell$-order Legendre polynomial of the cosine of the angle to the LOS of each random object, 
\begin{equation}
W_\ell(s) = \frac{(2\ell+1)}{ I_{2}\alpha_{\rm ran}^{-2}} \sum_{i,\,j>i}^{N_{\rm ran}}\frac{w_{\rm tot}({\bf x}_i)w_{\rm tot}({\bf x}_j+{\bf s})}{2\pi s^2\Delta s}\mathcal{L}_\ell( {\bf \hat x}_{\rm los}\cdot \bf{\hat s}),
\label{eq:wl}
\end{equation}
where, following the same convention used for the power spectrum estimator, we assign $\bf{ x}_{\rm los}={\bf x}_1$. The pair-count is divided by the associated volume under a linear binning, $\propto s^2 \Delta s$. Note that the summation avoids pair-repetitions as it is performed only over $j>i$ and consequently the actual volume associated to those pairs within $s\pm\Delta s $ separation is $2\pi s^2\Delta s$.  Eq. \ref{eq:wl} is normalised in such a way that $\lim_{s\rightarrow0}W_0(s)=1$. One can impose this normalisation by dividing the function by its value in the first $s$-bin of $W_0(s)$. However, if the random catalogue is not sufficiently dense with respect to the typical small-scale variations induced by the selection function, one would propagate such variations in the normalisation of the window, that will eventually impact the measurement, in particular for $f\sigma_8$ or $b_1\sigma_8$, though the BAO peak position is insensitive to the overall normalisation factor. Similarly, the same problem appears when computing the factor $I_2$ when normalising the measured power spectrum in Eq. \ref{eq:P0}-\ref{eq:P4}. As suggested by \cite{de_mattia_integral_2019} we follow a consistent normalisation of both window and power spectrum by the same quantity, $I_{2}$ and therefore our final measurements are independent of this arbitrary choice. 
Note that since $I_2$ is associated to the densities of the galaxy catalogue, but Eq. \ref{eq:wl} is performed over the random catalogue, we need to include the factor $\alpha_{\rm ran}^{-2}$ in the normalisation. In Fig.~\ref{Fig:window} we show the shape of the window functions of Eq. \ref{eq:wl} for the survey geometry of the combined CMASS + eBOSS LRGs, for both NGC (solid lines) and SGC (dashed lines), where the different colours display different $\ell$-multipoles. 

In appendix \ref{sec:window} we explicitly write how the selection effect is included in the power spectrum model.

\section{Methodology}\label{sec:methodology}
In this paper we perform two parallel analyses: the analysis of the position of the BAO peak in the anisotropic power spectrum (hereafter BAO analysis), and on the RSD and Alcock-Paczynski effect using the full shape information in the power spectrum (hereafter Full Shape analysis or simply FS analysis).
\begin{itemize}
\item The BAO analysis consists of using a fixed and arbitrary template to compare the relative BAO peak positions in the power spectrum multipoles. Such analysis can be performed on both pre- and post-reconstruction catalogues.  The analysis performed on the reconstructed catalogue measurements has a higher probability of providing a larger significance detection, and consequently, smaller error-bars than the pre-reconstruction measurement. The BAO peak position along and across the LOS direction is then linked to the expansion history and angular diameter distance at the redshift-bin of the measurement.  
\item The FS analysis consists of a full modelling of the shape and amplitude of the power spectrum multipoles, taking into account non-linear dark matter effects, galaxy bias and RSD, and is only performed over the pre-reconstructed catalogues. In order to do so, we choose an underlying linear power spectrum template at fixed cosmological parameters and infer the scale dilations and the amplitudes of the power spectrum multipoles. With this we are able to infer not only the expansion history and angular diameter distance, but as well the logarithmic growth of structure times the fluctuations of the dark matter field filtered by a top-hat function of $8\,\mpcoh$, $f\sigma_8$. 
\end{itemize}

Unlike $\Lambda$CDM-model based analyses, the previously described FS and BAO analyses  do not guarantee a consistent relation between the expansion history and the angular diameter distance within a $\Lambda$CDM model. In this sense, our analysis goes beyond such assumption and can be used to actually test the validity of the model. 

Pre- and post-reconstruction catalogues are considered to contain independent, although correlated, cosmological information. In this fashion we maximise the amount of cosmological information if we combine them with the appropriate covariance. 
\subsection{Modelling the BAO signal}\label{sec:BAO}

We model the anisotropic power spectrum signal in order to measure the BAO peak position and marginalise over the broadband information. We take into account the BAO signal both in the radial- and transverse-to-LOS directions. Accordingly, we define the dilation scales across and along the LOS as,

\begin{eqnarray}
\label{eq:apara}\alpha_\parallel(z) &=& \frac{D_H(z)r_{\rm drag}^{\rm ref}}{D_H^{\rm ref}(z)r_{\rm drag}},\\
\label{eq:aperp}\alpha_\perp(z) &=& \frac{D_M(z)r_{\rm drag}^{\rm ref}}{D_M^{\rm ref}(z)r_{\rm drag}},
\end{eqnarray}
where $D_H\equiv c/H(z)$, $H(z)$ is the Hubble expansion parameter, $c$ the speed of light, $D_M(z)$ the comoving angular diameter distance at given redshift $z$,\footnote{The angular diameter distance, $D_A(z)$ and the comoving angular diameter distance are related by $D_M(z)=(1+z)D_A(z)$.} $r_{\rm drag}$ is the comoving sound horizon at $z=z_{\rm drag}$, where $z_{\rm drag}$ is the redshift at which the baryon-drag optical depth equals unity \citep{HuSugiyama1996}, and the $\rm ``ref"$ superscript stands for the values corresponding to the reference cosmology (in the standard approach this will be the fiducial cosmology, $\boldsymbol{\Theta}_{\rm fid}$). 

As the BAO peak position in the power spectrum monopole is affected by the reference cosmology chosen to convert redshifts into distance, as well as by the value of $r_{\rm drag}$ of this reference template, $r_{\rm drag}^{\rm ref}$, one can infer the shift in the expected BAO peak position with respect to the reference $\Lambda$CDM model and therefore infer the actual cosmology of the Universe.\footnote{In this paper these two reference cosmologies, the cosmology chosen to convert redshift into distance and the cosmology chosen for the model-template, are chosen to be the same for simplicity.} This measurement is known as an isotropic BAO measurement and is sensitive to the isotropic BAO distance $D_V$,
\begin{equation}
\frac{D_V(z)}{r_{\rm drag}}=\alpha_0\left( \left[\frac{D_M^{\rm ref}(z)}{r_{\rm drag}^{\rm ref}}\right]^2 \frac{D_H^{\rm ref}(z)z}{r_{\rm drag}^{\rm ref}} \right)^{1/3},
\label{eq:dv}
\end{equation}
where $c$ is the speed of light, and $\alpha_0=(\alpha_\perp^2\alpha_\parallel)^{1/3}$ is the isotropic BAO scale dilation. 
Additionally, we can also make a comparison of the BAO peak position in the radial direction relative to the transverse direction. Under the cosmological principle we assume that the Universe is isotropic and homogeneous and therefore the BAO should be a symmetric structure along all spacial directions. In this case, any excess in the relative BAO scales along and across the LOS must be due to the difference between the reference cosmology and the true cosmology of the Universe. This apparent anisotropy is known as the Alcock-Paczynski effect (hereafter AP effect; \citealt{AP}) and is parametrised as, 
\begin{equation}
F_{\rm AP}(z)=F^{-1}_\epsilon(z)D_M(z)^{\rm ref}/ D_H(z)^{\rm ref}
\end{equation}
where $F_\epsilon=\alpha_\parallel/\alpha_\perp$. $F_{\rm AP}$ is a relative parameter which does not depend on the sound horizon scale, $r_{\rm drag}$ and is therefore measured independently of CMB physics. Alternatively, other parametrisations also use the variable $\epsilon\equiv F_\epsilon^{1/3}-1$.

The AP effect distorts the true wave numbers of power spectrum: the observed wave number along and across the LOS, $k_\parallel$ and $k_\perp$, are related to the true wave numbers $k'_\parallel$ and $k'_\perp$ as, $k'_\parallel=k_\parallel/\alpha_\parallel$ and $k'_\perp=k_\perp/\alpha_\perp$, respectively. In terms of the absolute wave number $k'=\sqrt{{k'}_\parallel^2+{k'}_\perp^2}$, and the cosine of the angle between the wave number vector and the LOS direction, $\mu$, one can write the relations,
\begin{eqnarray}
\label{eq:AP1}k' &=& \frac{k}{\alpha_\perp}\left[1+\mu^2\left(\frac{1}{F_\epsilon^2}-1\right)\right]^{1/2},   \\
\label{eq:AP2}\mu'\ & = & \frac{\mu}{F_\epsilon}\left[ 1+\mu^2\left(\frac{1}{F_\epsilon^2}-1\right)\right]^{-1/2}.
\end{eqnarray}
We highlight that in Eq. \ref{eq:AP1} and \ref{eq:AP2} the $F_\epsilon$ and $\alpha_\perp$ dependence implies that the scale constraint comes exclusively from the BAO peak position. This is true for the BAO-type of analysis. However, for the FS type of analysis the scale constraints come partly from the BAO-shift and partly from the modification of the shape of the smoothed power spectrum. Since this shape is close to be a power law in the FS range of analysis, $0.02<k\,[\hompc]<0.15$, most of the scale constraint will effectively come from the BAO-shift. However, analysis of next generation data will have to deal consistently with these two types of re-scalings in order to obtain an accurate interpretation of cosmology data. 

In order to model the BAO peak position in a $\mu$-dependent power spectrum we follow the model proposed by \cite{beutler_clustering_2017},
\begin{eqnarray}
\nonumber P(k,\mu)&=&B(1+R\beta\mu^2)^2P_{\rm lin}(k)\left\{ 1+\left[\mathcal{O}_{\rm lin}(k)-1  \right] \right. \\
\label{eq:baolin}&\times& \left. e^{-\frac{1}{2}k^2(\mu^2\Sigma_{\parallel}^2+(1-\mu^2)\Sigma_{\perp})} \right\},
\end{eqnarray}
where the dark matter linear power spectrum $P_{\rm lin}(k)$ is enhanced with the Kaiser factor, $B(1+R\beta\mu^2)^2$, where $B$ is a free parameter which under certain conditions could be interpreted as the linear bias squared, $b_1^2$, $\beta$ is the redshift space distortion parameter and is also treated as free and nuisance parameter in this analysis.\footnote{We do not attempt any physical interpretation of $\beta$ as the ratio of the logarithmic growth of structure $f$ and the linear bias parameter $b_1$.} $R$ is a parameter which stands for the redshift-space distortion suppression due to reconstruction. In this analysis, it is fixed to $R=1$ for pre-reconstructed catalogues and to $R=1-\exp(-k^2\Sigma_s^2/2)$, where $\Sigma_s$ is the smoothing scale used during the reconstruction process. The $\mathcal{O}_{\rm lin}$ is the linear BAO template defined as $\mathcal{O}_{\rm lin}\equiv P_{\rm lin}/P_{\rm lin}^{(\rm sm)}$ where $P_{\rm lin}^{(\rm sm)}$ is a {\it smoothed} power spectrum with no BAO signal. In this paper we infer $P^{\rm (sm)}_{\rm lin}$ following the methodology described by \cite{Kirkby13} where the BAO peak in configuration space is replaced by a smoothed non-BAO template. Other approaches such as the one by \cite{eisenstein_baryonic_1998} are also possible producing equivalent results for the given precision of the BOSS and eBOSS data. The parameters $\Sigma_{\parallel}$ and $\Sigma_{\perp}$ describe the smoothing of the BAO along and across the LOS due to non-linear bulk motions. These parameters can be estimated for the pre-reconstructed catalogues as, $\Sigma_\perp=10.4D(z)\sigma_8$, where $D(z)$ is the linear growth factor,  and $\Sigma_\perp=(1+f)\Sigma_\parallel$ \citep{SeoEis07}, where $\Sigma_\parallel>\Sigma_\perp$ due to RSD induced by the logarithmic growth factor $f$. Such damping terms reduce the amplitude of BAO oscillations of the linear power spectrum template of $\mathcal{O}_{\rm lin}$, and make the BAO feature less prominent and consequently more difficult to detect. For the post-reconstruction catalogues the non-linear bulk motions are removed above a certain smoothing scale and therefore the effective values of $\Sigma_\parallel$ and $\Sigma_\perp$ are expected to be reduced. In order to determine $\Sigma_\parallel$ and $\Sigma_\perp$ we fit them as free parameters to the mean of the \textsc{EZmocks},\footnote{When analysing other type of mocks, such \textsc{Nseries} or \textsc{OuterRim}-derived mocks, we set $\Sigma_\parallel$ and $\Sigma_\perp$ to the best-fitting values of the mean of these mocks, respectively} and use these best-fitting values when determining the BAO peak of the individual mocks, and consequently on the data as well. We choose the \textsc{EZmocks} to determine the best-fitting values of $\Sigma_\parallel$ and $\Sigma_\perp$ for the data, as these are the only mocks with a clustering signal very similar to the data. In section \S\ref{sec:results} we check that allowing for certain freedom on the values of these parameters does not impact the final BAO results significantly. 

We integrate the template of Eq. \ref{eq:baolin} weighting it by the Legendre polynomials of $\mu$, $\mathcal{L}_\ell$ and add a number of broadband nuisance parameters to get the $\ell-$multipole of the power spectrum, 
\begin{equation}
\label{eq:baoaniso}P^{(\ell)}(k)=\frac{2\ell+1}{2}\int_{-1}^{1}d\mu \mathcal{L}_{\ell}(\mu)P[k'(k,\mu),\mu'(\mu)] + \sum_{i=1}^n A^{(\ell)}_ik^{2-i},
\end{equation}
where $A_i$ are the parameters which allow us to marginalise over non-linear effects of the broadband. Note that the non-linear part of the broadband is not assumed to be dependent of the AP effect in this model, unlike the FS type of templates. For the BAO fits we take as the  standard analysis $n=3$ as the broadband parameter maximum order. We have checked that this order is a good compromise between speed and precision, given the statistical error bars of the sample. 

We fit the data by considering independent NGC and SGC broadband and bias parameters, both on the power spectra monopole and quadrupole. Thus, in the standard fit we consider 2 physical parameters, $\{\alpha_\parallel,\alpha_\perp \}$ and 15 nuisance parameters,  $\{ \beta, B_{\rm N}, {A_i^{(0)}}_{\rm N},  {A_i^{(2)}}_{\rm N}, B_{\rm S}, {A_i^{(0)}}_{\rm S},  {A_i^{(2)}}_{\rm S}  \}$, where $i=1,\ldots,n$, and ${\rm N},\,{\rm S}$ stand for NGC and SGC, respectively. 

Alternatively to the template described above we also check the performance of the following isotropic template \citep{gil-marin_clustering_2016-1},

\begin{equation}
P^{(\ell)}(k)=P^{(\ell)}_{\rm sm}(k)\left\{1+\left[\mathcal{O}_{\rm lin}(k/\alpha_\ell)-1  \right]e^{-\frac{1}{2}k^2\Sigma_{{\rm nl}\, \ell}^2} \right\},
\label{eq:iso}
\end{equation}
where,
\begin{equation}
P^{(\ell)}_{\rm sm}(k)=B^{(\ell)}P^{\rm (sm)}_{\rm lin}(k)+\sum_{i=1}^n A^{(\ell)}_ik^{2-i}.
\label{eq:Psm}
\end{equation}
For $\ell=0$ one fits the monopole, $P^{(0)}$ in order to constrain $\alpha_0=\alpha_\parallel^{1/3}\alpha_\perp^{2/3}$ as in Eq. \ref{eq:dv}. The first anisotropic moment, $\ell=2$ is not the quadrupole but a linear combination of monopole, and quadrupole, the $\mu^2$-moment, which constrains the variable $a_2=\alpha_\parallel^{3/5} \alpha_\perp^{2/5}$ \citep{Rossetal:2015}. In this fashion one also can extend this to the next anisotropic moment for $\ell=4$, the $\mu^4$-moment, which constrains  $a_4=\alpha_\parallel^{5/7} \alpha_\perp^{2/7}$. Such moments are defined such that,
\begin{eqnarray}
P^{(\mu^0)}&=&P^{(0)},\\
\label{eq:mu2moment} P^{(\mu^2)}&=&P^{(0)}+2/5P^{(2)}, \\
P^{(\mu^4)}&=&P^{(0)}+4/7P^{(2)}+8/64P^{(4)}.
\end{eqnarray}
Typically, most of the BAO information is contained by the two first moments, and by adding $\mu^4$ one does not gain much extra information (see fig. 3 of \citealt{Rossetal:2015}).

The main difference between the above isotropic template and the anisotropic template of Eqs. \ref{eq:baolin} and  \ref{eq:baoaniso} is the effect of the BAO damping parameter $\Sigma_{\rm nl}$. In the anisotropic template the exponential argument has an explicit $\mu$-dependence through the damping terms along and across the LOS, $\Sigma_\perp$ and $\Sigma_\parallel$. In this case, the monopole and quadrupole contain an effective weighted-averaged damping parameter, $\Sigma_0$,  $\Sigma_2$ and $\Sigma_4$. The main advantage of the isotropic template is that {\it i)} it is faster to evaluate, as it does not require an integration over the LOS, and {\it ii)} the broadband parameters are in linear combination and therefore an analytical solver can be applied without the need of running an Monte Carlo Markov Chain (\textsc{mcmc}) solver to explore the likelihood. The drawback is that the BAO damping is not as accurately described as in the anisotropic BAO template, especially for the anisotropic signal.

In \S\ref{sec:results} we test this effect on the mean of the mocks and in Table~\ref{table:BAOresults} we present an alternative analysis using this template. We show that the differences observed among these templates are sufficiently small to not be relevant for the precision of the measurements of this paper. 

\subsection{Modelling the redshift space distortions and galaxy bias}\label{sec:RSD}

The FS analysis model employed  to describe the power spectrum multipoles is the same to the one used in previous analyses of the BOSS survey for the redshift range $0.15<z<0.70$ (\citealt{BispectrumDR11}, \citealt{gil-marin_clustering_2016}) and for DR14 eBOSS quasars $0.8\leq z \leq 1.2$ \citep{gil-marin_clustering_2018}, so we briefly present it here to avoid repetition. 

\subsubsection{Galaxy bias model}
We follow the Eulerian non-linear bias model presented by \cite{McDonald_Roy:2009}. The model consists of four bias parameters: the linear galaxy bias $b_1$, the non-linear galaxy bias $b_2$, and two non-local galaxy bias parameters, $b_{s2}$ and $b_{3{\rm nl}}$.  We always consider the local biases $b_1$ and $b_2$ as nuisance and free parameters of the model. Unless stated otherwise, the non-local bias parameters are constrained by assuming the local bias relations from Lagrangian space, $b_{s^2}=-4/7\,(b_1-1)$ \citep{Baldaufetal:2012} and $b_{3{\rm nl}}=32/315\,(b_1-1)$ \citep{Saitoetal:2014}. 

\subsubsection{Real space spectra}
The real space dark matter auto- and cross-power spectra, density-density, $P_{\delta\delta}$, density-velocity, $P_{\delta\theta}$ and velocity-velocity $P_{\theta\theta}$ are given by the 2-loop re-summation perturbation theory. In particular we follow the approach described in \cite{GM12} (hereafter GM12) where these moments are given by, 
\begin{equation}
P_{ij}(k)= {\mathcal{N}_{ij}}^{2}(k)\left [P_{\rm lin}(k)+P_{ij}^{1L}(k)+P_{ij}^{2L}(k) \right]
\label{eq:2loop}
\end{equation}
where $i,\,j=\delta\, {\rm or}\, \theta$, $\mathcal{N}_{ij}(k)$ is the resummed propagator of order 2 (given by eq. B39 of GM12), $P_{ij}^{nL}(k)$ is the full $n$-loop coupling (see eq. A5 for $n=1$ and eq. B29 for $n=2$, of GM12). These moments accurately describe the clustering of dark matter up to $k\simeq 0.15$ at $z=0.5$; $k\simeq 0.20$ at $z=1.0$; and $k\simeq 0.30$ at $z=1.5$ (see fig. 2 of GM12). Using the expressions given above, we express the galaxy density-density, density-velocity, and velocity-velocity power spectra as \citep{beutler_clustering_2014}, 
\begin{eqnarray}
\nonumber P_{g,\,\delta\delta}(k)&=&b_1^2P_{\delta\delta}(k)+2b_2b_1P_{ b2,\,\delta}(k)+2b_{s2}b_1P_{bs2,\,\delta}(k)+\\
\nonumber&\quad&b_2^2P_{b22}+2b_2b_{s2}P_{b2s2}(k)+b^2_{s2}P_{bs22}(k)+\\
&\quad& 2b_1b_{3\rm nl}\sigma_3^2(k)P_{\rm lin}(k) \\
\nonumber P_{g,\,\delta\theta}(k)&=&b_1P_{\delta\theta}(k)+b_2P_{ b2,\,\theta}(k)+b_{s2}P_{bs2,\,\theta}(k)+\\
&\quad&b_{3\rm nl}\sigma_3^2(k)P_{\rm lin}(k)\\
P_{g,\,\theta\theta}(k)&=&P_{\theta\theta}(k)
\label{eq:Pthetatheta}
\end{eqnarray}
where no velocity bias is being assumed. The bias 1-loop correction, $P_{bX}$ and $\sigma_3^2$ terms can be found in eq. B2- B7 of \cite{BispectrumDR11}. Note that there is an implicit scaling $\propto\sigma_8^2$ on all the terms which depend on $P_{\rm lin}$ or  $\sigma_3^2$; a scaling $\propto\sigma_8^4$ on the terms $P_{ij}^{1L}$ and on the bias terms, $P_X$, which are all 1-loop corrections; and finally a scaling $\propto\sigma_8^6$ on $P_{ij}^{2L}$. The propagator $\mathcal{N}_{ij}$ also depends on $\propto\sigma_8^2$ and $\propto\sigma_8^4$ through the ratios of $P_{ij}^{(13)}/P_{\rm lin}$ and $P_{ij}^{(15)}/P_{\rm lin}$, respectively.

\subsubsection{Redshift Space Distortions}
We include the effect of RSD following the approach proposed by \cite{SC04} and extended by \cite{TNS}. Thus, we write the redshift space galaxy power spectrum as, 

\begin{eqnarray}
\nonumber P^{(s)}_{g}(k,\,\mu)&=&D_{\rm FoG}(k,\,\mu)\left[P_{g,\,\delta\delta}(k)+2f\mu^2P_{g,\,\delta\theta}(k) +\right. \\
\nonumber&\quad&f^2\mu^4P_{\theta\theta}(k)+b_1^3A^{\rm TNS}(k,\mu,f/b_1)+\\
&\quad&\left. b_1^4B^{\rm TNS}(k,\,\mu,f/b_1) \right].
\label{eq:TNSterms}
\end{eqnarray}
The galaxy real space quantities $P_{g\,ij}$ are computed using the prescriptions described above assuming a fixed $P_{\rm lin}$ template at the reference cosmology computed using \textsc{camb} \citep{lewis_efficient_2000}. The power spectrum multipoles encode the coherent velocity field through the redshift space displacement and the logarithmic growth of structure parameter. The effect of this parameter is to increase the clustering along the LOS with respect to the transverse direction, boosting the amplitude of the isotropic power spectrum and generating an anisotropic component. The $D_{\rm FoG}$ term accounts for the Finger-of-God (hereafter FoG) effect along the LOS direction. The physical origin of this term is the velocity dispersion of the satellite galaxies inside the host dark matter haloes, which damps the power spectrum at small scales. In this paper we test both Lorentzian and Gaussian ans\"atze, 
\begin{eqnarray}
D^{\rm Lor}_{\rm FoG}(k,\mu;\sigma_P)&=&(1+[k\mu\sigma_P]^2/2)^{-2},\\
D^{\rm Gau}_{\rm FoG}(k,\mu;\sigma_P)&=&\exp{(-[k\mu\sigma_P]^2/2)},
\label{eq:GaussFog}
\end{eqnarray}
where $\sigma_P$ is a free parameter to marginalise over. We assume $D_{\rm FoG}^{\rm Lor}$ as the standard modelling approach. The $A^{\rm TNS}$ and $B^{\rm TNS}$ are second order corrections and their form is given by eq. A3 and A4 of \cite{TNS}. 
Finally, the AP effect is added in the same way as in Eq. \ref{eq:baoaniso} when computing the multipoles,
\begin{equation}
\label{eq:rsdaniso}P_g^{(\ell)}(k)=\frac{2\ell+1}{2\alpha_\parallel\alpha_\perp^2}\int_{-1}^{1}d\mu \mathcal{L}_{\ell}(\mu)P_g^{(s)}[k'(k,\mu),\mu'(\mu)]
\end{equation}
where $k'(k,\mu)$ and $\mu'(\mu)$ are given by Eq. \ref{eq:AP1} and \ref{eq:AP2}, respectively. In the above Eq. the term $1/(\alpha_\parallel\alpha_\perp^2)$ accounts for the volume rescaling caused by the differences in cosmology. This is an approximation as the actual volume rescaling should also include a pre-factor $(r^{\rm ref}_{\rm drag}/r_{\rm drag})^3$. In practice we account for such difference by assuming that the reference cosmology, $r^{\rm ref}_{\rm drag}$ should be close to the actual value, $r_{\rm drag}$, measured by Planck with $\sim0.02\%$ precision. We test the impact of such approximation in \S\ref{sec:systematics}, where templates of cosmologies with different values of $r^{\rm ref}_{\rm drag}$ are used to measure the actual cosmology of {\it N}-body galaxy mocks.

We also consider that the shot noise contribution to the power spectrum monopole may differ from the Poisson sampling prediction. We parametrise this potential deviation through a free parameter, $A_{\rm noise}$, which modifies the amplitude of shot noise, but without introducing any scale dependence. By default our measured power spectrum monopole has a fixed Poissonian shot noise contribution subtracted, $\hat{P}^{(0)}=P^{(0)}_{\rm meas.}-P_{\rm Poisson}$, whereas the higher order multipoles do not, $\hat{P}^{(\ell>0)}=P^{(\ell>0)}_{\rm meas.}$.Thus, from Eq. \ref{eq:rsdaniso} we add the non-Poissonian contribution on our model in the following way,
\begin{equation}
\label{eq:noise}P_g^{(0)}(k)  \rightarrow P_g^{(0)}(k) +P_{\rm Poisson}\left[ \frac{A_{\rm noise}}{\alpha_\parallel\alpha_\perp^2}-1 \right]
\end{equation}
where the factor $\alpha_\parallel\alpha_\perp^2$ accounts for the change in density as a result of the isotropic dilation. Note that the $A_{\rm noise}=\alpha_\parallel\alpha_\perp^2$ correspond to the exact Poissonian case, whereas $A_{\rm noise}>\alpha_\parallel\alpha_\perp^2$ is an over-Poissonian shot noise and $A_{\rm noise}<\alpha_\parallel\alpha_\perp^2$ a sub-Poissonian shot noise. Also note that the higher order multipoles are slightly affected by this parameter $A_{\rm noise}$ through the window function coupling.  $P_{\rm Poisson}$ is computed as,
\begin{eqnarray}
\label{eq:Pnoise} P_{\rm Poisson}&=&\sum_{i-{\rm gal}}w^2_{\rm FKP}({\bf r}_i)w^2_{\rm col}({\bf r}_i)w^2_{\rm sys}({\bf r}_i)\\
 &+&\alpha^2\sum_{i-{\rm ran}}w^2_{\rm FKP}({\bf r}_i)w^2_{\rm col}({\bf r}_i)w^2_{\rm sys}({\bf r}_i),
\end{eqnarray}
under the assumption that all the collided pairs do contribute to shot noise (all collided pairs are not true pairs). For the CMASS+eBOSS LRG sample the shot noise values are $13,071(\,\mpcoh)^3$ and $12,622\,(\mpcoh)^3$ for NGC and SGC, respectively.

\subsection{Parameter inference}\label{sec:parameter_inference}

We define the likelihood distribution, $\mathcal{L}_G$, of the data vector of parameters, $p$, as a multi-variate Gaussian distribution,
\begin{equation}
\mathcal{L}_G(p)\propto e^{{-\chi^2(p)/2}},
\end{equation}
where $\chi^2(p)$ is defined as,
\begin{equation}
\chi^2(p)\equiv { D_p} C^{-1}  { D_p}^T,
\end{equation}
where $D_p$ represent the difference between the data and the model for a given $p$-parameters, and $C$ is the covariance matrix of the data vector, which we approximate to be independent of the $p$-set of parameters and the same for different realisations of the Universe. 

In this paper we infer the covariance matrix from 1000 realisations of the \textsc{EZmocks} \citep{zhao20a}.  Due to the finite number of mock catalogues when estimating the covariance, we expect a noise term arising when inverting the covariance. We apply the corrections described in \cite{Hartlap07} which for the current sample is $\sim 6\%$ factor in the $\chi^2$ values for BAO analysis and $4\%$ for FS when the hexadecapole is used.  Extra corrections, such as the ones described in \cite{Percival14}, have a minor contribution to the final errors. They represent a $2\%$ and $1.4\%$ increase for the BAO and FS analyses, respectively. We include them only on the last stage of the analysis along with other systematic contributions.

In order to explore the full likelihood surface of a given set of parameters, we run Markov-chains (\textsc{mcmc}-chains). We use \textsc{Brass}\footnote{Bao and Rsd Algorithm for Spectroscopic Surveys, \href{https://github.com/hectorgil/brass}{github.com/hectorgil/Brass}.} based on the Metropolis-Hasting algorithm with a proposal covariance and ensure its convergence performing the Gelman-Rubin convergence test, $R-1<0.005$, on each parameter. We apply the flat priors listed in Table~\ref{table:generalpriors} otherwise stated.

\begin{table}
\caption{Flat prior ranges on the parameters used in the \textsc{mcmc} analyses. The priors on $\alpha_\parallel$ and $\alpha_\perp$ applies both to FS and BAO analyses. $\beta$, $B$ and $A_i$ priors correspond to BAO type of analyses; whereas, $f$, $b_1$, $b_2$, $\sigma_P$ and $A_{\rm noise}$ correspond to FS type. For the $A_{\rm noise}$ term we try two type of priors, as we describe in \S\ref{sec:results}. }
\begin{center}
\begin{tabular}{|c|c|}
\hline
\hline
Parameter & flat-prior range \\
\hline
$\alpha_{\rm \parallel}$ & $[0.5,\,1.5]$ \\
$\alpha_{\rm \perp}$ & $[0.5,\,1.5]$ \\
\hline
$\beta$ & [0,\,,30]\\
$B$ & $[0,\,20]$ \\
$A_i\times 10^{-3}[(\mpcoh)^{5-i}]$ & [-20,\,+20]\\
\hline
$f$ & $[0,\,10]$ \\
$b_1$ & $[0,\,30]$ \\
$b_2$ & $[-10,\,10]$ \\
$\sigma_P\,[\mpcoh]$ & $[0,\,20]$ \\
$A_{\rm noise}$ & $[-5,\,5]$ or $[0.5,\,1.5]$ \\

\hline
\end{tabular}
\end{center}
\label{table:generalpriors}
\end{table}%

For the FS type of fit, we let free the cosmological parameters, $\{ \alpha_\parallel,\, \alpha_\perp,\, f\}$ and the galaxy bias parameters, $\{ b_1, b_2, A_{\rm noise}, \sigma_{\rm FoG}\}$ which we treat differently for NGC and SGC; in total 11 free parameters. $\sigma_8$ is kept fixed to its fiducial value during the likelihood exploration. Then, $f$ and $b_1$ are re-scaled by a fixed $\sigma_8$ value eventually just reporting $f\sigma_8$ and $b_1\sigma_8$. The fixed $\sigma_8$ value is not just the one obtained from filtering the linear power spectrum with a top-hat function at $8\,\mpcoh$, but we include an additional correction due to the isotropic BAO-shift between the template and the data,
\begin{equation}
\sigma_8^2(\alpha_{0})\equiv \sigma_8^2=\frac{1}{\alpha_0^3}\int_{\rm 0}^{\infty} dk k^2 P_{\rm lin}(k/\alpha_0) W_{\rm TH}^2(s_8 k)
\label{eq:sigma8res}
\end{equation}
where the smoothing scale is set to $s_8=8\,\mpcoh$ and $W_{\rm TH}$ is the FT of the top-hat function. $\alpha_{0}$ is inferred from the best-fitting parameters $\alpha_\parallel$ and $\alpha_\perp$ on the same (pre-reconstructed) catalogue. Note that since the integration limits are unaffected by the change of variables $q\equiv k/\alpha_0$, one could write Eq. \ref{eq:sigma8res} as the usual $\sigma_8$ expression just rescaling $s_8$ by $\alpha_0$. This is an alternative approach to the recently proposed $\sigma_{12}$-parametrisation \citep{Sanchez:2020}, where the smoothing scale is set to $12\, {\rm Mpc}$ instead of $8\, \mpcoh$, in order to obtain growth of structure measurements in a template-independent way. We later test in \S\ref{sec:systematicbudget} how the Eq. \ref{eq:sigma8res} re-scaling makes the $f\sigma_8$ variable stable under aggressive changes of the reference cosmology. Note that for those templates whose $\alpha_0$ is sufficiently close to unity, this correction has a negligible effect, which is the reason why it is not usually included in the other FS-analysis. Also, one should apply this re-scaling iteratively as $\alpha_0$ changes within the \textsc{mcmc} chain (or during the likelihood exploration), to properly account for the cross-correlation coefficients between the rescaled $f\sigma_8$ and $\alpha_0$ (or $f\sigma_8$ and $\alpha_\parallel$, $\alpha_\perp$ in this case). However, we have found that the shape of the whole likelihood barely changes with respect to the case of applying a global $\sigma_8$ rescaling based on the mean inferred value of $\alpha_0$. This simplifies the treatment of our data and also opens the possibility of rescaling other datasets based only on their Gaussian likelihoods.  

Since $\sigma_8$ is very degenerate with $f$ and $b_1$ this is equivalent to treat the terms $f\sigma_8^{n+1}$ in the Eqs. \ref{eq:2loop}-\ref{eq:Pthetatheta}, as two independent parameters:\footnote{In these Eqs. $\sigma_8$ is not explicitly written, but it is hidden within the linear power spectra,  $P_{\rm lin}\propto\sigma_8^2$.} $f\sigma_8$ and $\sigma_8^n$, where $f\sigma_8$ is freely fit, whereas $\sigma_8^n$ is kept fixed. For large scales $n=0$, so this approach is exact. At smaller scales $n>0$ terms arise, but the systematic effect of fixing this part to a constant is very small. We have checked that varying $\sigma_8$ on the $\sigma_8^n$ terms by 15\%, only shifts $f\sigma_8$ by 0.2\%. In section \S\ref{sec:results} we present a fit to the data where both $f$ and $\sigma_8$ are varied freely and we show how this has no effect on the final results, although the convergence time for such runs is larger.

For the standard BAO case we apply Eq. \ref{eq:baoaniso} and leave free $\{ \alpha_\parallel,\, \alpha_\perp,\beta\}$ and the broadband parameters $\{B,A_i^{(\ell)} \}$, which we fit separately for NGC and SGC. This corresponds to 17 free parameters. In some cases we also leave the damping terms, $\Sigma_\parallel$ and $\Sigma_\perp$, free, and treat them as independent. 

For both BAO and FS cases the covariances from NGC and SGC are drawn from two independent sets of mocks and are assumed to be fully independent, as these are two disconnected patches of the Universe. In this fashion the total likelihood is just the product of NGC and SGC likelihoods: $\mathcal{L}=\mathcal{L}_{\rm NGC}\times\mathcal{L}_{\rm SGC}$. We expect that only for very large modes ($k$ much smaller than $0.02\, \mpcoh$) this assumption loses validity . 

In this paper we report the mean of the \textsc{mcmc} chain when converged, $R-1<0.005$, except for the burn-in part which we discard (the first $10^4$ steps), and report its {\it rms} as the $1\sigma$ error. This matches the 68\% confident level in case of having a Gaussian distribution. For the mocks we run 6 independent sub-chains where after convergence we concatenate and treat as a single chain when calculating the mean and {\it rms}. We also test that running different set of chains on the same dataset report the same values within the statistical precision required, which indicates that the chain noise is below the statistical precision of the sample. In Appendix~\ref{sec:gauss} we show how the contours drawn from the \textsc{mcmc} chain of the data are in very good agreement with the inferred Gaussian contours. 

The isotropic BAO template described by Eq. \ref{eq:iso} and \ref{eq:Psm} can be solved analytically for most of its parameters using the least squares method. Given a fixed $\alpha_\parallel$, $\alpha_\perp$ and $\Sigma_{{\rm nl},\ell}$, the rest of variables,  $B$ and $A_i^{(\ell)}$ can be solved analytically so a full \textsc{mcmc} run is not required.
One therefore only needs to perform subsequent fits changing $\alpha_\parallel$, $\alpha_\perp$ and $\Sigma_{{\rm nl},\ell}$ within a fixed array, in order to resolve the likelihood shape, and then interpolate to find the best-fit and its error.

\section{Results}\label{sec:results}
In this section we describe the results obtained when applying the BAO  and FS pipeline described in the previous section. We perform these two analyses separately, and later in \S\ref{sec:consensus} we discuss how to combine them. The error-bars reported in this section only contain the statistical error budget. Later in \S\ref{sec:systematics} we discuss qualitatively and quantitatively the systematic error budget of such approaches. 

\subsection{Baryon Acoustic Oscillation analysis}
\begin{figure*}
\includegraphics[scale=0.3]{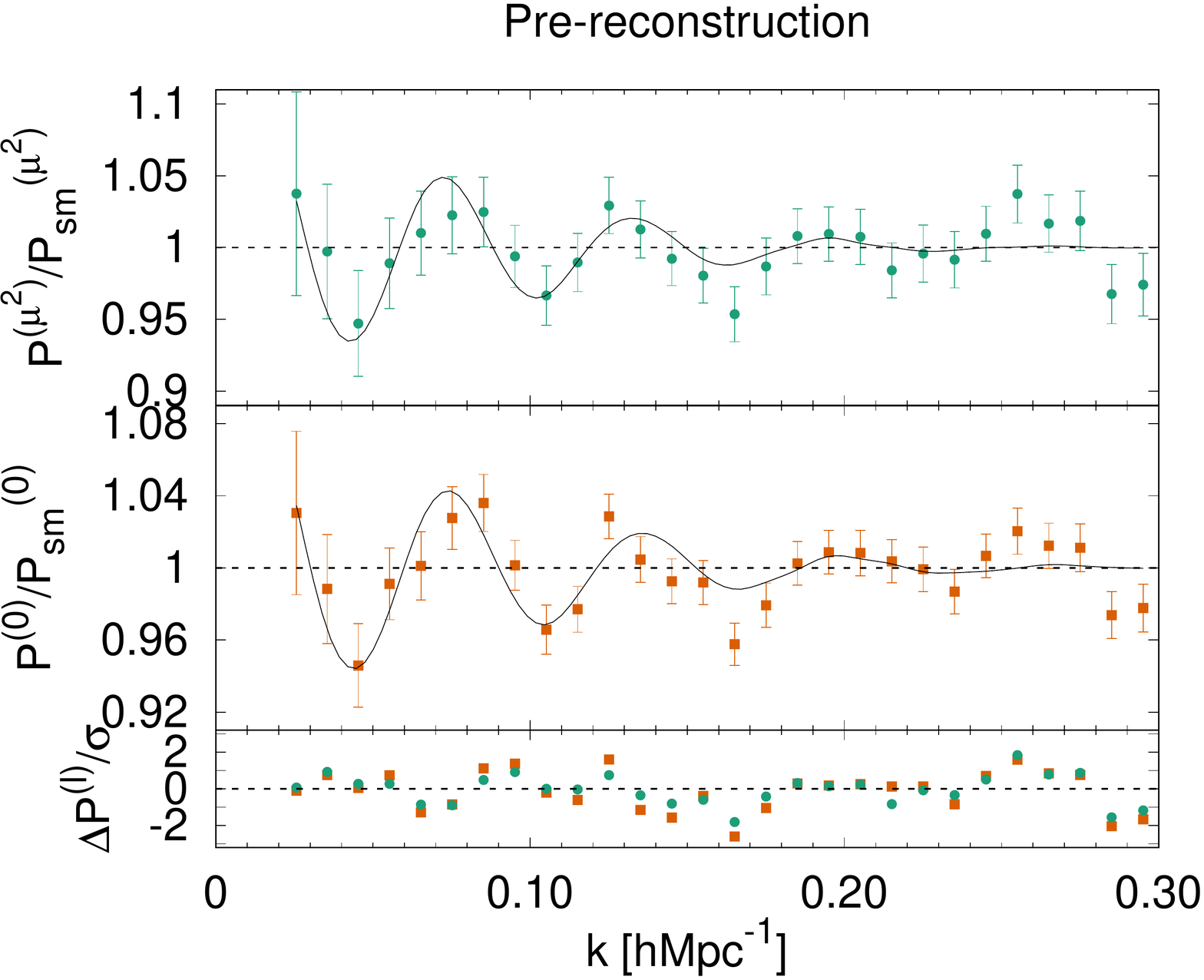}
\includegraphics[scale=0.3]{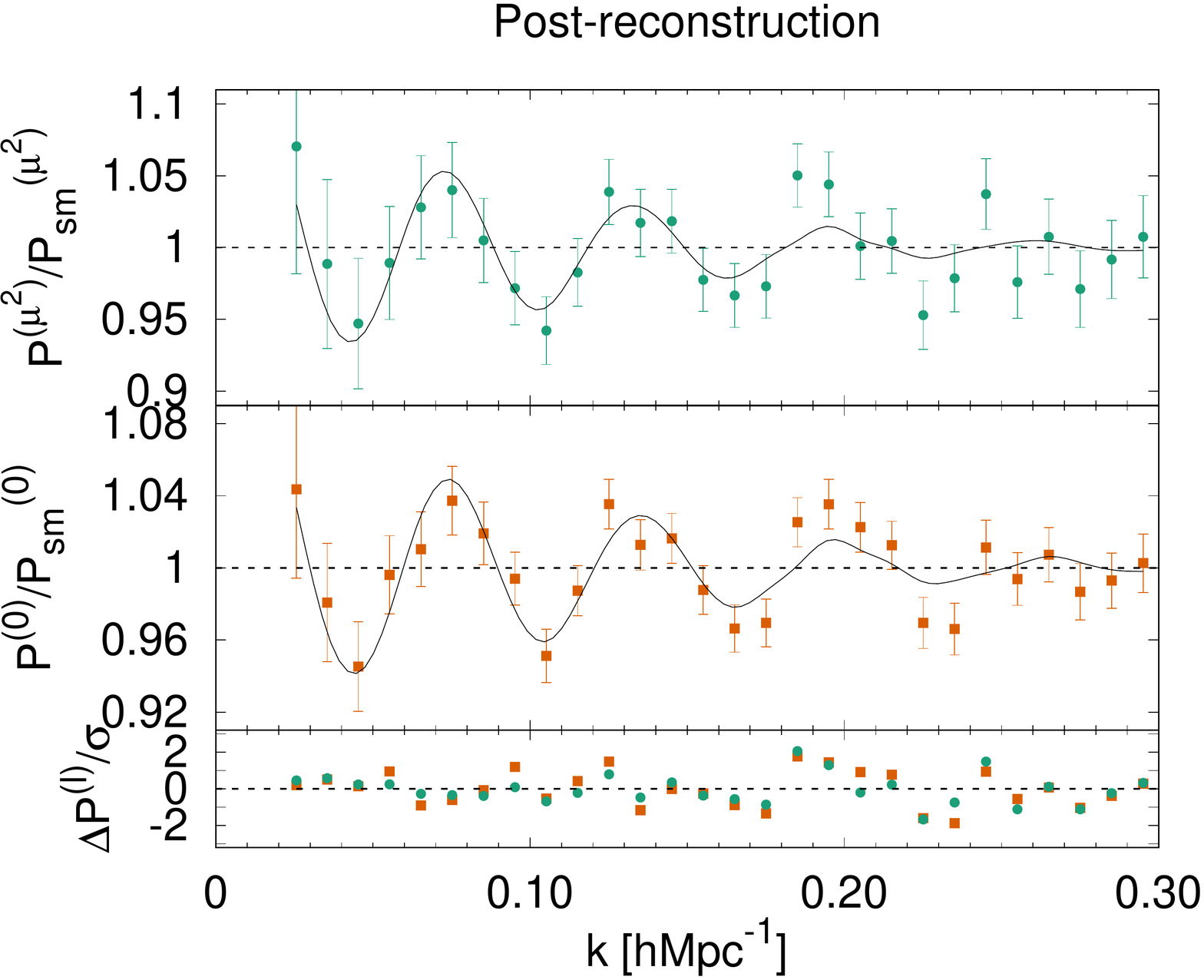}
\caption{DR16 CMASS+eBOSS LRG power spectrum measurements for the pre- (left panel) and post-reconstructed catalogue (right panel). The orange points display the power spectrum monopole and the green points the $\mu^2$-moment (see Eq. \ref{eq:mu2moment} for definition). The associated errors are drawn from the covariance of 1000 mocks and the black solid line represent the best-fitting solution (quoted in Table~\ref{table:BAOresults} using the anisotropic templated at the fixed values of $\Sigma_\parallel=7.0\,\mpcoh$ and $\Sigma_\perp=2.0\,\mpcoh$ for post-recon and  $\Sigma_\parallel=9.4\,\mpcoh$ and $\Sigma_\perp=4.8\,\mpcoh$ for pre-recon). The bottom sub-panels show the difference between model and measurement divided by the 1-$\sigma$ errors.}
\label{fig:BAOresults}
\end{figure*}

Fig.~\ref{fig:BAOresults} displays the BAO oscillatory features measured from the CMASS+eBOSS LRG data with respect to the broadband, for the isotropic signal, in orange symbols, and the anisotropic $\mu^2$-moment, in green symbols. The black solid lines represent the best-fit and the lower panel the model-data deviations in units of statistical $1\sigma$-error. The left panel displays the pre-reconstructed results and the right panel the post-reconstructed results. Reconstruction enhances significantly the BAO signal both in the isotropic and anisotropic power spectrum signal. Note that the actual BAO analysis is performed on the monopole and quadrupole, although we visually report the $\mu^2$-moment, as defined by Eq. \ref{eq:mu2moment} instead of the quadrupole, as the BAO feature is more evident there. 

Table~\ref{table:BAOresults} presents the main results from the BAO analysis of the data in terms of the scaling parameters,  $\alpha_\parallel$ and $\alpha_\perp$. We perform the BAO analysis keeping the $\Sigma_\parallel$ and $\Sigma_\perp$ variables fixed at their best-fitting values on the mean of the pre- and post-reconstructed mocks. These values are $\Sigma_\parallel=9.4\,\mpcoh$ and $\Sigma_\perp=4.8\,\mpcoh$ for the pre-reconstructed and $\Sigma_\parallel=7.0\,\mpcoh$ and $\Sigma_\perp=2.0\,\mpcoh$ for the post-reconstructed catalogues.\footnote{When the reference template is modified, these values are accordingly changed.} The first two rows of Table~\ref{table:BAOresults} report the BAO analysis on the pre- and post-reconstructed data in the Fourier space (matching the performance displayed by Fig.~\ref{fig:BAOresults}) and in configuration space of the same dataset (presented in \citealt{LRG_corr}). Along with those the consensus between Fourier and configuration space is also presented. The technique used to infer this value is described later in \S\ref{sec:consensus}.  The rest of the rows represent the values obtained from the pre- or post-reconstructed analysis on Fourier space with variations of the standard pipeline analysis, to show the sensitivity of the results under certain assumptions. Among these cases we present analyses when: NGC and SGC are the only-fitted regions, ignoring the effect of the selection function in the modelling (no-mask case), turning off the systematic and collision weights on the data (no-$w_{\rm sys}w_{\rm col}$), using the isotropic template of Eq. \ref{eq:iso} with 3- (Isotropic template) and 5-parameter broadband (Isotropic template order-5), using the anisotropic template of Eq. \ref{eq:baolin} with 5 parameters (Order-5), allowing $\Sigma_\parallel$ and $\Sigma_\perp$ to be free parameters ($\Sigma_{\parallel,\,\perp}$ Free ), or free but with a Gaussian prior, $\bar{x}\pm\sigma_x$,\footnote{Here $\bar{x}$ and $\sigma_x^2$ represent the mean and the variance, respectively, of the normal distribution used as a prior.}  $\Sigma_\parallel= 7\pm3$ and $\Sigma_\perp= 2\pm3$ ($\Sigma_{\parallel,\,\perp}$ Gaussian prior), using the hexadecapole along with the monopole and quadrupole on the BAO fit (+hexadecapole), using a different reference cosmology for the BAO fitting template ($\boldsymbol{\Theta}_{\rm OR}$, $\boldsymbol{\Theta}_X$, $\boldsymbol{\Theta}_Y$ and $\boldsymbol{\Theta}_{Z}$; see Table~\ref{tab:cosmo} and the top panel of  Fig~\ref{fig:olin} for a description of these cosmologies) and finally using only 500 realisations of the EZmocks to estimate the covariance (500 real.). When using different $x$-reference cosmologies we re-scale the obtained $\alpha$-parameters by the appropriate factor, $(D_{H,M}^x/r_{\rm drag})/(D_{H,M}^{\rm fid}/r_{\rm drag}^{\rm fid})$, to match the results one would have obtained if a fiducial cosmology would have been used as reference cosmology instead. In this way, all the $\alpha$-parameters of the different rows are comparable, regardless of the template cosmology used.

\begin{table*}
\caption{Impact of different parameters and data-vectors choices when performing a BAO analysis on the DR16 CMASS+eBOSS LRG dataset using the pipeline described in \S\ref{sec:BAO}. The Fourier space post-recon represent the main BAO results of this paper and correspond to the model displayed in the right panel of Fig~\ref{fig:BAOresults}. The configuration space results correspond to the analysis described in \citealt{LRG_corr}. The rest of cases (see text for a full description) represent variations of the standard pipeline. For each case we only report the physical BAO scaling parameters and their corresponding $\chi^2$. For the cases where the $\Sigma_\parallel$ and $\Sigma_\perp$ are varied, we find that when these are treated as free parameters (with a wide uninformative prior) we obtain $\Sigma_\parallel^{\rm free}=2.2\pm1.7$, $\Sigma_\perp^{\rm free}=2.3\pm1.7$; whereas under the Gaussian prior we find $\Sigma_\parallel^{\rm Gauss}=3.5 \pm 1.9$ (Gaussian prior: $7\pm3$) and $\Sigma_\perp^{\rm Gauss}= 2.0 \pm 1.4$ (Gaussian prior: $2\pm3$). The error-bars correspond to $1\sigma$ and only include the statistical error budget.   }
\begin{center}
\begin{tabular}{|c|c|c|c}
case & $\alpha_\parallel$ & $\alpha_\perp$ & $\chi^2/{\rm d.o.f.}$ \\
\hline
\hline
$P_k$ pre-recon & $0.939 \pm 0.036$& $1.043 \pm 0.032$ & $96/(112-17)$\\
$P_k$ post-recon  & $0.956 \pm 0.024$ & $1.025 \pm 0.019$ & $108/(112-17)$ \\
$\xi_s$ pre-recon & $0.954 \pm 0.035 $ & $1.034 \pm 0.025$ & $41/(40-9)$\\
$\xi_s$ post-recon & $0.958 \pm 0.026$ & $1.024 \pm 0.019$ & $41/(40 - 9)$ \\
($P_k+\xi_s$) post-recon & $0.956 \pm 0.024$ & $1.024 \pm 0.018$ & $-$ \\
\hline
NGC-only pre-recon &  $ 0.932 \pm 0.046$ & $ 1.054 \pm 0.043$ & $46/(56-10)$ \\
NGC-only post-recon & $0.947 \pm 0.026$ & $1.042\pm 0.024$ & $65/(56-10)$ \\
SGC-only pre-recon & $0.928 \pm 0.088$ & $1.058 \pm 0.091$ & $46/(56-10)$ \\
SGC-only post-recon & $0.996 \pm 0.113$ & $ 0.992 \pm 0.038$  & $40/(56-10)$ \\
no-mask post-recon & $0.953\pm 0.022$ & $1.030 \pm 0.016$ & $109/(112-17)$ \\
no-$w_{\rm sys}w_{\rm col}$ post-recon & $0.950 \pm 0.027$ & $1.023 \pm 0.020$ & $87/(112-17)$ \\
Isotropic template post-recon & $0.941 \pm 0.027$ & $1.030 \pm 0.023$ & $126/(112-18)$ \\
Isotropic template order-5 post-recon & $0.941 \pm 0.027$ & $1.027 \pm 0.024$& $102/(112-26)$ \\
Order-5 post-recon & $0.959 \pm 0.024$ & $1.018 \pm 0.021$ & $99/(112-25)$ \\
$\Sigma_{\parallel,\,\perp}$ Free post-recon & $0.949 \pm 0.019$ & $1.027 \pm 0.019$ & $101/(112-19)$\\
$\Sigma_{\parallel,\,\perp}$ Gaussian prior post-recon & $0.950 \pm 0.020$ & $1.027 \pm 0.019$ & $100/(112-19)$\\
+ Hexadecapole pre-recon & $0.914\pm0.035$  & $1.054 \pm 0.031$ & $190/(168-22)$ \\
+ Hexadecapole post-recon &  $0.949 \pm 0.026$ & $1.025 \pm 0.020$ & $157/(168-22)$ \\
$\boldsymbol{\Theta}_{\rm OR}$ (re-scaled to fiducial) & $ 0.962 \pm  0.026$ & $ 1.009 \pm 0.018$ & $120/(112-17)$\\
$\boldsymbol{\Theta}_{X}$ (re-scaled to fiducial) & $0.959 \pm 0.025$ & $1.022 \pm 0.020$ & $109/(112-17)$\\
$\boldsymbol{\Theta}_{Y}$ (re-scaled to fiducial) & $0.962 \pm 0.025$ & $1.024 \pm 0.020$ & $106/(112-17)$\\
$\boldsymbol{\Theta}_{Z}$ (re-scaled to fiducial) & $0.956 \pm 0.024$ & $1.017 \pm 0.017$ & $112/(112-17)$\\
500 real. in covariance & $0.955 \pm 0.025$ & $1.029 \pm 0.019$ & $106/(112-17)$\\
\end{tabular}
\end{center}
\label{table:BAOresults}
\end{table*}%

In general we see that most of these arbitrary choices produce no significant variation ($<0.5\sigma$) with respect to the standard pipeline, demonstrating a strong robustness on the BAO results. Some exceptions are when the data-vector is different (pre- vs. post-) or when the NGC and SGC are analysed independently. However, in these cases the cosmic variance has a much larger impact and therefore a larger shift is expected. The highest shift we observe (when the data-vector is unchanged) is on the variable $\alpha_\perp$ when the reference cosmology is varied from $\boldsymbol{\Theta}_{\rm fid}$ to $\boldsymbol{\Theta}_{\rm OR}$. In such case $\alpha_\perp$ changes by $0.85\sigma$. Note that these $\alpha-$values have been re-scaled after the actual fit to be both with respect to the same reference cosmology, so in the absence of noise and systematics both $\alpha$-value should be the same. Later in \S\ref{sec:fidcosmoBAO} and in Table~\ref{tab:sys} we investigate such effect using the \textsc{EZmocks} and the \textsc{Nseries} mocks, and find no strong shift when the template cosmology is changed, concluding that the difference we observe for the data is exclusively due to a statistical fluctuation. 

The reader could think that the results obtained by adding the hexadecapole to the standard monopole plus quadrupole analysis should have reported larger BAO information and smaller statistical error component. Previously, \cite{ross_information_2015} demonstrated that the amount of BAO information that higher-than-quadrupole moments add in terms of anisotropic BAO is very small. Indeed we report as well such findings later in Table~\ref{tab:sys} when we apply our analysis to the mocks. However, this is not the case for a FS analysis, where the hexadecapole is key to break degeneracies between the anisotropy generated by AP and RSD. We therefore conclude that the difference between the BAO analyses with and without hexadecapole are exclusively due to noise fluctuations, and do not correspond to any significant extra BAO information. Because of this, we take as our main BAO results those in which only the monopole and quadrupole are analysed. 

\begin{figure}
\centering
\includegraphics[scale=0.8]{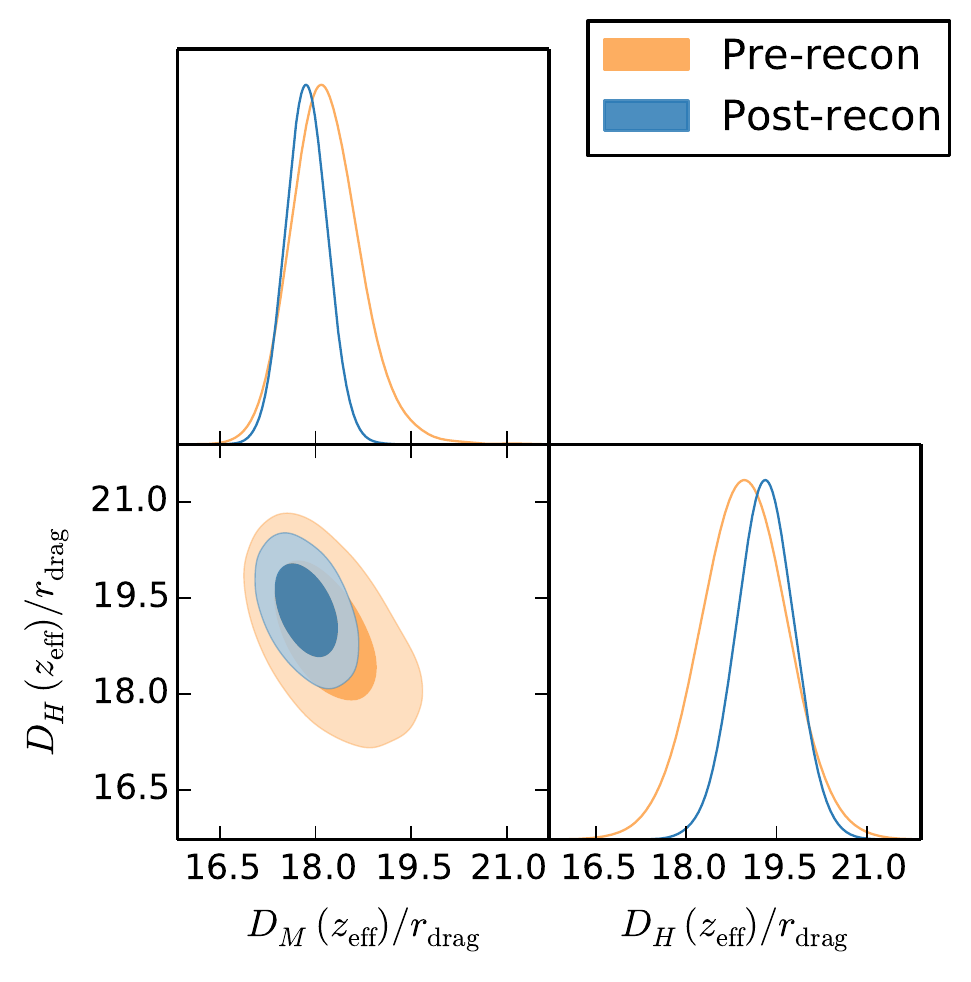}
\caption{Likelihood posterior for $1-$ and $2-\sigma$ contours (only statistical contribution), from the BAO type of analysis on the DR16 CMASS+eBOSS LRG data for the pre-reconstructed catalogues (in orange) and the post-reconstructed catalogues (in blue) in terms of $D_M(z_{\rm eff})/r_{\rm drag}$ and $D_H(z_{\rm eff})/r_{\rm drag}$ variables, at $z_{\rm eff}=0.698$. Results corresponding to the first two rows of Table~\ref{table:BAOresults}. }
\label{fig:BAOlikelihoods}
\end{figure}

Fig.~\ref{fig:BAOlikelihoods} displays the likelihood posteriors for 1 and $2\sigma$ for the BAO analysis using the pre- (orange) and post-reconstruction (blue) catalogues, in terms of the physical variables, $D_M/r_{\rm drag}$-$D_H/r_{\rm drag}$. In both cases the agreement is very good.  
The statistical errors on the cosmological parameters inferred from the post-reconstructed catalogues present are a factor of 1.5 smaller than those obtained from the pre-reconstructed catalogues. Later in \S\ref{sec:systematics} we study how typical this gain factor is by using the results from individual mocks. 

\subsection{Full Shape analysis}
We run the FS pipeline on the power spectrum monopole, quadrupole and hexadecapole measured from the CMASS+eBOSS galaxies for the $k$-range $0.02\leq k\,[\hompc] \leq 0.15$, as described in \S\ref{sec:RSD}. The covariance among $k$-bins is estimated from the analysis of 1000 \textsc{EZmocks}. 

\begin{figure}
\centering
\includegraphics[scale=0.3]{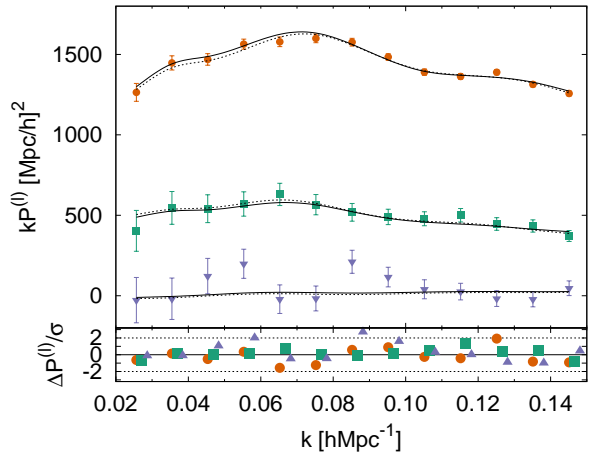}
\caption{Power spectrum multipoles measured from the DR16 CMASS+eBOSS LRG sample (weight-averaged between NGC and SGC), monopole (circular orange symbols), quadrupole (square green symbols) and hexadecapole (triangle purple symbols), along with the error-bars predicted by the {\it rms} of the 1000 \textsc{EZmocks}. The solid and dashed black lines represent the FS best-fit model (weight-averaged between NGC and SGC) when the monopole and quadrupole only are  fitting (black dashed lines) and when the hexadecapole is also used (black solid line). In the bottom sub-panel the differences between the measurement and the model, relative to the value of $1\sigma$ error-bar, are also displayed using the same colour notation. The results for the best-fitting parameters are reported in Table~\ref{tab:resultsPk} for the narrow prior on the amplitude of shot noise, $0.5\leq A_{\rm noise} \leq 1.5$.}
\label{Fig:FS}
\end{figure}

Fig.~\ref{Fig:FS} displays the monopole (round orange symbols), quadrupole (square green symbols) and hexadecapole (triangle purple symbols) for the results of the CMASS+eBOSS LRG data used for the FS analysis. The black solid lines display the performance of the best-fitting model when the three multipoles are simultaneously fitted, whereas the black dashed lines when only the monopole and quadrupole are used. 

Table~\ref{table:FSresults} displays the results for FS analysis under different cases. The first two rows represent the case where the monopole (M), quadrupole (Q) and hexadecapole (H) are fitted up to a $k_{\rm max}=0.15\,\hompc$ with a wide and flat uninformative prior on the amplitude of shot noise (first row) and with a more restrictive prior allowing for such amplitude to vary within $50\%$ of its Poisson prediction (second row). The third row displays the result from the configuration space analysis reported in \cite{LRG_corr} and the fourth row the consensus between Fourier and configuration space, as described in \S\ref{sec:consensus}. The rest of the rows are variations of the above pipeline (with a wide uninformative prior on the amplitude of shot noise as a default option): using only the monopole and quadrupole (M+Q), fitting to NGC- and SGC-only (NGC, SGC M+Q+H), fitting to the weighted mean signal of NGC and SGC (NGC+SGC M+Q+H), setting a hard prior on $b_2$ to be positive ($b_2>0$ prior), using the monopole, quadrupole and hexadecapole with a different $k$-range, computing the hexadecapole using a different decomposition on the LOS (hexadecapole as $\mathcal{L}_4(\hat{\bf k}\cdot\hat{\bf r}_1)\mathcal{L}_0(\hat{\bf k}\cdot\hat{\bf r}_2)$), ignoring the $A^{\rm TNS}$ and $B^{\rm TNS}$ in the modelling of Eq. \ref{eq:TNSterms} (no-TNS terms), only using 1-loop correction in the modelling of Eq. \ref{eq:2loop} (only 1-loop terms), using SPT predictions instead of RPT for the terms of Eq. \ref{eq:2loop} (SPT 2-loop), using the Gaussian form of Eq. \ref{eq:GaussFog} for FoG (FoG Gaussian), setting the fiducial $\sigma_8$ to a 15\% higher value than the predicted by the reference cosmology at $z=0.70$ ($\sigma_8$ 15\% high), treating $f$ and $\sigma_8$ as free independent parameters ($\sigma_8$ free), using a different cosmology as a reference cosmology ($\boldsymbol{\Theta}_{\rm OR}$, $\boldsymbol{\Theta}_{\rm X}$, $\boldsymbol{\Theta}_{\rm Y}$ and $\boldsymbol{\Theta}_{\rm Z}$\footnote{As in Table~\ref{table:BAOresults} the obtained $\alpha$-parameters are re-scaled after the fit to match the prediction of the fiducial cosmology when used as a reference cosmology.}), turning off the systematics and/or collision weights ($w_{\rm sys}$ off, $w_{\rm col}$ off,  $w_{\rm sys}w_{\rm col}$ off), and using only 500 \textsc{EZmock} realisation to estimate the covariance (500 real. in covariance). 

\begin{table*}
\caption{Impact of different parameters and data-vector choices when performing a FS analysis on the DR16 CMASS+eBOSS LRG dataset using the pipeline described in \S\ref{sec:RSD}. The second row with the $0.5\leq A_{\rm noise} \leq 1.5$ prior on the amplitude of shot noise represent the main result of FS analysis of this paper and correspond to the model (in solid black lines) displayed in Fig~\ref{Fig:FS}. The configuration space results are main results reported by \citealt{LRG_corr}. The rest of cases (see text for a full description) represent variations of the standard pipeline or data-vector choices. For each case we do not report all the nuisance parameters, only the physical BAO scaling parameters, $f\sigma_8$, and their corresponding $\chi^2$.  The error-bars correspond to $1\sigma$ and only include the statistical error budget. }
\begin{center}
\begin{tabular}{|c|c|c|c|c|}
case & $\alpha_\parallel$ & $\alpha_\perp$ & $f\sigma_8$  & $\chi^2/{\rm dof}$ \\
\hline
\hline
$P_k$,\, $k_{\rm max}=0.15$, M+Q+H  & $1.017 \pm 0.045$ & $1.006 \pm 0.027$ & $0.469\pm0.046$ & $77/(78-11)$\\
$P_k$,\,  $0.5\leq A_{\rm noise}  \leq 1.5$ prior & $0.999 \pm 0.036$ & $1.003 \pm 0.027$ & $0.454\pm0.042$ &  $77/(78-11)$  \\
$\xi_s$ & $1.016 \pm 0.029$ & $ 1.004 \pm 0.019$ & $0.461 \pm 0.042$ & --  \\
$P_k+\xi_s$ & $1.008\pm0.027$ & $1.002\pm0.018$ & $0.449 \pm 0.039$ & -- \\
\hline
M+Q & $0.977 \pm 0.054$ & $1.032 \pm 0.037$ & $0.511\pm0.065$ & $38/(52-11)$\\
M+Q, $0.5\leq A\leq 1.5$ prior  & $0.972 \pm 0.050$ & $1.026 \pm 0.036$ & $0.496\pm0.062$ & $38/(52-11)$\\
NGC M+Q+H & $0.983\pm0.049$ & $1.045\pm0.036$ & $0.495\pm0.059$ & $28/(39-7)$\\
SGC M+Q+H&  $1.174\pm 0.109$ & $0.955 \pm 0.042$ & $0.375\pm0.093$ & $45/(39-7)$\\
NGC+SGC M+Q+H & $ 1.007 \pm 0.044$ & $1.004 \pm 0.027$ & $0.459\pm0.046$ & $34/(39-7)$\\
  $0\leq b_2$ prior & $1.002 \pm 0.036$ & $1.007 \pm 0.026$ & $0.455\pm0.043$ & $78/(78-11)$  \\
$k_{\rm max}=0.20$, M+Q  & $1.006 \pm 0.045$ & $1.013 \pm 0.027$ & $ 0.499\pm0.053$  & $58/(72-11)$\\
$k_{\rm max}=(0.20, {\rm M+Q})+(\,0.15,\,{\rm H})$ & $ 1.025 \pm 0.035$ & $ 0.999 \pm 0.021$ & $ 0.481\pm0.041$ & $97/(98-11)$ \\
$k_{\rm max}=(0.15, {\rm M+Q})+(\,0.10,\,{\rm H})$ & $1.052 \pm 0.062$ & $0.989 \pm 0.029$ & $0.449\pm0.052$& $68/(68-11)$ \\
$k_{\rm max}=0.20$, M+Q+H& $1.041 \pm 0.033$ & $0.989 \pm 0.021$ &  $0.450\pm0.041$ & $124/(108-11)$ \\
Hexadecapole as $\mathcal{L}_4(\hat{\bf k}\cdot\hat{\bf r}_1)\mathcal{L}_0(\hat{\bf k}\cdot\hat{\bf r}_2)$ & $1.017 \pm 0.046$ & $1.006 \pm 0.028$ & $0.471\pm0.047$ & $78/(78-11)$ \\
no-TNS terms & $1.007 \pm 0.034$ & $0.996 \pm 0.025$ & $0.446\pm0.040$ & $76/(78-11)$\\
only 1-loop terms. & $1.009 \pm 0.042$ & $1.007 \pm 0.027$ & $0.470\pm0.046$ & $76/(78-11)$ \\
SPT 2-loop & $1.012 \pm 0.043$ & $1.007 \pm 0.027$ & $0.466\pm0.047$ & $77/(72-11)$ \\
FoG Gaussian & $1.020 \pm 0.046$ & $1.003 \pm 0.028$ & $0.467\pm0.045$ & $76/(78-11)$ \\
$\sigma_8$ free & $0.989 \pm 0.039$ & $1.005 \pm 0.026$ & $0.452 \pm 0.043$ & $77/(78-12)$ \\
$\sigma_8$ free,  $0.5\leq A\leq 1.5$ prior & $0.979\pm0.034$ & $1.005 \pm 0.026$ & $0.445 \pm 0.039$ & $77/(78-12)$ \\
$\sigma_8$ 15\% high & $1.021\pm 0.044$ & $1.008 \pm 0.028$ & $0.468 \pm 0.047$ & $76/(78-11)$\\
$\boldsymbol{\Theta}_{\rm OR}$ ($\alpha$s re-scaled to fiducial)  & $1.008 \pm 0.038$ & $ 1.013\pm 0.026$ & $0.453\pm0.040$ & $85/(78-11)$ \\
$\boldsymbol{\Theta}_{\rm X}$ ($\alpha$s re-scaled to fiducial) & $1.015\pm 0.042$ & $1.008 \pm  0.028$ & $0.472\pm0.053$ &  $77/(78-11)$ \\
$\boldsymbol{\Theta}_{\rm Y}$ ($\alpha$s re-scaled to fiducial) & $1.041\pm 0.048$ & $1.015 \pm  0.029$ & $0.474\pm0.053$ &  $74/(78-11)$ \\
$\boldsymbol{\Theta}_{\rm Z}$ ($\alpha$s re-scaled to fiducial) & $1.026\pm 0.048$ & $1.005 \pm  0.027$ & $0.495\pm0.058$ &  $75/(78-11)$ \\
$w_{\rm sys}$ off & $1.037 \pm 0.049$ & $0.992 \pm 0.028$ & $0.426\pm0.047$ & $82/(78-11)$ \\
$w_{\rm col}$ off & $1.002 \pm 0.040$ & $ 1.003 \pm 0.028$ & $0.468\pm0.048$ & $ 64/(78-11)$ \\
$w_{\rm sys}w_{\rm col}$ off & $ 1.019 \pm 0.044$ & $0.990 \pm 0.028$ & $0.431\pm0.047$ & $69/(78-11)$ \\
500 real. in covariance & $1.020 \pm 0.042$ & $0.996 \pm 0.027$ & $0.464\pm0.047$ & $79/(78-11)$ \\
\end{tabular}
\end{center}
\label{table:FSresults}
\end{table*}

Except for some extreme cases, such as those when the systematic weights are not applied,  we do not observe any strong dependence of the inferred cosmological parameters with any of the studied variations. In particular we observe very mild changes when the underlying linear power spectrum template is changed. We also note the change in the size of errors of $\alpha_\parallel$ and $f\sigma_8$ when the 50\%-prior on $A_{\rm noise}$ is set (first vs. second row). In this last case the $2\sigma$ contours do hit the higher boundary of $A_{\rm noise}=1.5$, and therefore the reduction of error is a direct consequence of this. The amplitude of shot noise is very correlated with $b_2$ which is poorly constrained without higher-order moments as the bispectrum \citep{gil-marin_clustering_2017}. In particular, the $A_{\rm noise}\simeq1$ solution also corresponds to $b_2>0$, whereas $A_{\rm noise}>1.5$ corresponds to $b_2<0$ (see Fig.~\ref{plot:fullvector_Anoise}). From the power spectrum and bispectrum analysis of BOSS DR12 CMASS sample $0.43<z<0.70$ we expect that $b_2$ for these type of galaxies is close to the value reported in \cite{gil-marin_clustering_2017},  $b_2\sigma_8=0.606\pm 0.069$, which agrees with the solution of shot noise being close to the Poisson prediction. Thus we find plausible that the shot noise should not differ by more than 50\% (which is already quite a large amount) from the Poissonian prediction and we decide to take as the FS analysis main result of this paper the cosmological parameters inferred when this $50\%$-prior is applied.  In appendix \ref{sec:Anoise} we further comment on this effect (see also Fig. \ref{fig:contoursRSDBAO_Anoise}).

\begin{figure}
\centering
\includegraphics[scale=0.6]{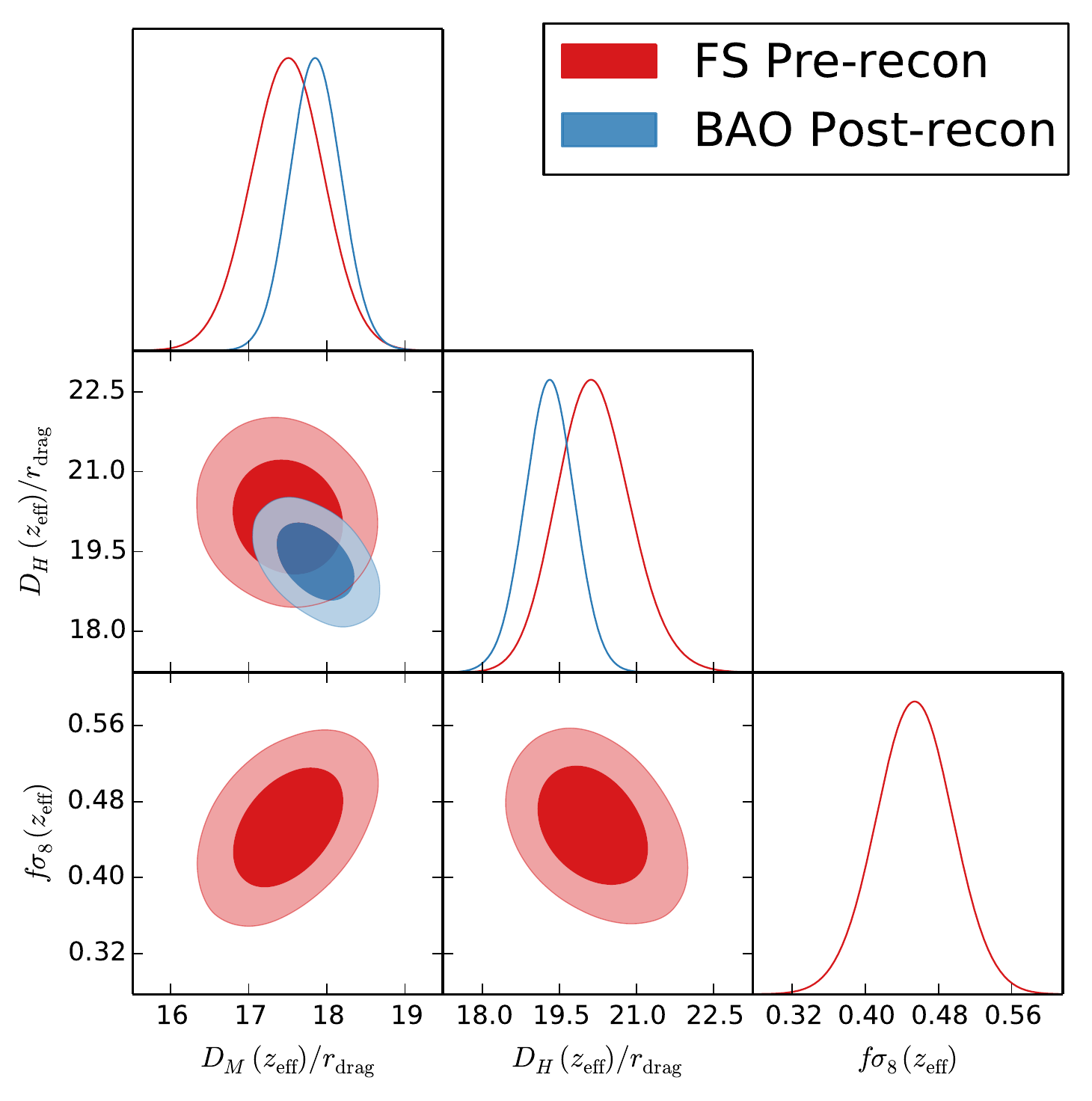}
\caption{Comparison of the cosmological inferred parameters from the FS  and BAO analysis (on post-reconstructed catalogues), respectively, from the power spectrum multipoles. In this case the FS derived results have been computed from the power spectrum monopole, quadrupole and hexadecapole under the analysis with $50\%$-priors on $A_{\rm noise}$ reported by the second row of Table~\ref{table:FSresults}. The BAO reconstructed results comes from the standard analysis on power spectrum monopole and quadrupole, as it is reported by the second row of Table~\ref{table:BAOresults}. In all cases the contours represent only the statistical contribution.}
\label{fig:consensusPk}
\end{figure}

In Fig.~\ref{fig:consensusPk} we display the derived $D_H/r_{\rm drag}$, $D_M/r_{\rm drag}$ and $f\sigma_8$ cosmological parameters from the FS and BAO analysis on the pre- and post-reconstructed catalogues, respectively. Both measurements rely on very correlated pre- and post-recon catalogues, and therefore it is not straightforward to resolve the level of agreement between them. Later in \S\ref{sec:consensus} we come back to this question and also compare these findings with the quantities inferred from configuration space. For now, we just note that the reconstructed-BAO analysis provides tighter constrains on both $D_H/r_{\rm drag}$ and $D_M/r_{\rm drag}$ than the FS analysis. This feature is actually expected due to the enhancement that reconstruction provides in the measurement of the BAO peak oscillatory features. We also note that the reconstructed-BAO analysis favours higher values of both $D_M/r_{\rm drag}$ and $D_H/r_{\rm drag}$ with respect to the FS analysis on pre-reconstructed data. In fact, if we were looking at the pair of variables $\alpha_{\rm 0}$ and $\epsilon$ we would notice that such difference arises from the AP variable, $\epsilon$  or $F_\epsilon$, where the results inferred from reconstructed data present a $\sim2\sigma$ deviation from the null-AP behaviour, which is what we observe for the pre-reconstructed catalogue. We will fully discuss these differences later in \S\ref{sec:consensus}.

\section{Systematic Tests}\label{sec:systematics}
In this section we aim to run the BAO  and FS pipeline analyses on different sets of mocks to check the performance and to identify potential systematic errors. In total we use $N_{\rm EZ}=1000$ realisations of the \textsc{EZmocks}, $N_{\rm Nseries}=84$ realisations of the \textsc{Nseries} mocks and $N_{\rm OR}=27$ realisations of the \textsc{OuterRim-HOD} mocks.

\subsection{Baryon Acoustic Oscillation systematics}\label{sec:BAOsys}
We start by running the BAO pipeline described in  \S\ref{sec:BAO} on the pre- and post-reconstructed \textsc{EZmocks}, \textsc{Nseries} and \textsc{OuterRim} +`Hearin-Threshold-2', +`Leauthaud-Threshold-2' and +`Tinker-Threshold-2' HOD mocks.  We run the BAO pipeline on the power spectrum monopole and quadrupole for $0.02\leq k\,[\hompc]\leq0.30$. Smaller and larger scales do not contain relevant BAO information. 

In this section we aim to,
\begin{itemize}
\item Check how typical the data is with respect to the \textsc{EZmocks}. 
\item Determine the systematic budget of the pipeline.
\item Check whether the arbitrary choice of the BAO reference template has an impact on the inferred cosmological parameters.
\item Determine whether the underlying galaxy HOD has an impact on the recovered parameters.
\end{itemize}

\begin{figure*}
\includegraphics[scale=0.31]{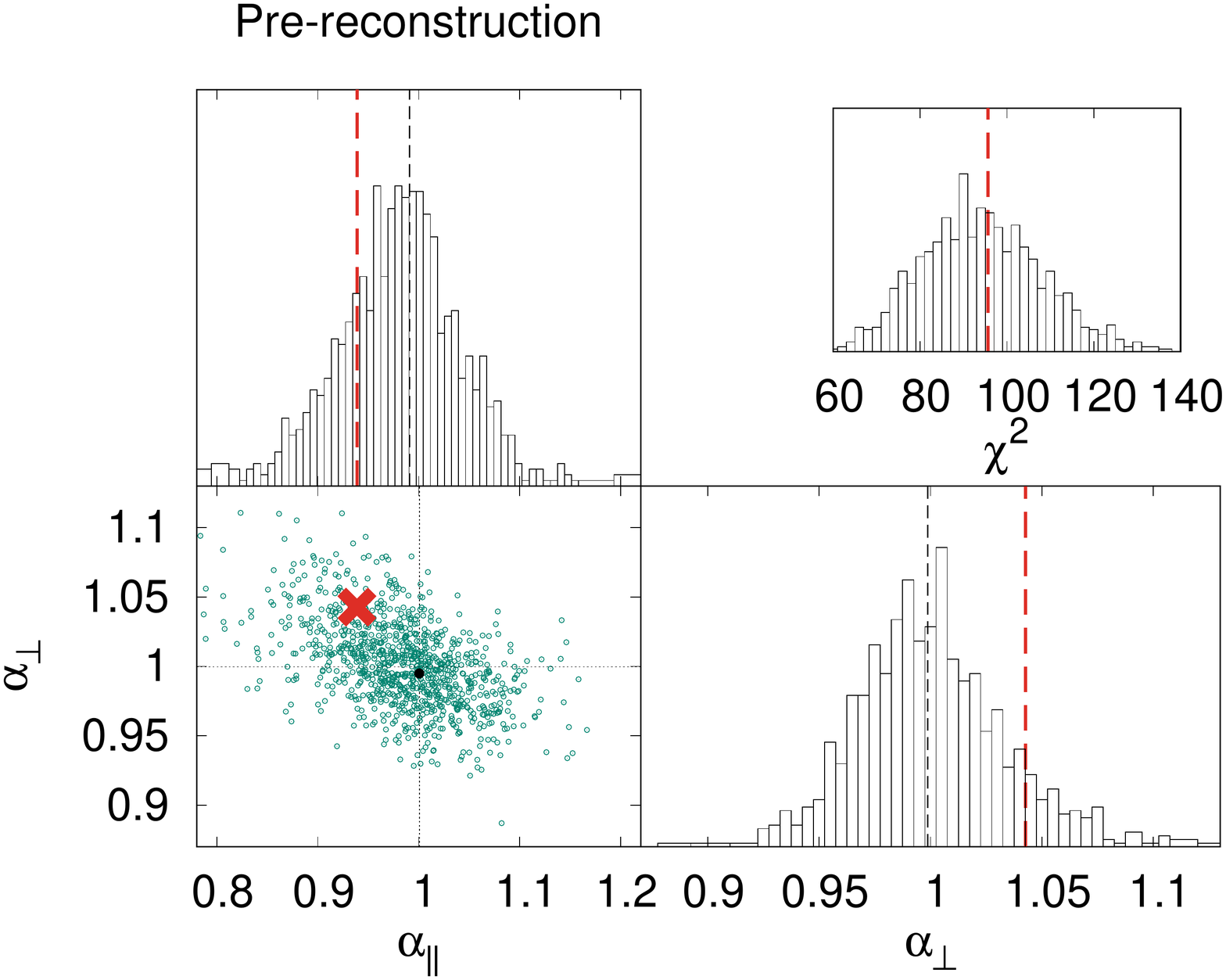}
\includegraphics[scale=0.31]{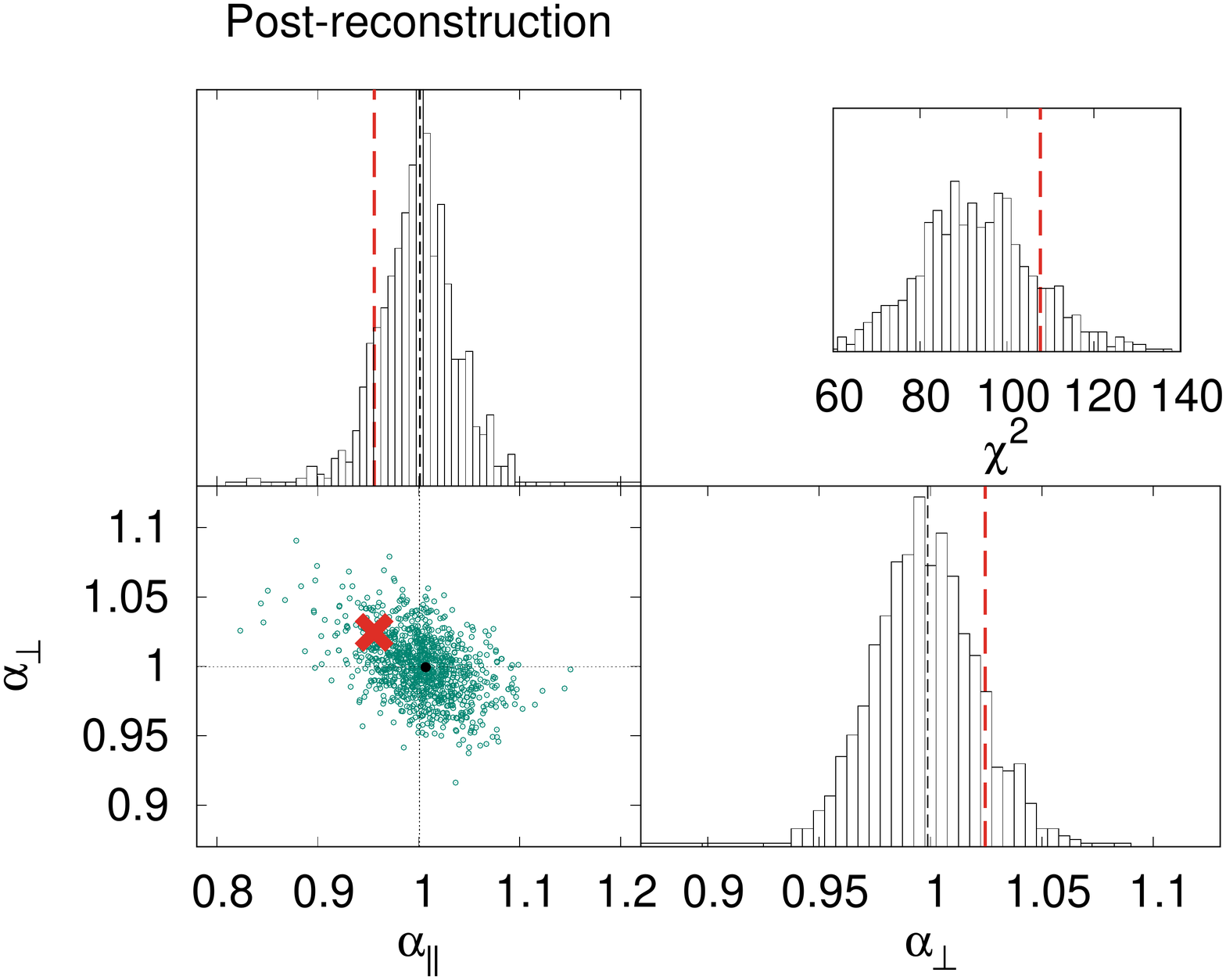}

\includegraphics[scale=0.31]{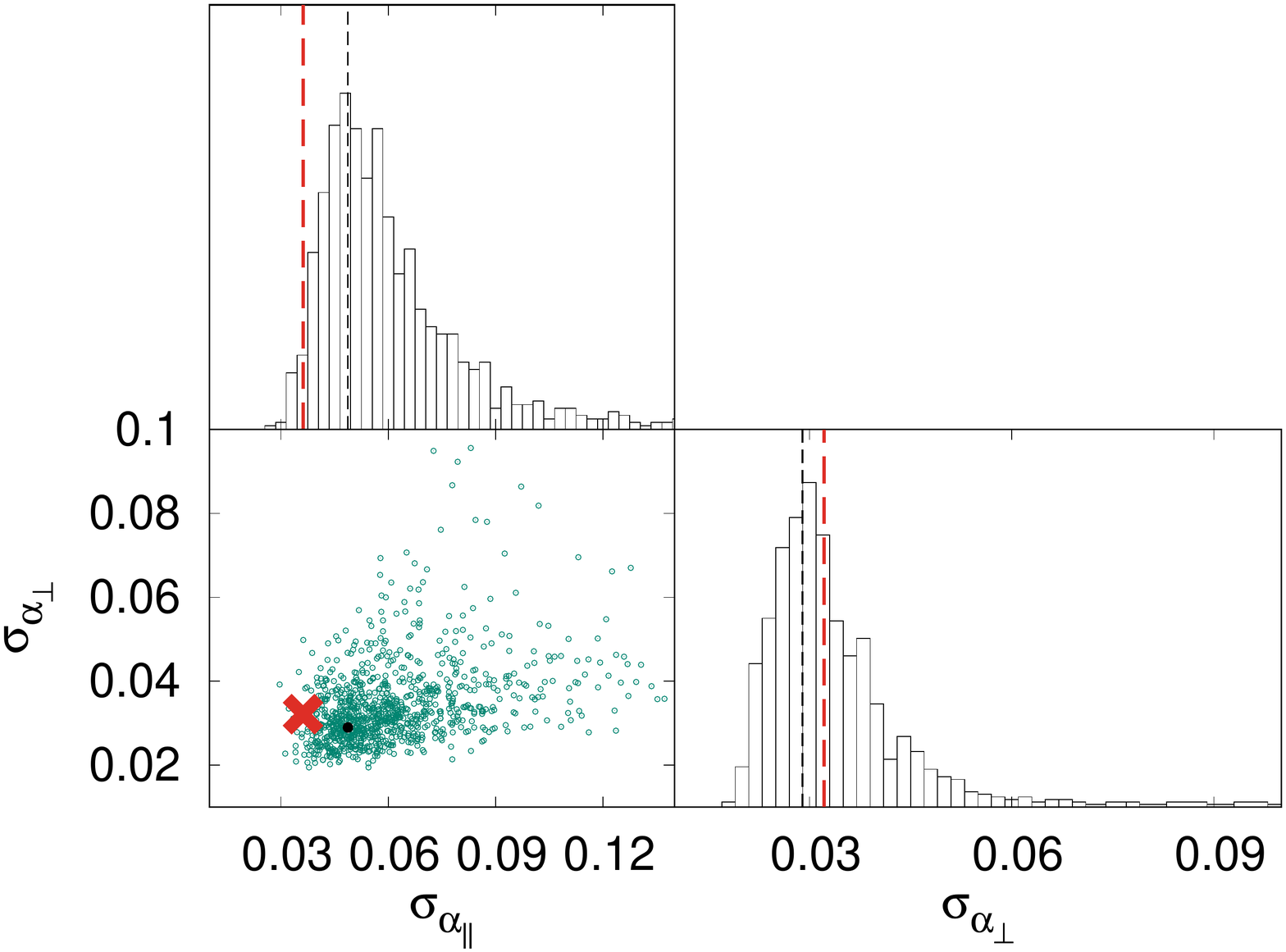}
\includegraphics[scale=0.31]{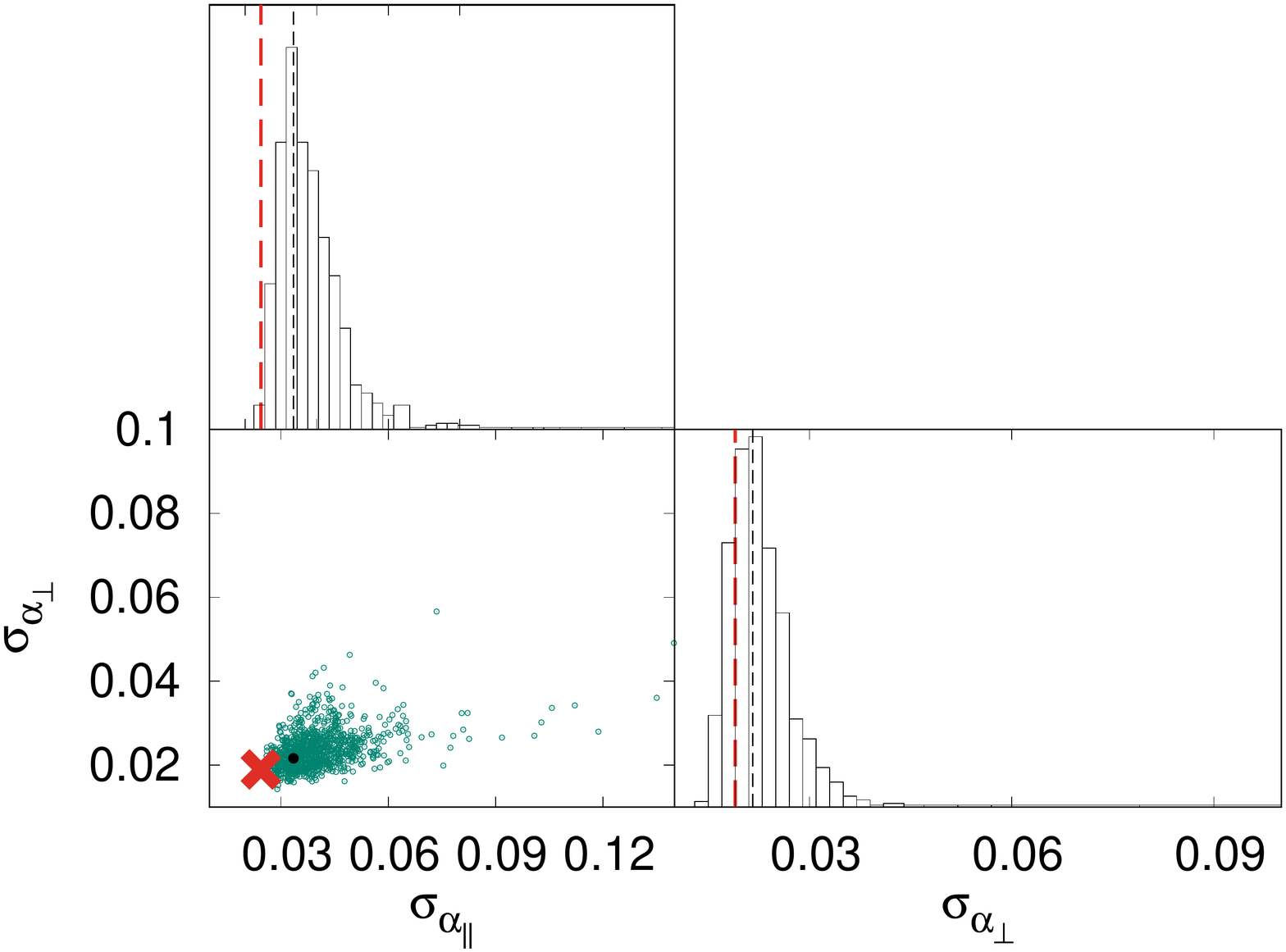}

\caption{The top sub-panels display the distribution of best-fitting $\alpha_\parallel$ and $\alpha_\perp$, for the pre- (left panels) and post-reconstructed (right panels) power spectrum monopole and quadrupole for $0.02\leq k\,[\hompc] \leq 0.30$, for the 1000 realisations of the \textsc{EZmocks} catalogues (green points), for the DR16 CMASS+eBOSS LRG data catalogue (red cross) and for the best-fit to the mean of the 1000 mock power spectra (black dot). The horizontal and vertical black dotted lines represent the expected $\alpha$ values for the \textsc{EZmocks}. The distribution of $\chi^2$ values is also shown for each case. The bottom sub-panels display analogous plots for the errors of $\alpha_\parallel$ and $\alpha_\perp$, instead. In this case the error of the mean has been re-scaled by the square root of number of realisations. }
\label{Fig:bao_ezmocks1}
\end{figure*}

The top panels of Fig.~\ref{Fig:bao_ezmocks1} display the recovered $\alpha_\parallel$ and $\alpha_\perp$ scaling parameters on the pre- (left panels) and post-reconstructed (right panels) 1000 \textsc{EZmocks} realisations (green points). The corresponding bottom panels display the distribution of errors inferred from the {\it rms} of the individual \textsc{mcmc} chains. In addition we represent with a red cross the values for the actual data catalogue, and with a black dot the values obtained when fitting the average power spectrum of 1000 \textsc{EZmocks} realisations. The error of this last case is expected to scale with the square root of the total volume, and therefore we re-scale it by the $\sqrt{N_{\rm EZ}}$ factor in order to match the value of the error of a single realisation. For all the cases the results inferred from the mean of the mocks are in excellent agreement with the results of the individual cases, suggesting that the mean of the fits is close to the fit of the mean (shown later in Table~\ref{tab:sys}). We find that the values of $\alpha_\parallel$ and $\alpha_\perp$ inferred from the data catalogue are also consistent with the intrinsic scatter observed from the mocks. However we obtain atypically small errors when analysing the data catalogue, both for the pre- and post-reconstructed cases of $\alpha_\parallel$. In particular, for the post-reconstruction case we have only found a total of $\sim10/1000$ realisations whose error on $\alpha_\parallel$ is comparable to the one found in the data, which certainly suggest a $\sim1\%$ probability of being in such situation. As we show later (see Fig.~\ref{fig:baocomp} in \S\ref{sec:consensus}), this result is perfectly compatible with what we find in the complementary BAO analysis in configuration space performed in \cite{LRG_corr}. The $\chi^2$ value of the data is not small with respect to the typical value obtained by the mocks, which suggests that this small BAO error on $\alpha_\parallel$ may be caused by noise fluctuations, which enhance the BAO signal in the data along the LOS, with respect to the typical noise level predicted by the mocks. Given the number of physical parameters ($\alpha_\parallel$, $\alpha_\perp$, pre- and post-reconstructed catalogues, $f\sigma_8$ and their corresponding errors: 10 variables in total) it is not very unlikely that at least one of them is atypical at $3\sigma$ level, which is known as `the look-elsewhere effect'. 

\begin{figure}

\centering
\includegraphics[scale=0.3]{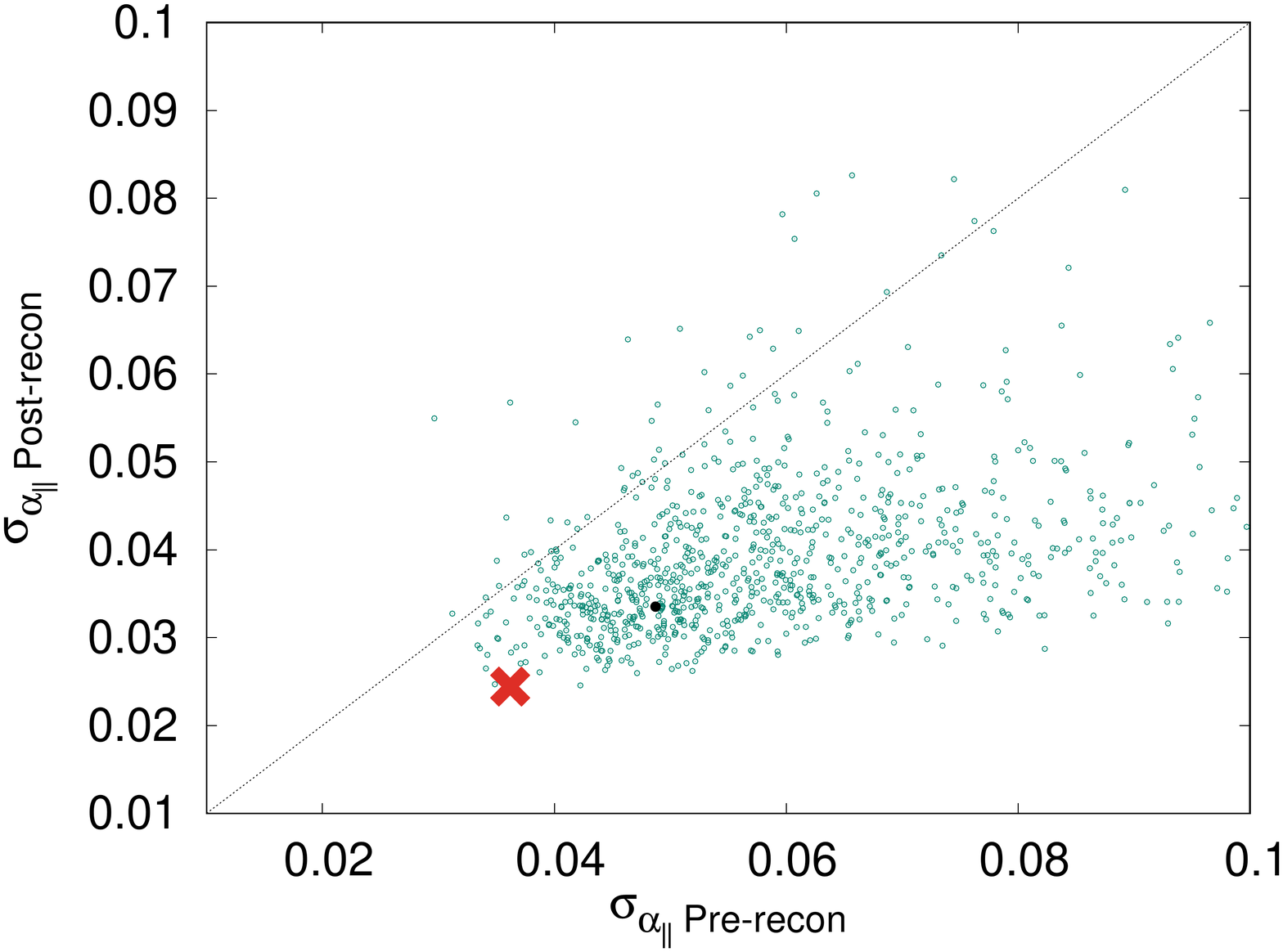}

\includegraphics[scale=0.3]{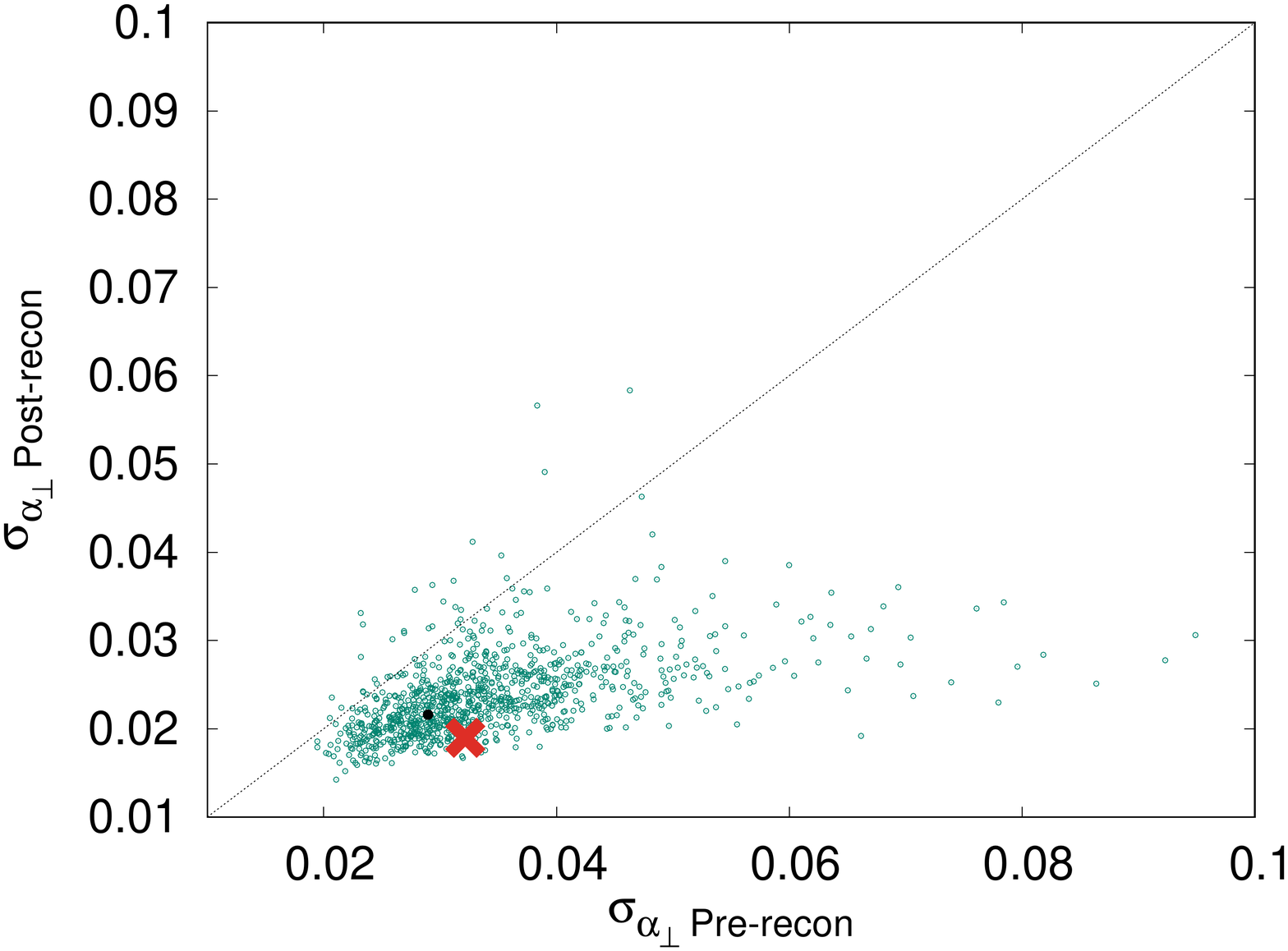}

\caption{Performance of the reconstruction on the DR16 CMASS+eBOSS LRG catalogues, the green symbols represent the 1000 realisations of the \textsc{EZmocks}, the red cross the actual DR16 CMASS+eBOSS LRG data catalogue. The $x$-axis represent the pre-reconstructed quantity and the $y$-axis the post-reconstructed quantity. For reference a black dashed line, $x=y$ is also shown. The top and bottom panels display the 1-$\sigma$ errors of $\alpha_\parallel$ and $\alpha_\perp$, respectively. The black dot represent the mean of 1000 realisations re-scaled by the square root of the 1000 realisations.}

\label{Fig:bao_ezmocks2}
\end{figure}

The panels in Fig.~\ref{Fig:bao_ezmocks2} display how the reconstruction algorithm performs on the \textsc{EZmocks}, for the errors of $\alpha_\parallel$ and $\alpha_\perp$. Reconstruction significantly helps to improve the determination of the $\alpha$s in almost all the realisations of the mocks, where the typical improvement (ratio between pre- and post-recon errors) is $\sim 40\%$ for $\alpha_\parallel$ and $20\%$ on $\alpha_\perp$. This behaviour is expected for cosmic variance limited samples, like the LRGs and ELGs, unlike other more sparse samples like the quasars. We find that for the data, the improvement on $\alpha_\perp$ is expected and typical with respect to what is observed in the mocks, whereas for $\alpha_\parallel$ the values are atypical as we have commented above, but the improvement ratio is typical. 

\subsubsection{Performance of BAO template}
We  start by applying the BAO anisotropic template described by Eq. \ref{eq:baoaniso} on the different sets of mocks. Table~\ref{tab:sys} displays the results when fitting the mean power spectra of all available realisations for a given type of mock (rows labeled as ``Mean''); and the mean of the fits of individual realisations (rows labeled as ``Individual''). 

\begin{figure*}
\centering
\includegraphics[scale=0.5]{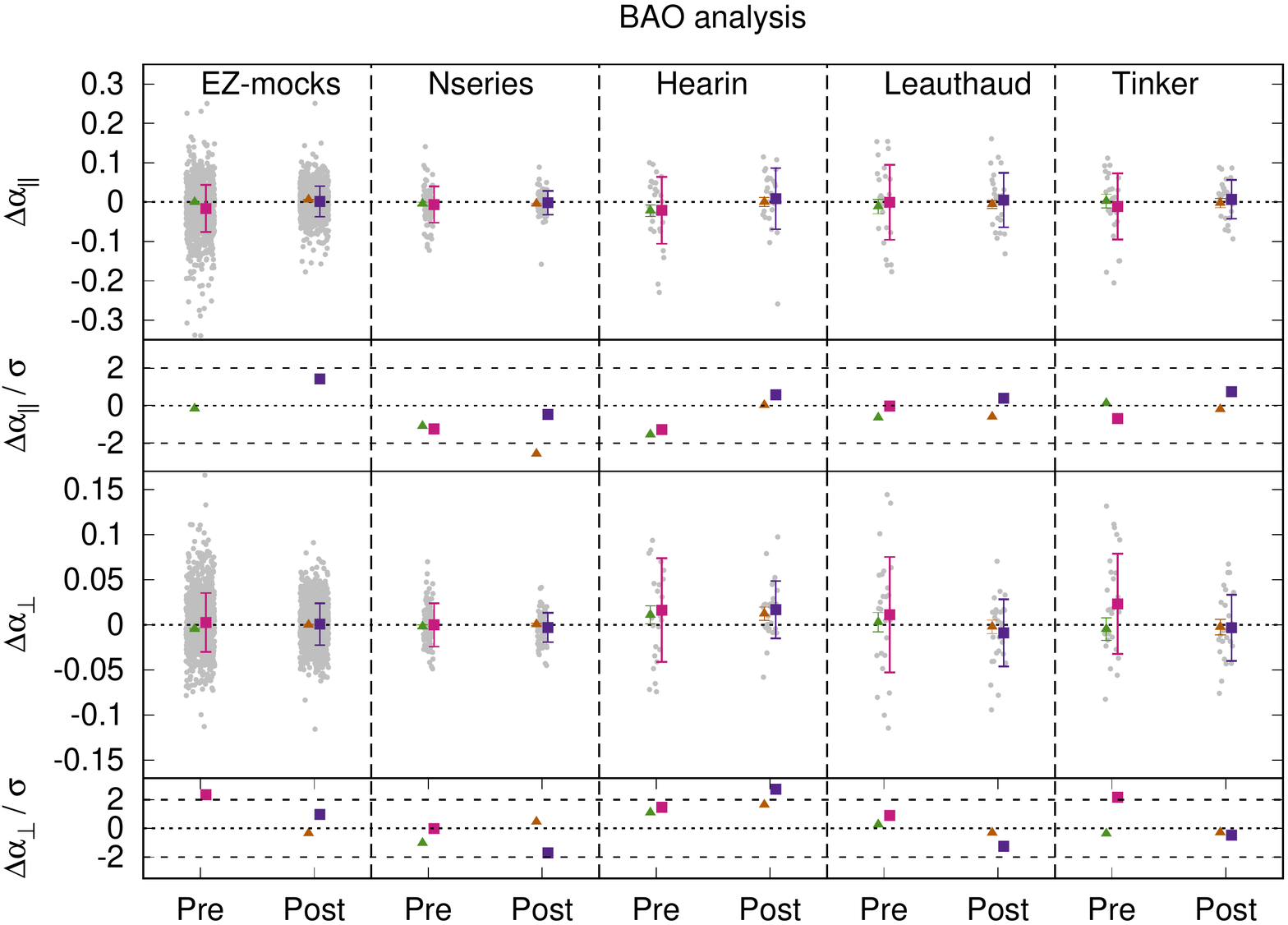}
\caption{Performance of the BAO type of analysis for $\alpha_\parallel$ (top) and $\alpha_\perp$ (bottom) as indicated. Each of the 5 vertical sub-panels corresponds to the results on the \textsc{EZmocks},  \textsc{Nseries} and the \textsc{OuterRim-HOD} mocks by Hearin+threshold2, Leauthaud+threshold2 and Tinker+threshold2, as indicated. A padding has been applied to the original non-periodic \textsc{OuterRim-HOD} cubic sub-box in order avoid spurious non-periodic effects. For each $\alpha$ we display the panels $\Delta\alpha\equiv\alpha-\alpha^{\rm exp}$ and $\Delta\alpha/\sigma$. For each of these sub-panels we display the pre- (left) and post-recon results (right). The grey points correspond to the individual best-fit parameters. The green (for pre-recon) and orange (for post-recon) triangles symbols correspond to the results of fitting the mean of the mocks, and its error corresponds to the error of the mean. The pink (for pre-recon) and purple (for post-recon) squares symbols display the results of taking the mean of $N_{\rm det}$ individual fits, and the reported error is its {\it rms}. Consequently the error of the mean of the fits is $\sqrt{N_{\rm det}}$ times the error of the mean. In the $\Delta\alpha/\sigma$ sub-panels $\sigma$ represents the error of the mean for the fit to the mean (orange and green triangles), and the {\it rms}/$\sqrt{N_{\rm det}}$ for the mean of the individual fits. The numerical results of this plot are also listed in Table~\ref{tab:sys}.}
\label{fig:sys}
\end{figure*}

Fig.~\ref{fig:sys} graphically represents the data contained in Table~\ref{tab:sys}.  For each sub-panel the difference between the measured $\alpha_\parallel$ and $\alpha_\perp$, and their expected value are inferred for the pre- (left) and post-reconstructed (right) catalogues. The individual results on the mocks are shown in grey, the results of the mean of the mocks are represented by a $\bigtriangleup$-symbol in green (for the pre-reconstructed catalogues) and in orange (for post-reconstructed catalogues). The associated errors are consistently the errors of the mean, obtained by re-scaling the covariance by a factor the number of realisations, $N_{\rm EZ}$, $N_{\rm Nseries}$ and $N_{\rm OR}$. Therefore, these errors are a factor $\sqrt{N_i}$ smaller than the error we obtain for a single realisation of these mocks.  The mean of the individual fits is represented by a \sq-symbol in pink (for pre-reconstructed catalogues) and in purple (for post-reconstructed catalogues). In this case the error associated is the {\it rms} of all the individual fits, which is $\sim\sqrt{N_i}$ larger than the error associated to the mean. The sub-panels show the difference between the measured and the expected value of $\alpha_\parallel$ and $\alpha_\perp$ in terms of number of statistical $\sigma$ of the error of the mean, and the ${ rms}/\sqrt{N_i}$. Note that for the \textsc{EZmocks} the effective volume is so large ($\simeq 2,650\,{\rm Gpc}^3$) that in some cases the result is totally dominated by systematics and the symbols are off the scale of $\pm3\sigma$. 

The \textsc{EZmocks} `Mean' results on post-reconstruction catalogues reveal that the $\alpha_\parallel$ variable is significantly shifted by $0.6\%$ with respect to their expected quantity, which correspond to $6\sigma$ deviation from the expected value; whereas for $\alpha_\perp$, mocks and model agree to within $0.068\%$ ($1\sigma$ level). 
It is also worth mentioning that at this sub-percent level of precision, we would require full {\it N}-body mocks to actually validate this kind of systematic shifts, as the \textsc{EZmocks} have not been designed to be accurate at this level of precision. Therefore we cannot discern whether this observed $0.6\%$ shift in $\alpha_\parallel$ is due to a limitation of the model of Eq. \ref{eq:baoaniso}, a limitation of the \textsc{EZmocks} themselves,  or an effect arising from the reconstruction technique. From the remaining {\it N}-body mocks we do not observe any significant BAO peak position shift with respect to their corresponding expected value in any of the post-reconstructed catalogues analysed. The BAO pipeline is able to deal with different kinds of HOD models. We do see some fluctuations, but these are always below $\pm2.5\sigma$ limit, so we do not take them as significant shifts.  However, the statistical errors associated to these catalogues are not as small as those corresponding to the \textsc{EZmocks}, so we can only state that we have not detected any systematic above the statistical threshold of $1-2\%$ for \textsc{OuterRim-HOD}, and $0.1-0.5\%$ for \textsc{Nseries}. Such upper limits are below the statistical precision of our sample: for post-reconstructed catalogues we obtain a statistical precision of $\sim2.4\%$ for $\alpha_\parallel$ and  $\sim1.9\%$. for $\alpha_\perp$. From the \textsc{Nseries} results we conclude that there are no strong modelling-systematic errors associated when determining $\alpha$s,  which validates our modelling pipeline, including the reconstruction technique. From the \textsc{OuterRim-HOD} results we conclude that we do not detect any relative systematic due to different HOD modelling, although the statistical precision reached on these mocks is comparable to the statistical precision of our sample. 

\begin{table*}
\caption{Performance of the BAO template of Eq \ref{eq:baoaniso} in different set of mocks. For The EZmocks expected values are $\alpha_\parallel^{\rm exp}=1+8.853\cdot10^{-4}$ and $\alpha_\perp^{\rm exp}=1-3.650\cdot10^{-4}$ as the mocks are not analysed in their underlying true cosmology (except for those where the cosmology is explicitly varied, $\boldsymbol{\Theta}_{\rm OR,X,Y,Z}$). For the rest of the mocks these values are $1$, as they are respectively analysed in their own underlying true cosmology . For each set of mocks the results from both pre- and post-recon catalogues are presented. We display the results of fitting the mean of all the mocks, indicated with `Mean', and the mean of the individual fits on the mocks, indicated with `Individual'. For the fit to the mean the error quoted is the $1\sigma$ of the error on this fit, where in order to do so, we have scaled the covariances by the number of realisations used to take the mean, which in all the cases is the maximum number of realisations available, $N_{\rm tot}$, 1000, 84 and 27 for \textsc{EZmocks},  \textsc{Nseries} and \textsc{OuterRim-HOD} mocks, respectively. For the \textsc{OuterRim-HOD} type of mocks only the `Threshold2' flavour is represented, where a padding has been applied to the original non-periodic cubic sub-box in order avoid spurious non-periodic effects. For the mean of individual best-fits the error quoted is {\it rms} divided by $\sqrt{N_{\rm det}}$, where $N_{\rm det}$ is the number of BAO detections (those fits with $\alpha_\parallel$ and $\alpha_\perp$ were both between 0.8 and 1.2, for  \textsc{EZmocks},  \textsc{Nseries}, and 0.7 and 1.3 for \textsc{OuterRim-HOD}). Fig.~\ref{fig:sys} visually displays the results of this table. }
\begin{center}
\begin{tabular}{|c|c|c|c|c|}
Mock name & catalogue & $ \alpha_\parallel -\alpha_\parallel^{\rm exp}$ & $\alpha_\perp -\alpha_\perp^{\rm exp}$ & $N_{\rm det}/N_{\rm tot}$\\
\hline
\hline
Mean \textsc{EZmocks} & pre-recon & $-0.0003\pm0.0015$ & $-0.00480\pm0.00092$ & $1/1$\\
Individual \textsc{EZmocks} & pre-recon & $-0.0165 \pm 0.0019$  & $0.0025\pm0.0010$  & $982/1000$\\
Mean \textsc{EZmocks} & post-recon & $0.0060\pm0.0010$ &  $-0.00024\pm 0.00068$ &  $1/1$ \\
Individual \textsc{EZmocks} & post-recon & $0.0017\pm0.0012$  & $0.00071 \pm0.00073$ & $999/1000$\\
\hline
Mean \textsc{EZmocks} (+Hexadecapole) & pre-recon & $-0.0039 \pm 0.0015$ & $-0.00946 \pm 0.00096$ & 1/1 \\
Mean \textsc{EZmocks} (+Hexadecapole) & post-recon & $-0.0022 \pm 0.0011$ & $ -0.00367 \pm 0.00066$ & 1/1 \\
\hline
Mean \textsc{EZmocks} ($\boldsymbol{\Theta}_{\rm OR}$) & post-recon & $-0.0015 \pm 0.0013$ & $0.00118 \pm 0.00075$ & 1/1 \\
Mean \textsc{EZmocks} ($\boldsymbol{\Theta}_{\rm X}$) & post-recon & $0.0044 \pm 0.0013$ & $-0.00401 \pm 0.00076$ & 1/1 \\
Mean \textsc{EZmocks} ($\boldsymbol{\Theta}_{\rm Y}$) & post-recon & $0.0032 \pm 0.0012$ & $0.00089 \pm 0.00078$ & 1/1 \\
Mean \textsc{EZmocks} ($\boldsymbol{\Theta}_{\rm Z}$) & post-recon & $-0.0004 \pm 0.0011$ & $-0.00838 \pm 0.00072$ & 1/1 \\
\hline
Mean  \textsc{Nseries} & pre-recon & $-0.0045\pm 0.0041$  & $-0.0021\pm 0.0020$ &  $1/1$\\
Individual  \textsc{Nseries} & pre-recon & $-0.0062 \pm 0.0051$   & $0.0000 \pm0.0026$  & $84/84$\\
Mean  \textsc{Nseries} & post-recon & $-0.0048\pm 0.0019$ &  $0.0005\pm 0.0010$   &  $1/1$\\
Individual  \textsc{Nseries} & post-recon &  $-0.0016\pm 0.0033$  & $-0.0030\pm 0.0018$  & $84/84$\\
\hline
Mean \textsc{OuterRim-HOD}-Hearin & pre-recon & $-0.022 \pm 0.014$ & $0.0108 \pm 0.0099$ &  $1/1$\\
Individual \textsc{OuterRim-HOD}-Hearin & pre-recon & $-0.021\pm 0.016$ & $0.016 \pm0.011$ &  $27/27$\\
Mean \textsc{OuterRim-HOD}-Hearin & post-recon & $0.000 \pm 0.011$ & $0.0122 \pm 0.0075$ &  $1/1$\\
Individual \textsc{OuterRim-HOD}-Hearin & post-recon & $0.009\pm 0.015$ & $0.0167\pm 0.0061$  & $27/27$\\
\hline
Mean \textsc{OuterRim-HOD}-Leauthaud & pre-recon & $-0.011 \pm 0.018$  & $0.003 \pm 0.011$ & $1/1$\\
Individual \textsc{OuterRim-HOD}-Leauthaud & pre-recon & $0.000 \pm0.018$  & $0.011 \pm0.012$ & $27/27$\\
Mean \textsc{OuterRim-HOD}-Leauthaud & post-recon & $-0.006 \pm 0.010$ &  $-0.0024 \pm 0.0075$ & $1/1$\\
Individual \textsc{OuterRim-HOD}-Leauthaud & post-recon & $0.002 \pm0.013$  & $-0.0093 \pm0.0074$  & $27/27$\\
\hline
Mean \textsc{OuterRim-HOD}-Tinker & pre-recon & $0.002 \pm 0.018$ & $-0.005 \pm 0.012$ &  $1/1$\\
Individual \textsc{OuterRim-HOD}-Tinker & pre-recon & $-0.011 \pm0.016$  & $0.023 \pm0.011$  & $27/27$\\
Mean \textsc{OuterRim-HOD}-Tinker & post-recon & $-0.002 \pm 0.012$ & $-0.0025 \pm 0.0088$ &  $1/1$\\
Individual \textsc{OuterRim-HOD}-Tinker & post-recon & $0.0038 \pm0.0097$  & $-0.0006 \pm0.0072$ & $27/27$\\
\end{tabular}
\end{center}
\label{tab:sys}
\end{table*}%

\subsubsection{Effect of reference cosmology on BAO}\label{sec:fidcosmoBAO}
\begin{figure}
\centering
\includegraphics[scale=0.29, trim={0 0 0 0}, clip=false]{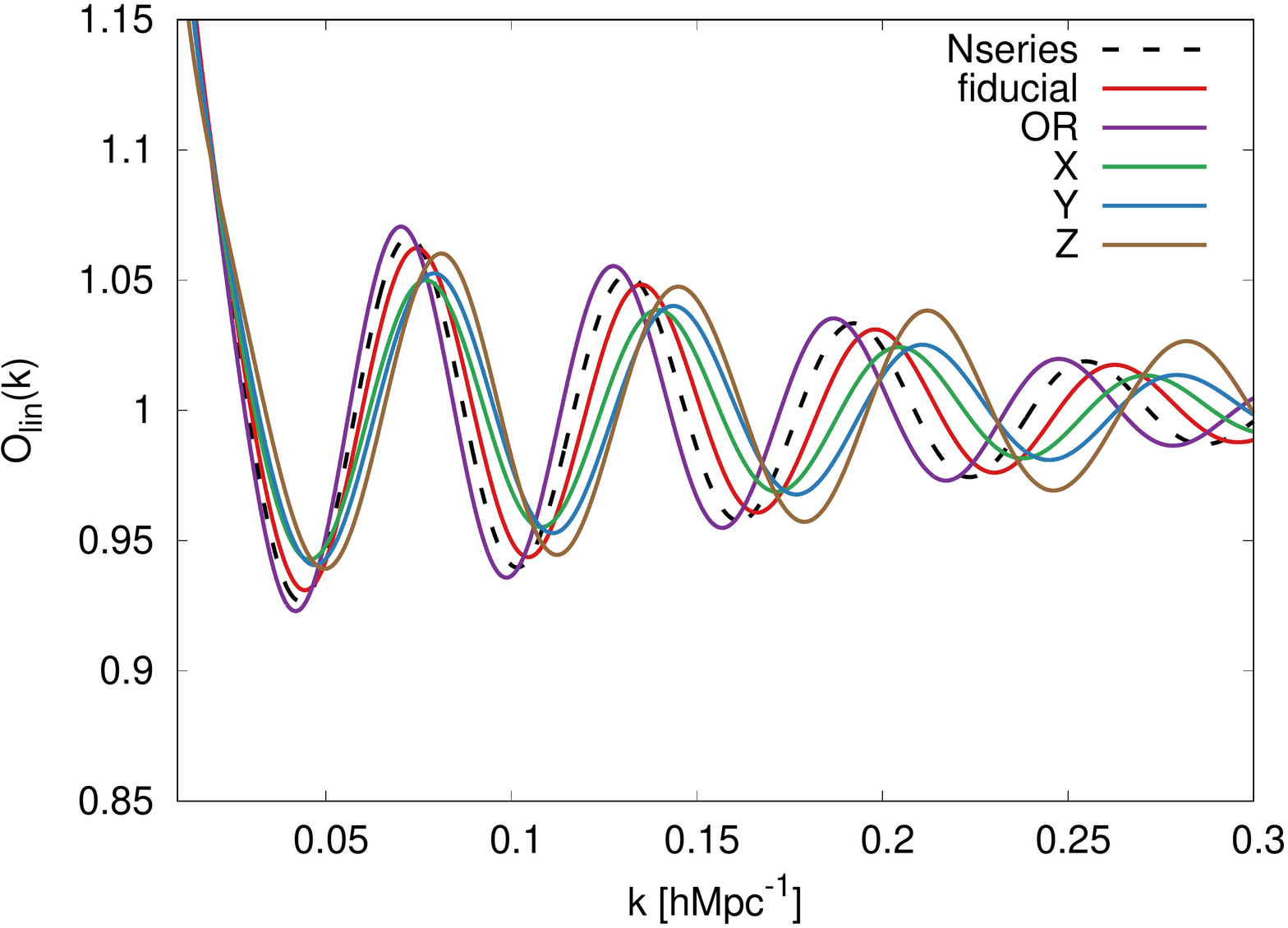}
\includegraphics[scale=0.29, trim={0 40 0 0}, clip=false]{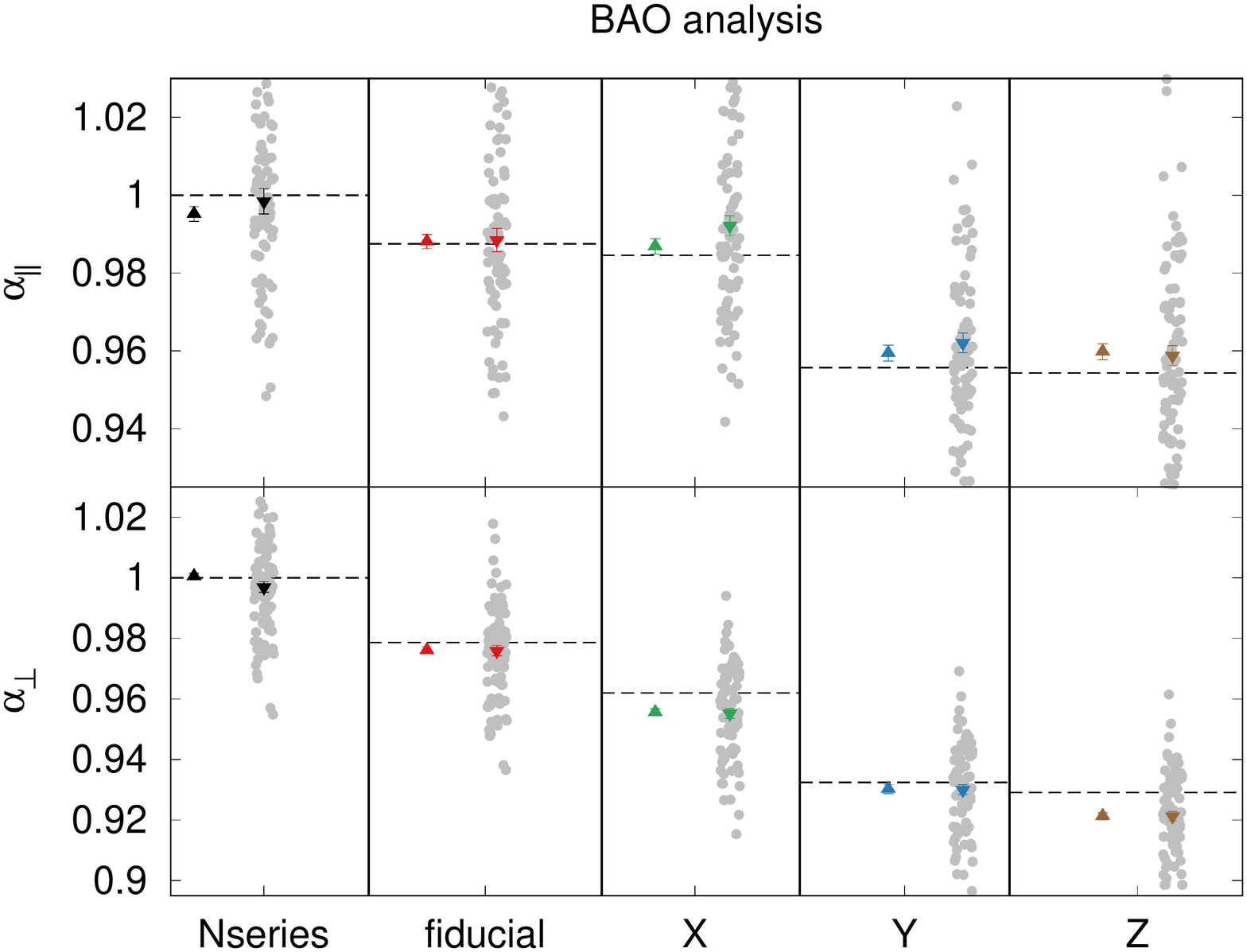}
\includegraphics[scale=0.29, trim={0 40 0 0}, clip=false]{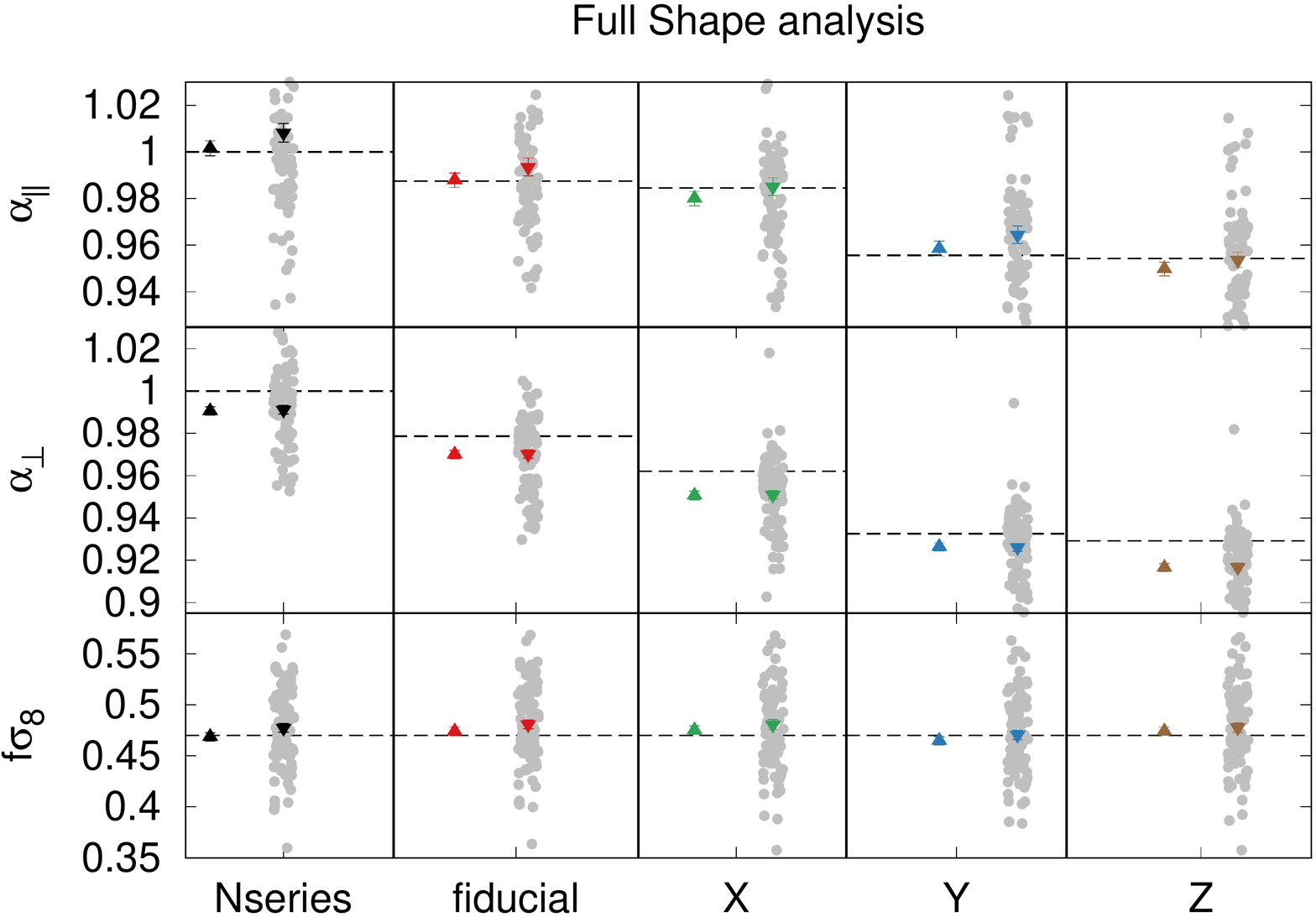}
\caption{{\it Top panel:} BAO signal, $\mathcal{O}_{\rm lin}(k)$ of the cosmologies listed in Table~\ref{tab:cosmo} (except for $\boldsymbol{\Theta}_{\rm EZ}$), used to test the cosmology dependence of the BAO  and FS analyses with the reference cosmology displayed in the middle panel (BAO analysis on post-recon catalogues and listed in Table~\ref{tab:cosmodependence}) and in the bottom panel (FS analysis on pre-recon catalogues listed in Table~\ref{tab:cosmodependenceRSD}). {\it Middle panel}: BAO scale factors along ($\alpha_\parallel$) and across ($\alpha_\perp$) the LOS measured on the reconstructed  \textsc{Nseries} mocks. The $\bigtriangleup$ symbols display the fit on the mean of the 84 realisations, the grey symbols display the fit on the individual 84 realisations, and the $\bigtriangledown$ symbols the mean of the 84 individual fits. The error associated is the error of the mean and the {\it rms} among the 84 realisations divided by $\sqrt{84}$, respectively. {\it Bottom panel}: same notation than the middle panel applied to FS analysis on the same mocks.  The $x$-axis shows the results for different cosmologies used for the template (and for mapping redshifts into comoving distances when computing the power spectrum). The horizontal dashed lines mark the expected value of $\alpha_{\parallel,\,\perp}$ and $f\sigma_8$.}
\label{fig:olin}
\end{figure}

We are interested in testing the potential impact of the arbitrary choice of reference cosmology. For simplicity, we use the same cosmology to {\it i)} produce the BAO template, and {\it ii)} convert redshifts into distances in both random and galaxy catalogues. The BAO analysis measures relative differences between the BAO peak position in the power spectrum with respect to the template. Therefore, \emph{a priori} the specific choice of reference cosmology should not impact this result. In this section we explicitly check this by analysing the \textsc{Nseries} mocks in 5 different cosmologies: their own  cosmology, $\boldsymbol{\Theta}_{\rm Nseries}$; the fiducial cosmology which is used for the baseline analysis of the actual data catalogue, $\boldsymbol{\Theta}_{\rm fid}$; and three extra cosmologies with a higher value of $\Omega_m$: $\boldsymbol{\Theta}_X$, $\boldsymbol{\Theta}_Y$ and $\boldsymbol{\Theta}_Z$, all listed in Table~\ref{tab:cosmo}. The oscillatory features of these cosmologies are plotted in the top panel of  Fig.~\ref{fig:olin}. The expected values of $\alpha_\parallel$ and $\alpha_\perp$ are therefore different among these cosmologies. We do not analyse these mocks on the $\boldsymbol{\Theta}_{\rm OR}$ cosmology, as this cosmology is very similar to the $\boldsymbol{\Theta}_{\rm Nseries}$. Inputing the values of the true $\boldsymbol{\Theta}_{\rm Nseries}$ and these 4 cosmologies into Eq. \ref{eq:apara} and \ref{eq:aperp} we determine the  expected values of the scaling factors. For the $\boldsymbol{\Theta}_{\rm fid}$ cosmology we find that $\alpha^{\rm exp}_\parallel=0.9875$ and $\alpha_\perp^{\rm exp}=0.9787$; for the $\boldsymbol{\Theta}_X$ cosmology we find that  $\alpha_\parallel^{\rm exp} = 0.9846$ and $\alpha_\perp^{\rm exp} = 0.9620$;  for the $\boldsymbol{\Theta}_Y$ cosmology we find that  $\alpha_\parallel^{\rm exp} = 0.9543$ and $\alpha_\perp^{\rm exp} = 0.9325$;  for the $\boldsymbol{\Theta}_Z$ cosmology we find that  $\alpha_\parallel^{\rm exp} = 0.9557$ and $\alpha_\perp^{\rm exp} = 0.9291$; and of course when the mocks are analysed in their own cosmology the expected values are unity. 

The results of the fits on the mean of the mocks and on the individual fits are reported in Table~\ref{tab:cosmodependence}, for post-reconstruction analyses, where reconstruction has been performed using the true value of $f$ and $b_1$. We follow this approach because in this test we aim to check the impact of the arbitrary choice of the reference cosmology when recovering the BAO parameters, rather than to test the efficiency of reconstruction as a function of the assumed parameters (see \citealt{Carter19} for such a study). The middle panel of Fig.~\ref{fig:olin} displays the post-reconstruction results from Table~\ref{tab:cosmodependence}, where the horizontal dashed lines represent the expected values for $\alpha$s in each cosmology. For each case we display the fit to the mean of the $N_{\rm Nseries}$ realisations along with the error of the mean, the individual fit to these $N_{\rm Nseries}$ realisations, and mean of the fit of these $N_{\rm Nseries}$ realisations, along with the {\it rms} divided by $N_{\rm Nseries}^{1/2}$. When studying the post-reconstructed catalogues we find that for both $\alpha_\parallel$ and $\alpha_\perp$ the highest shift, relative to those parameters inferred from $\boldsymbol{\Theta}_{\rm Nseries}$, are those inferred using the templates of $\boldsymbol{\Theta}_Z$, which show a deviation of $1\%$ in $\alpha_\parallel$ and $0.8\%$ for $\alpha_\perp$.  We note that $\boldsymbol{\Theta}_Z$ represents a very distinct cosmology with respect to the true cosmology of \textsc{Nseries}, with shifts of $\Delta\Omega_m=0.08$ and $\Delta\Omega_b=0.019$, which are 10 and 50 sigma away, respectively, from the results reported by Planck \citep{planck_collaboration_planck_2018}. On the other hand, if a closer-to-standard $\Lambda$CDM reference cosmology is used, such as $\boldsymbol{\Theta}_{\rm fid}$, these shifts reduce to $0.5\%$ on $\alpha_\parallel$ and $0.3\%$ for $\alpha_\perp$.

\subsubsection{Effect of non-periodicity on BAO measurements}\label{sec:BAOnonPeriodic}

The \textsc{OuterRim-HOD} mocks comes from a single \textsc{OuterRim} dark matter simulation, split into 27 non-periodic cubic sub-boxes and populated with different types of HOD models and flavours. In this section we aim to quantify the impact of analysing the non-periodic sub-boxes when  DFT algorithms are used to obtain the power spectra, which implicitly assume periodic boundary conditions. In this case, and across the paper we only consider wave numbers between $0.02\leq k\,[\hompc]\leq 0.30$ for BAO analyses. In order to test this impact within such scale-range we only focus on the set of HOD types and flavours closer to the LRG galaxy sample. 
We refer to such padded catalogues as `sky-cuts'. In Table~\ref{table:periodicity} we display the results on the mean of the 27 mocks, for both cubic non-periodic and sky-cut case. Note that the non-periodic cubic and sky-cut contain the same galaxies, therefore, the information content should be the same in the absence of spurious non-periodic effects.  We do not observe any significant changes in $\alpha_\perp$ parameter, but a consistent shift of $\alpha_\parallel$ by $2-3\%$: for the non-periodic cubic box we find an excess in the value of $\alpha_\parallel$ with respect to what is expected, whereas for the sky-cut case the results are in agreement (within $2\sigma$ error-bars) with the expected values. We conclude that 1) the non-periodic effects are important when determining $\alpha_\parallel$, but not $\alpha_\perp$ 2) we do not observe relative shifts in any of the $\alpha$ parameters when the HOD model or flavour is varied. 

\subsubsection{Impact of HOD modelling in BAO}
\begin{figure}
\centering
\includegraphics[scale=0.3]{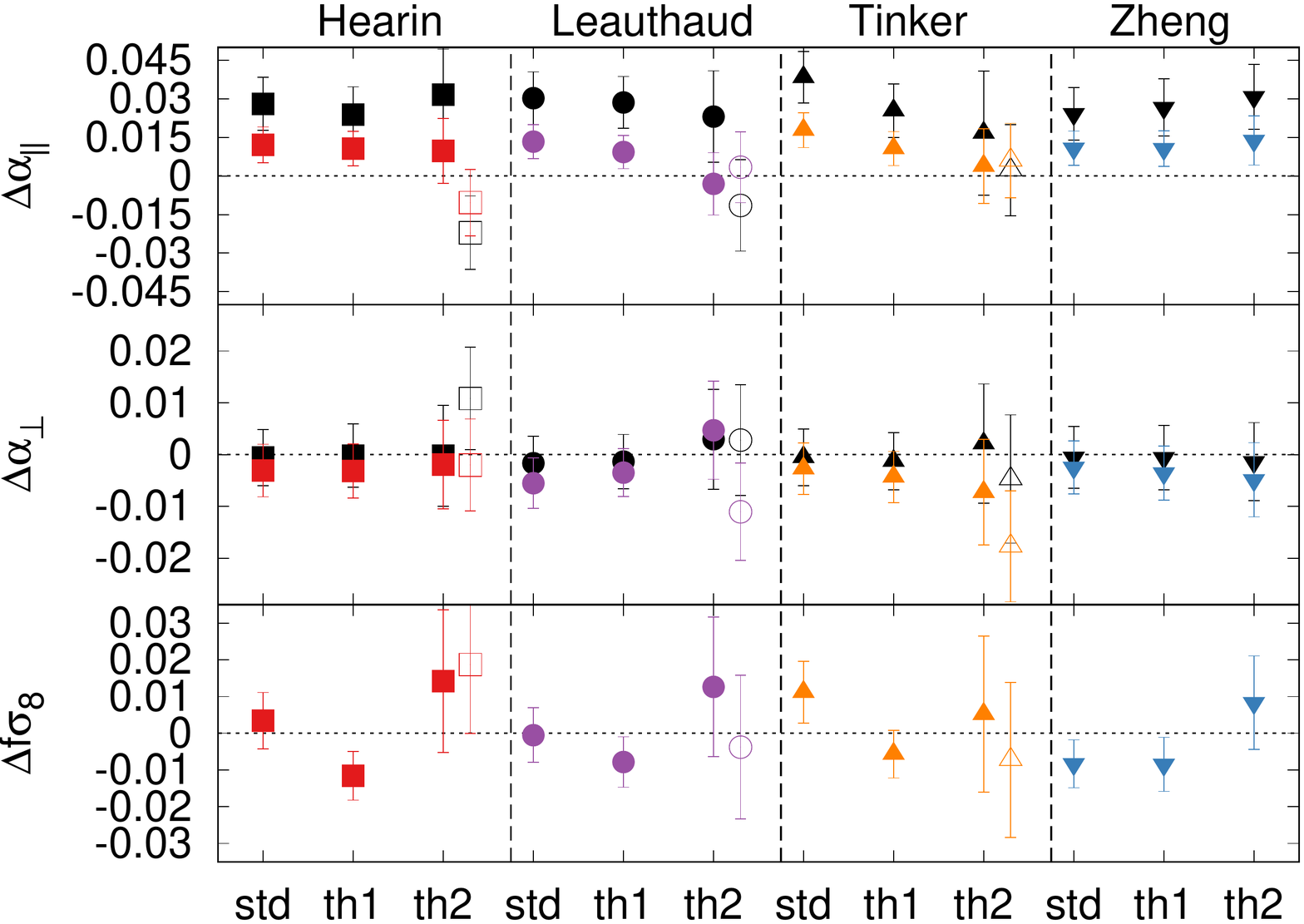}

\caption{Inferred scaling parameters, $\alpha_\parallel$ and $\alpha_\perp$, and $f\sigma_8$ using a BAO  (black symbols) and FS (coloured symbols) type of analysis from the pre-reconstructed \textsc{OuterRim-HOD} mocks, for different types of HOD models and flavours (listed in the $x$-axis). Red, purple, orange and blue colours correspond to Hearin, Leauthaud, Tinker and Zheng HOD models for the FS type of analysis, respectively. For each of these models 3 flavours have been implemented: standard (std), threshold1 (th1) and threshold2 (th2), as labeled. The filled symbols correspond to $1\,h^{-1}{\rm Gpc}$-size cubic box without periodic boundary conditions. The empty symbols correspond to the results obtained from a padded and larger box corresponding to $3\,h^{-1}{\rm Gpc}$-size, where the non-periodicity impact is negligible. All results correspond to fitting the mean of 27 mocks, and the reported error is $1\sigma$ of the error of the mean. }
\label{fig:HOD4BAO}
\end{figure}

In this section we aim to explore the impact of different HODs and flavours when recovering the scaling parameters. Ideally we should pad as well around these catalogues in order to remove the effect of non-periodicity. However, since the galaxy catalogues can be very large for some of the HODs studied (5 million galaxies for threshold1), and the random catalogues need to be at least 20 to 50 times larger, this is not feasible. 
Since the spurious effects of non-periodicity have a geometrical origin, they should be independent of the intrinsic clustering, and we are only interested in the relative effect of the HOD modelling, we opt to analyse the \textsc{OuterRim-HOD} mocks as if they were periodic boxes, and compare only the relative recovered values among them, bearing in mind that we expect a $\sim2-3\%$ offset on $\alpha_\parallel$. These results are listed in Table~\ref{tab:HOD4BAO} for the mean of 27 mocks and graphically represented in Fig.~\ref{fig:HOD4BAO}, where the filled black symbols are the results for the cubic boxes (affected by the non-periodicity) and the empty black symbols the results where the boxes have been padded (not affected by non-periodicity). 
For all the HODs we consistently see that there is the expected $3\%$ offset on $\alpha_\parallel$ as an effect of the non-periodicity of the box, whereas $\alpha_\perp$ is well recovered in all the cases. We therefore conclude that within the statistical errors of $1-2\%$ the AP parameters recovered from the BAO type of analysis are not affected by the HOD model for LRGs. These findings are in agreement with the recent study by \cite{Yutong_Eisenstein19}.

\subsection{RSD systematics}\label{sec:RSDsys}

We repeat the same strategy used for BAO systematics in the previous \S\ref{sec:BAOsys}. We run the standard FS pipeline described previously in \S\ref{sec:RSD} on the pre-reconstructed mocks. We aim to check the typicality of the data with respect to the \textsc{EZmocks} under the FS analysis,  how such analysis responses to change in the HOD of the mocks, the impact of the arbitrary choice of the reference template used to compute the FS model, and to calibrate  the optimal $k$-range to be used on FS in order to maximise the statistical error and minimise the systematic budget. 

\begin{figure}
\centering
\includegraphics[scale=0.3]{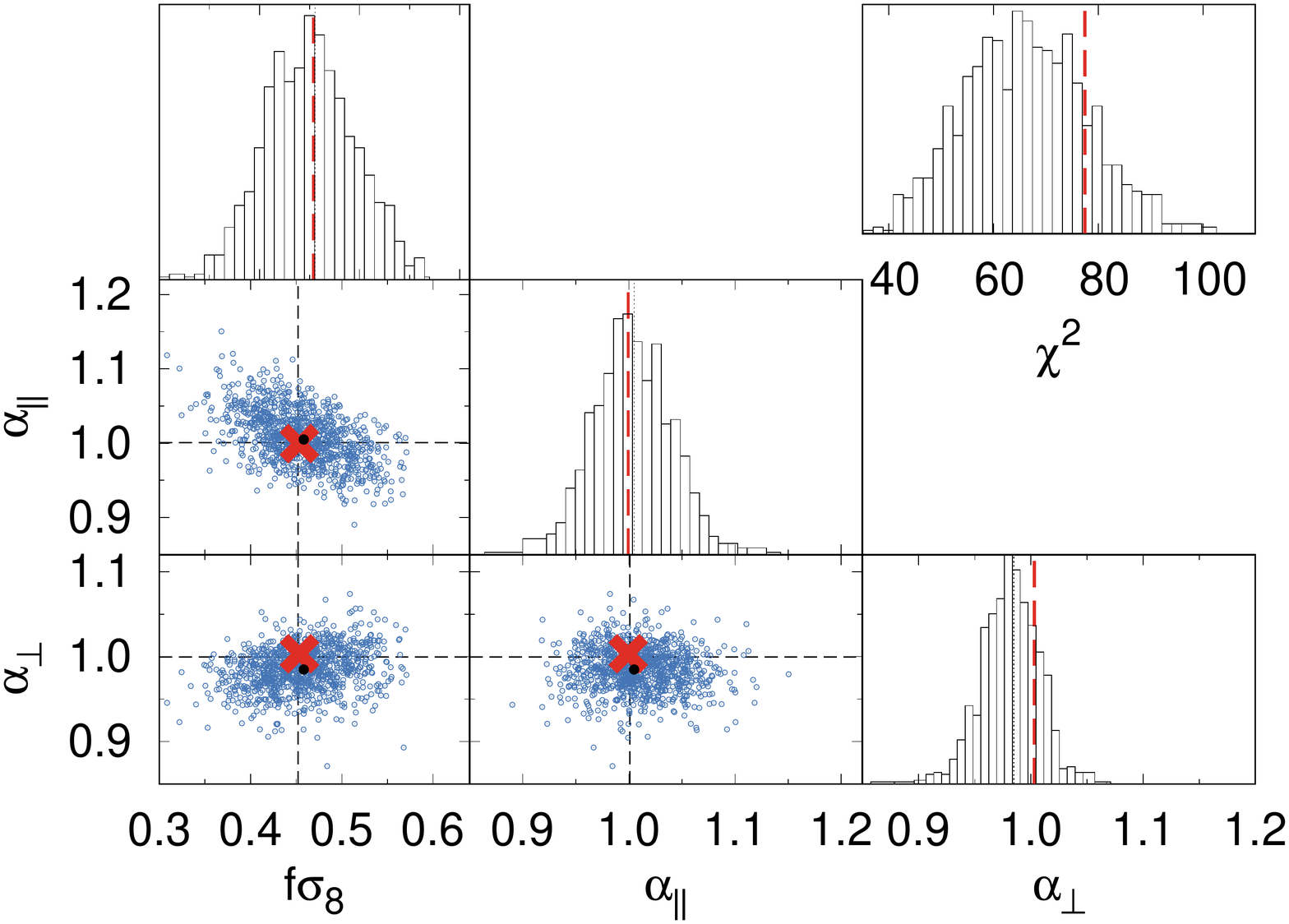}

\includegraphics[scale=0.3]{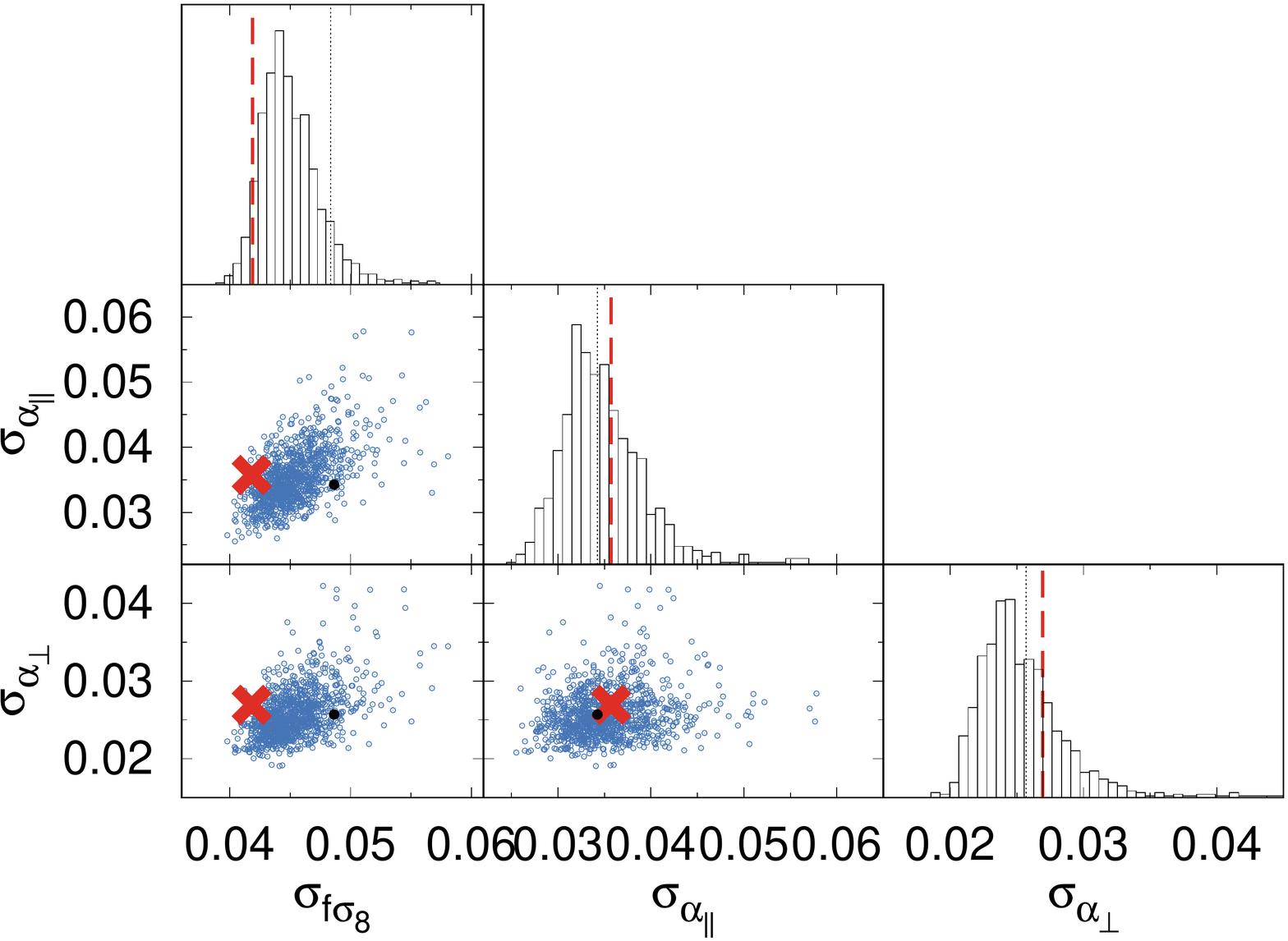}
\caption{The top sub-panels display the distribution of inferred $\alpha_\parallel$, $\alpha_\perp$ and $f\sigma_8$ parameters using a FS analysis of the power spectrum monopole, quadrupole and hexadecapole for the 1000 realisations of the \textsc{EZmocks} catalogues (blue points), DR16 CMASS+eBOSS LRG data catalogue (red cross) and the best-fit to the mean of the 1000 mock power spectra. The horizontal and vertical black dashed lines represent the expected $\alpha$s and $f\sigma_8$ values for the \textsc{EZmocks}. The scales fitted are in all cases $0.02\leq k\,[\hompc]\leq 0.15$. In each panel the distribution of $\chi^2$ values is also shown. The bottom sub-panels display an analogous plots for the errors of $\alpha_\parallel$, $\alpha_\perp$ and $f\sigma_8$ instead. In this case the error of the mean have been re-scaled by the square root of number of realisations.}
\label{Fig:RSD}
\end{figure}

The panels of Fig.~\ref{Fig:RSD} display the recovered parameters, $\alpha_\parallel$, $\alpha_\perp$ and $f\sigma_8$ (top panels), and their errors (bottom panels) from a FS analysis on the power spectrum monopole, quadrupole and hexadecapole using $0.02\leq k\,[\hompc]\leq 0.15$. Later  we justify the choice of these specific scales. As in Fig.~\ref{Fig:bao_ezmocks1}, the red cross displays the performance of the DR16 CMASS+eBOSS LRG data catalogue, and the black dots the performance on the mean of the 1000 realisations of \textsc{EZmocks}. In the bottom panels, the error of the mean has been re-scaled by the factor $\sqrt{1000}$ to match the typical error of one single realisation. We note that the re-scaled error of the mean of the mocks is displaced from the centre of the cloud of errors from individual mocks. This behaviour is caused by the 50\%-prior on $A_{\rm noise}$ on the individual mocks (this prior has no effect when fitting the mean of the mocks) which shrinks the distribution tail towards smaller errors, especially for $\alpha_\parallel$ and $f\sigma_8$. The values and errors of $\alpha_\parallel$, $\alpha_\perp$ and $f\sigma_8$ inferred from the data catalogue are consistent with the intrinsic scatter observed from the mocks; and also the $\chi^2$ value of the data is probable given the distribution of the mocks. 

\subsubsection{Optimal range of scales for the FS analysis}

\begin{figure}
\centering
\includegraphics[scale=0.3]{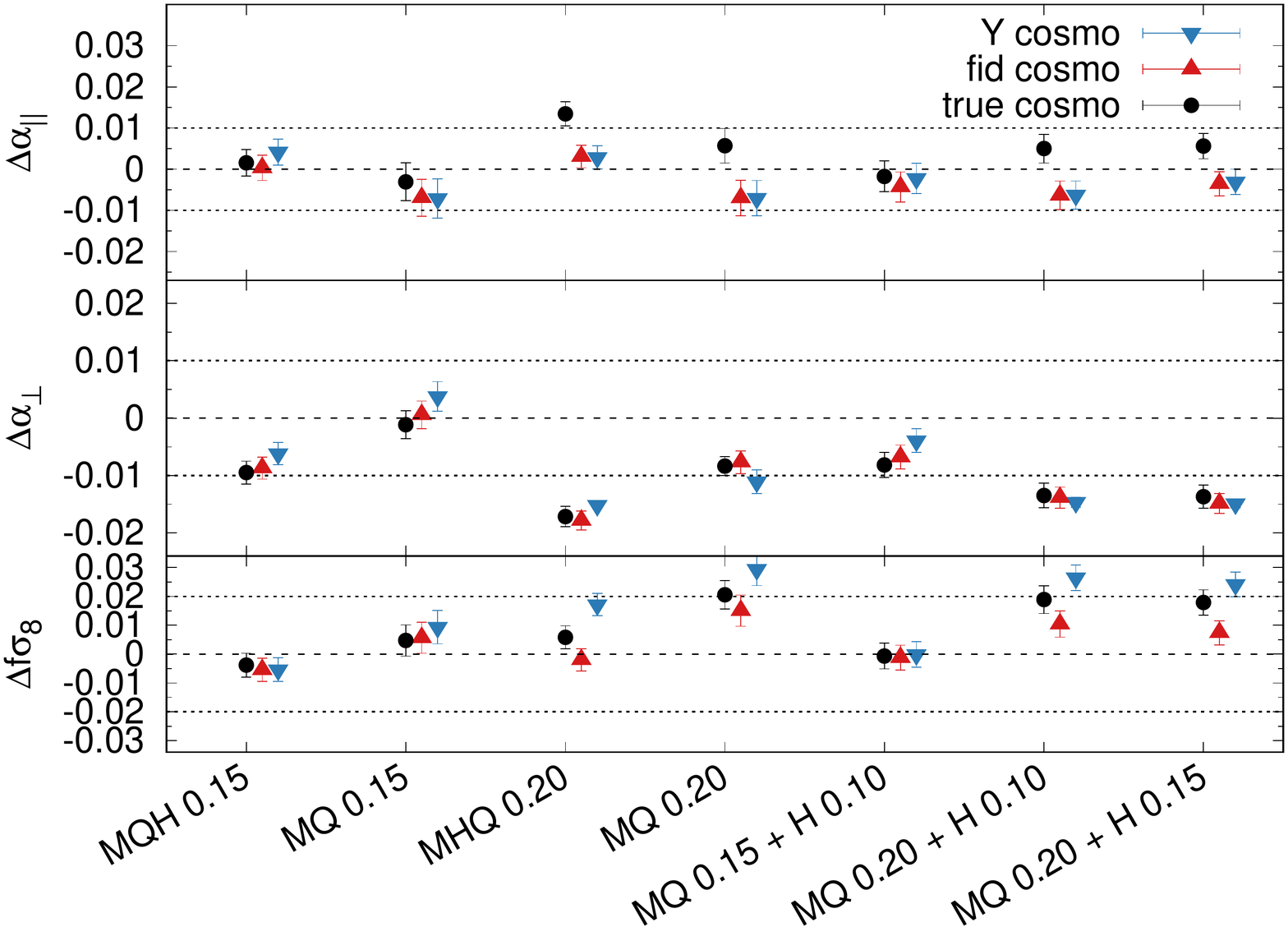}
\caption{Test of performance when recovering cosmological parameters for the FS analysis described in \S\ref{sec:RSD} as a function of the range of scales used in each power spectrum multipole, where $\Delta x\equiv x-x^{\rm exp}$ for $\alpha_\parallel$, $\alpha_\perp$ and $f\sigma_8$ for the three different panels. The symbols represent the response on the mean of the 84 realisations of the \textsc{Nseries} mocks when the fiducial cosmology is used as reference template (red $\bigtriangleup$ symbol), when the $\boldsymbol{\Theta}_Y$-cosmology template is used (blue $\bigtriangledown$ symbol), and when the true cosmology template is used (black $\bullet$ points). 
The $x$-axis labels indicate the combination of multipoles (monopole, quadrupole, hexadecapole) used, and the value of $k_{\rm max}$ in $\hompc$. 
The results do not show any particular trend with the template. Note that in some cases the value of $k_{\rm max}$ is different for the hexadecapole compared to that for the monopole and quadrupole. For all cases the value of $k_{\rm min}$ is set to $0.02\,\hompc$. The horizontal dotted lines represent the $1\%$ deviation for $\alpha$s and $2\%$ for $f\sigma_8$. The numerical results of this plot are displayed in Table \ref{tab:cosmodependenceRSD}.}
\label{fig:kmax}
\end{figure}

We aim to determine the range of scales we should be using when performing a FS analysis. Ideally, the wider this range, the smaller the statistical uncertainty in the inferred cosmological parameters should be. We expect that at very small scales the amount of information on cosmological parameters saturates, although the state-of-the art FS techniques have not reached that limit yet \citep{Handetal2017}. Thus, we set the small scale truncation limit based on the ability of the FS model to recover unbiased cosmological parameters. In order to determine this limit we apply the FS analysis on the full {\it N}-body \textsc{Nseries} mocks truncating the model at different scales. Fig.~\ref{fig:kmax} shows the response of $\alpha_\parallel$, $\alpha_\perp$ and $f\sigma_8$ on the truncation scale, $k_{\rm max}$, when the monopole and quadrupole are used (denoted MQ) and when the hexadecapole is also used (MQH). 

We study the following cases: using only the power spectrum monopole and quadrupole for $0.02\leq k\,[\hompc] \leq k_{\rm max}$ (labeled as MQ $k_{\rm max}$); using monopole, quadrupole and hexadecapole with the same scale truncation $0.02\leq k\,[\hompc] \leq k_{\rm max}$ for all three (labeled as MQH $k_{\rm max}$); and truncation of multipoles at different scales, $0.02\leq k\,[\hompc] \leq k_{\rm max}^{(1)}$ for monopole and quadrupole and  $0.02\leq k\,[\hompc] \leq k_{\rm max}^{(2)}$ for hexadecapole (labeled as MQ $k_{\rm max}^{(1)}$+H $k_{\rm max}^{(2)}$). 
The hexadecapole is more sensitive to RSD than the other multipoles, as the modes parallel to the LOS are relatively more weighted than the transverse ones. Therefore, a potential $\mu$-dependent systematic could appear as a parameter-shift when $k_{\rm max}$ grows for the hexadecapole, but not necessarily when it grows for lower multipoles. 

For completeness, we perform such analysis using three different templates, corresponding to $\boldsymbol{\Theta}_{\rm fid}$ at $z=0.70$ (red symbols), $\boldsymbol{\Theta}_Y$ at $z=0.55$ (blue symbols)  and to $\boldsymbol{\Theta}_{\rm Nseries}$ at $z=0.55$ (black symbols). The difference in redshift among these templates (and in particular for $z=0.70$ which is different from the effective redshift of the \textsc{Nseries} mocks, $z_{\rm eff}=0.55$) only enters in the model through the fixed $\sigma_8$ value in the second order terms of the model, and the difference between the $\sigma_8$ values at these two redshift is $\sim11\%$. From the results of Fig.~\ref{fig:kmax} there is no significant difference among these templates, suggesting that, {\it i}) the shape of the template for the reference cosmology has a negligible impact on the inferred cosmological parameters (we test this later in \S\ref{sec:rsdtemplate} for a wider range of templates); and {\it ii)} the redshift at which this template is computed (which solely regulates its amplitude) has an impact on $f\sigma_8$ which is $<1\%$ when the truncation scale is $\leq0.15\,\hompc$ and $\simeq1\%$ when the truncation scale is $0.20\,\hompc$. This happens due to the non-linear terms proportional to $f\times\sigma_8^n$ for $n>1$ in the non-linear terms of the model as already discussed in \S\ref{sec:parameter_inference}. The statistical error on the $f\sigma_8$ measurement is about 10\% for this sample and consequently this effect is completely negligible. 

The results reported in Fig.~\ref{fig:kmax} suggest that the effect of the truncation scale on $\alpha_\parallel$ is  of order $\leq1\%$ for the $k$-ranges studied here. For $\alpha_\perp$ we find that the effect of increasing the value of $k_{\rm max}$ and including the hexadecapole tend to under-predict its value by $1-1.5\%$ depending on the truncation option. On $f\sigma_8$ the effect of increasing $k_{\rm max}$ can be either over- and under-predict, depending on which template is being used, but the effects are always below $2-3\%$. 

As a fiducial choice, we opt to truncate the FS analysis at $k_{\rm max}=0.15\,\hompc$ for all three multipoles, (MHQ 0.15). This option shows no detected systematic on $\alpha_\parallel$ and $1\%$ systematic on $\alpha_\perp$ and $f\sigma_8$. Alternatively we could also have considered to use only the monopole and quadrupole at the same truncation scale (MQ 0.15), which shows no significant systematic in any of the three variables. However, not using the hexadecapole significantly increases the statistical errors (see Fig.~\ref{fig:contoursRSDBAO}), which does not compensate for the reduction of systematic errors. Therefore, our main analysis relies on the choice MQH 0.15. 

We have not shown any results with $k_{\rm max}$ below $0.15\,\hompc$. The reason is that doing this actually also introduces systematics. This paradoxical effect is caused by worsening the BAO detection. If $k_{\rm max}$ is reduced down to $0.10\,\hompc$, although the power spectrum and RSD behave closer to linear physics (which is better modelled), the lost BAO information between $0.10\,\hompc$ and $0.15\,\hompc$ makes it hard to distinguish the RSD signal from the AP signal, which ends up introducing very long degeneracy tails between $f\sigma_8$, $\alpha_\parallel$ and $\alpha_\perp$, which in the end introduce other type of systematics. This effect even appears in a BAO-only analysis, suggesting that there is an intrinsic effect of the data vector not related to the specific model, BAO or FS, used. On the other hand, we do not explore scales above $k_{\rm min}=0.02\,\hompc$, as they are usually more contaminated by large-scale systematics, and they barely contain extra information on $\alpha_{\parallel}$, $\alpha_\perp$ and $f\sigma_8$. 

In \cite{beutler_clustering_2017} DR12 CMASS LRG galaxies, and in particular those between $0.5\leq z\leq 0.75$ were analysed under the truncation scheme `MQ 0.15 H 0.10' using a very similar model. In this work, we have not found significant differences in terms of systematics between `MQ 0.15 H 0.10'  and `MQH 0.15', and therefore we have decided to also include those hexadecapole modes between $0.10\leq k\,[\hompc]\leq 0.15$.  

\subsubsection{Performance of the RSD modelling}\label{sec:RSDmodelling}
\begin{figure*}
\centering
\includegraphics[scale=0.5]{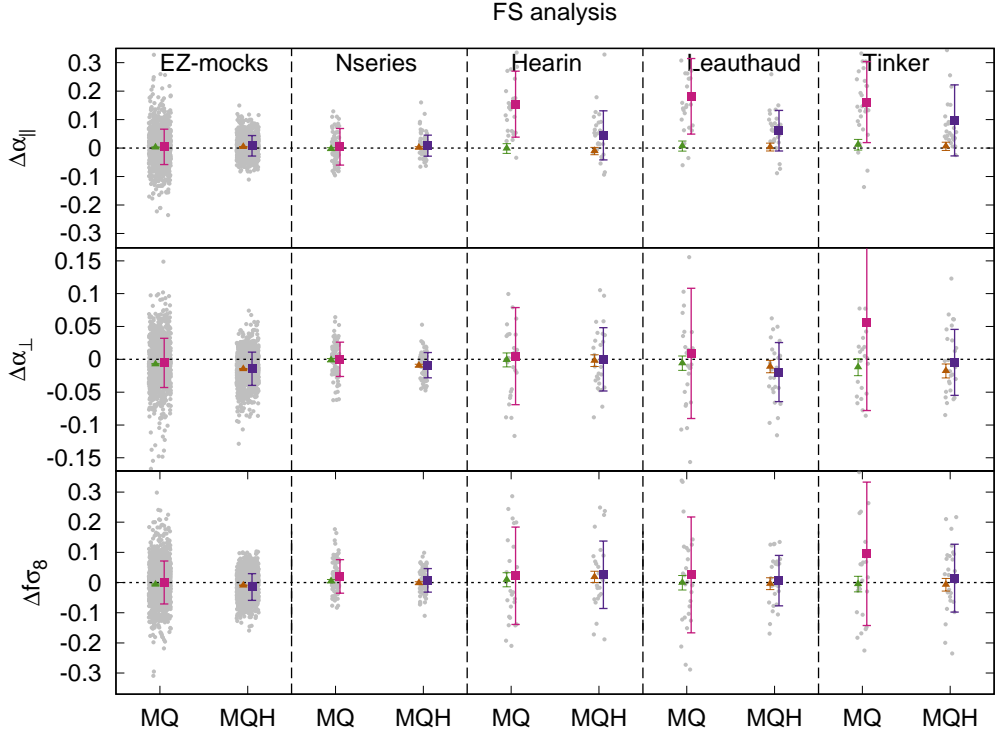}
\caption{Performance of the FS type of analyses on different type of mocks, with a similar notation as in Fig.~\ref{fig:sys}. In this case we analyse only pre-reconstruction catalogues using monopole and quadrupole only (MQ) and also including the hexadecapole (MQH). In all cases the range of scales used are $0.02\leq k\, [\hompc]\leq 0.15$, motivated the findings of Fig.~\ref{fig:kmax}. The corresponding numerical results for this figure can be found in Table~\ref{tab:sysRSD}. }
\label{fig:sysRSD}
\end{figure*}

We apply the FS template pipeline described in \S\ref{sec:RSD} for $0.02\leq k\,[\hompc] \leq 0.15$ on the different set of mocks: the results are displayed in Fig.~\ref{fig:sysRSD} for \textsc{EZmocks}, \textsc{Nseries} and \textsc{OuterRim-HOD} mocks (for threshold2 flavour only and with padding included), as labeled, when the monopole and quadrupole are used (MQ) and when the hexadecapole is also used (MQH). The green and orange symbols display the results when fitting the mean of the mocks and report the error of the mean. The pink and purple symbols display the average of the individual fits (displayed in grey) and report the {\it rms} in the error-bars. Consequently the error on the mean of the fits is a factor $\sqrt{N_{i}}$ larger than the error on the fit of the mean. We stress that the error bars associated to both measurements should be the same, as they are both inferred from the same volume (except for large scale modes and cases where the BAO is not detected), although the latter may suffer from extra systematic effects if the individual fits do not have a sufficiently high signal-to-noise ratio. In Fig.~\ref{fig:sysRSD} (as in Fig.~\ref{fig:sys}) we opt to display the {\it rms} for the mean of the individual fits as an error-bar, along with the individual fits in grey, for visualisation purposes. These results are also presented in Table~\ref{tab:sysRSD}, using the same notation and format of Table \ref{tab:sys}, where the error reported for the mean of individual fits is the {\it rms} divided by $\sqrt{N_{\rm det}}$, matching the error of the mean. 
The results displayed in Fig.~\ref{fig:sysRSD} for the \textsc{EZmocks} and \textsc{Nseries} present a good agreement between the fit of the mean and the mean of individual fits. For the \textsc{Nseries} mocks $\alpha_\parallel$ and $f\sigma_8$ are recovered with no systematics, for both MQ and MQH. 
For the \textsc{Nseries} mocks we are able to recover the expected $\alpha_\perp$ when the monopole and quadrupole are used, whereas when we add the hexadecapole there is a systematic offset of $\sim1\%$, as was reported in the previous section and in Fig.~\ref{fig:kmax}. As we have already mentioned, the \textsc{EZmocks} are not full {\it N}-body mocks, and therefore they should not be used to validate our pipeline in terms of systematics of the model. However, as a general trend we observe that the response these mocks have is very similar to the one observed for the \textsc{Nseries}, which serves as a validation of these mocks reproducing the clustering properties of full {\it N}-body mocks, and therefore for producing a reliable covariance. 

The results on the \textsc{OuterRim-HOD} mocks present a significant difference between the fit of the mean (green and orange symbols) and the mean of the fits (pink and purple symbols). In general for the fit of the mean the cosmological parameters $\alpha_\parallel$, $\alpha_\perp$ and $f\sigma_8$ are recovered within the statistical uncertainty, which is $1-3\%$. However the individual fits and its mean present systematic shifts, over-estimating $\alpha_\parallel$ and $f\sigma_8$. These shifts are larger for MQ than MQH. Since these effects are not shown in the mean of the fits, we conclude that they are caused by the low signal of individual realisations. Indeed the volume of the realisations of each of these mocks is not high enough to have good BAO detections, in particular for $\alpha_\parallel$ which is always the worse-detected scaling factor. In such low-signal-to noise conditions, the model tends to shift the BAO scale to larger scales (recall that $k'=k/\alpha$ in Fourier space, or $s'=s\alpha$ in configuration space), where the noise-per-$k$-mode is higher. This effect is partially mitigated by adding the hexadecapole, as the signal of the data-vector is increased. The reason why this effect is only present in the \textsc{OuterRim-HOD} mocks, and not in the \textsc{Nseries} or \textsc{EZmocks} has to do with the effective volume per mock, which is $1.10\,{\rm Gpc}^3$ for the  \textsc{OuterRim-HOD} mocks, $3.67\,{\rm Gpc}^3$ for the \textsc{Nseries}, and  $2.72\,{\rm Gpc}^3$ for the \textsc{EZmocks}.

In summary, {\it i}) the results displayed in Fig.~\ref{fig:sysRSD} show no systematic shift as a consequence to the change in the HOD models on any of the cosmological parameters; {\it ii}) the statistical limit of the previous statement only applies to potential systematics larger than the statistical limit of the \textsc{OuterRim-HOD} mocks, which is $1-3\%$; and {\it iii}) we observe a systematic shift of $\sim1\%$ on $\alpha_\perp$ when the \textsc{Nseries} mocks are analysed and no strong systematic shift on $\alpha_\parallel$ or $f\sigma_8$ above the statistical uncertainty, which for these mocks is $\sim0.5\%$. 

In Table~\ref{tab:sysRSD} we also display two additional results on potential observational systematics. We perform the FS analysis on the \textsc{EZmocks} before, labeled as \textsc{EZmocks} (raw), and after, labeled as \textsc{EZmocks}, applying the observational effects (which includes completeness and collision weights) followed by the corresponding correction applied in the data catalogue and described in \S\ref{sec:data}. These effects are redshift failures, close-pair collisions (only in the eBOSS LRGs) and completeness. The relative systematic shift of all these effects can be estimated by the difference of the inferred cosmological parameters from the \textsc{EZmocks} before and after applying them. For the fiducial case of MQH we obtain that these shifts are of order $0.7\%$ for $\alpha_\parallel$; of order $0.2\%$ for $\alpha_\perp$; and of order $0.6\%$ for $f\sigma_8$. Individually such shifts are sub-dominant with respect to the statistical error of the data.

We also check the effect of the radial integral constraint (hereafter RIC; see \citealt{demattia20a} for a description of this effect). We turn on and off this effect by computing the power spectra of the individual mocks with and without a random catalogue that matches the $n(z)$ of the galaxy catalogue. The best-fitting parameters to the mean of the mocks without the RIC is shown in Table~\ref{tab:sysRSD}. The effect of RIC on $\alpha_\parallel$ and $\alpha_\perp$ is about $0.3\%$ and  $1.6\%$ on $f\sigma_8$. 

\begin{table*}
\caption{Performance of the FS model of Eq \ref{eq:baoaniso} in different set of mocks. For The EZmocks $\alpha_\parallel^{\rm exp}=1+8.853\cdot10^{-4}$, $\alpha_\perp^{\rm exp}=1-3.650\cdot10^{-4}$ and $f\sigma_8^{\rm exp}=0.46781$. For the rest the $\alpha$ expected values are $1$ as they are respectively analysed in their own cosmology. For. the  \textsc{Nseries} mocks, $f\sigma_8^{\rm exp}=0.470166$, and for the \textsc{OuterRim-HOD} mocks, $f\sigma_8^{\rm exp}=0.4475$. For the \textsc{OuterRim-HOD} type of mocks only the threshold2 flavour is represented, where a padding has been applied to the original non-periodic cubic sub-box in order avoid spurious non-periodic effects. We follow the same notation presented in Table~\ref{tab:sys}. Fig.~\ref{fig:sysRSD} visually displays the results of this table. All results at $k_{\rm max}=0.15\,\hompc$.}
\begin{center}
\begin{tabular}{c|c|c|c|c|c}
Mock name & multipoles &$ \alpha_\parallel -\alpha_\parallel^{\rm exp}$ & $\alpha_\perp -\alpha_\perp^{\rm exp}$ & $f\sigma_8-f\sigma_8^{\rm exp}$ & $N_{\rm det}/N_{\rm tot}$\\
\hline
\hline
Mean \textsc{EZmocks} & M+Q+H & $0.0040\pm0.0011$ & $-0.01481\pm0.00081$ &$-0.0096\pm0.0015$ & $1/1$\\
Individual \textsc{EZmocks} & M+Q+H & $0.0079 \pm0.0011$  & $-0.01444 \pm0.00080$&  $-0.0124 \pm0.0014$ & $1000/1000$\\
Mean \textsc{EZmocks} & M+Q & $0.0025\pm0.0016$ & $-0.0075\pm0.0011$ &$-0.0053\pm0.0021$ & $1/1$\\
Individual \textsc{EZmocks} & M+Q& $0.0043 \pm 0.0020$  & $-0.0055\pm0.0012$&  $0.0003\pm 0.0023$ & $985/1000$\\
\hline
Mean \textsc{EZmocks} (raw) & M+Q+H & $ 0.0108\pm 0.0011$ & $-0.01680 \pm 0.00078$ & $ -0.0161 \pm 0.0014$ & $1/1$\\
Mean \textsc{EZmocks} (raw) & M+Q & $0.0011\pm0.0016$ & $-0.00568\pm0.00098$ & $-0.0016\pm0.0019$ & $1/1$\\
\hline
Mean \textsc{EZmocks} (no-RIC) & M+Q+H & $0.0073 \pm0.0011$ & $-0.01791\pm0.00081$ & $-0.0170\pm0.0015$ & $1/1$\\
Mean \textsc{EZmocks} (no-RIC) & M+Q & $0.0020\pm0.0016$ & $ -0.0080 \pm 0.0011$ & $  -0.0077 \pm 0.0020$ & $1/1$\\
 
\hline
Mean  \textsc{Nseries} & M+Q+H  & $0.0016\pm 0.0032$  & $-0.0095\pm 0.0020$ & $-0.0038\pm0.0041$  & $1/1$\\
Individual  \textsc{Nseries}  & M+Q+H& $0.0082 \pm 0.0040$  & $-0.0089 \pm 0.0021$  & 0.0073$\pm0.0043$ & $84/84$\\
Mean  \textsc{Nseries}& M+Q  & $-0.0031\pm 0.0046$  & $-0.0011\pm 0.0024$ & $0.0047\pm0.0054$  & $1/1$\\
Individual  \textsc{Nseries}  & M+Q& $0.0045 \pm 0.0070$  & $0.0001 \pm 0.0029$  & $0.0201\pm0.0061$ & $84/84$\\
\hline
Mean HOD-Hearin & M+Q+H & $-0.010 \pm 0.013$ & $-0.0020 \pm 0.0089$ & $0.019\pm0.019$ & $1/1$\\
Individual HOD-Hearin & M+Q+H & $0.045\pm 0.017$ & $0.0001 \pm0.0093$ & $0.026\pm0.022$ & $27/27$\\
Mean HOD-Hearin & M+Q& $-0.002 \pm 0.017$ & $-0.001 \pm 0.011$ & $0.009\pm0.023$ & $1/1$\\
Individual HOD-Hearin & M+Q & $0.154\pm 0.022$ & $0.005 \pm0.014$ & $0.022\pm0.031$ & $27/27$\\
\hline
Mean HOD-Leauthaud & M+Q+H & $0.003 \pm 0.014$  & $-0.0111 \pm 0.0094$ & $-0.004\pm0.020$ & $1/1$\\
Individual HOD-Leauthaud  & M+Q+H & $0.061 \pm  0.014$  & $-0.0195 \pm0.0087$ & $0.006\pm0.016$ & $27/27$\\
Mean HOD-Leauthaud & M+Q& $0.007 \pm 0.018$  & $-0.006 \pm 0.011$ & $-0.001\pm0.024$ & $1/1$\\
Individual HOD-Leauthaud  & M+Q & $0.182 \pm  0.026$  & $0.009 \pm0.019$ & $0.025\pm0.037$ & $27/27$\\
\hline
Mean HOD-Tinker & M+Q+H  & $0.006 \pm  0.014$ & $-0.018 \pm 0.011$ & $-0.007\pm0.021$ & $1/1$\\
Individual HOD-Tinker  & M+Q+H & $0.097 \pm0.024$  & $-0.0047 \pm0.0097$ & $0.014\pm0.022$ & $27/27$\\
Mean HOD-Tinker & M+Q  & $0.012 \pm  0.019$ & $-0.012 \pm 0.013$ & $-0.005\pm0.026$ & $1/1$\\
Individual HOD-Tinker  & M+Q & $0.162 \pm0.027$  & $0.056 \pm0.026$ & $0.095\pm0.046$ & $27/27$\\
\end{tabular}
\end{center}
\label{tab:sysRSD}
\end{table*}

\subsubsection{Impact of reference template}\label{sec:rsdtemplate}

In the bottom panel of Fig.~\ref{fig:olin} we display the recovered cosmological parameters for the \textsc{Nseries} mocks when they are analysed assuming 5 reference templates: $\boldsymbol{\Theta}_{\rm Nseries}$, $\boldsymbol{\Theta}_{\rm fid}$, $\boldsymbol{\Theta}_X$, $\boldsymbol{\Theta}_Y$ and $\boldsymbol{\Theta}_{Z}.$\footnote{We consistently map the redshifts into comoving distances in the catalogues using the cosmology of these templates.} As before,  we display both the fit on the mean, and the mean of the individual fits (which are also shown in grey). Table~\ref{tab:cosmodependenceRSD} lists these results. The horizontal dashed lines are the expected values, for each of these cosmologies. Note that the expected $\alpha_\parallel$ and $\alpha_\perp$ change with $\Omega_m$ and $h$, but $f\sigma_8$ does not, as the latter is an absolute variable which does not depend on the choice of template. 

For a displacement in the reference cosmology from $\boldsymbol{\Theta}_{\rm Nseries}$ and $\boldsymbol{\Theta}_{\rm fid}$ we do not observe any shift larger than the $1\sigma$ error-bar of one of the measurements. As the reference cosmology moves away from the true cosmology we observe a mild rise of systematics, which reaches $0.6\%$ and $0.4\%$ for $\alpha_\parallel$ and $\alpha_\perp$ respectively for $\boldsymbol{\Theta}_Z$, although the significance of detection given the statistics of \textsc{Nseries} is not very high. For $f\sigma_8$ the largest deviation appears for $\boldsymbol{\Theta}_X$, which reaches $\Delta f\sigma_8=0.009$, which is about a $2\%$ shift. As for the scaling variables, the significance of detection of this shift is well within a $2\sigma$ fluctuation. 

Bear in mind that the $\boldsymbol{\Theta}_X,\,\boldsymbol{\Theta}_Y,\,\boldsymbol{\Theta}_Z\,$ templates are very extreme cases with shifts of order $\Delta\Omega\simeq0.065$. In addition, for $\boldsymbol{\Theta}_Z$ the baryon density is $\sim50\sigma$ above the Planck-inferred value and for $\boldsymbol{\Theta}_Y$ the number of neutrino species is $N_{\rm eff}=4.046$. Even in such cases the systematics on the $\alpha$'s stay well below $1\%$.  These findings demonstrate how insensitive LSS structure data is to $\Omega_b$.
Alternatively one could put priors on $\Omega_b$ by assuming a $\Lambda$CDM model through the horizon scale, which also set tight constraints on the relation between $\alpha_\parallel$ and $\alpha_\perp$ \citep{Damicoetal2020,Ivanovetal:2020,Trosteretal:2020}.  By taking this approach one would increase the precision in measuring $\alpha_\parallel$ and $\alpha_\perp$ at the expense of assuming the functional form that a generic  $\Lambda$CDM imposes between these two scale factors.


\subsubsection{HOD and periodicity}

In \S\ref{sec:BAOnonPeriodic} we have already tested the impact of the non-periodicity of the 27 sub-boxes generated from the \textsc{OuterRim} mocks for a BAO type of analysis. In this section we do the same for the FS analysis. We use the same catalogues as before (Hearin, Leauthaud and Tinker with threshold2 flavour), and we compare how embedding the non-periodic $1\,h^{-1}{\rm Gpc}$ box into a $3\,h^{-1}{\rm Gpc}$ box changes the results. Table~\ref{tab:RSDperiodicity}  displays the results on performing a FS on the padded and unpadded catalogues. These results can also be seen in Fig.~\ref{fig:HOD4BAO}, where the empty symbols display the padded results, for FS analysis on the Hearin+threshold2 (red), Leauthaud+threshold2 (purple) and Tinker+threshold2 (orange). As was found for BAO fits (in black symbols in the same figure), only $\alpha_\parallel$ presents a significant shift for the Hearin case, whereas $\alpha_\perp$ and $f\sigma_8$ seem barely altered by this effect. In the same panel, and also along Table~\ref{tab:RSDHOD}, the results using other flavours (standard, and threshold1 as labeled), as well an extra HOD model (Zheng in blue symbols) are also displayed for FS. The full picture from BAO and RSD fits is that only $\alpha_\parallel$ is affected by the non-periodicity of the box. The systematic shift is of order $1.5\%$ for the FS analysis (unlike the $2-3\%$ for BAO analysis). This disparity can be due to the variation in the scales fitted, as well as the intrinsic modelling. As for the BAO analysis we do not detect any significant relative shift on the cosmological parameters when either the HOD model or the flavour is varied. Such results put constrains in the upper limit of systematic errors in the modelling as a result of different HOD models. Such upper limits are of order $0.5-1\%$ systematic shifts. 

\subsection{Systematic error budget}\label{sec:systematicbudget}
In this section we summarise all the potential systematic error contributions described above, for both BAO  and FS  analyses, and describe how the total systematic budget of the main results of this paper is computed. Additionally, we also quantify the systematic errors when a simultaneous BAO and FS fit is performed (see \S\ref{sec:consensus} for details on how the simultaneous fit is performed). We consider the BAO-type of analysis on the post-recon catalogue in the scale range $0.02\leq k\,[\hompc] \leq 0.30$, and the FS analysis on the pre-recon catalogue using monopole, quadrupole and hexadecapole in the scale range $0.02\leq k\,[\hompc] \leq 0.15$. We consider the following systematics.
\begin{itemize}
\item {\it Modelling systematics}, associated with the inaccuracies in the theoretical or phenomenological model used. We test these by comparing the inferred value from the \textsc{Nseries} mocks and the expected value, when the mocks are analysed using their own true cosmology as a reference cosmology. These results have been presented in Table~\ref{tab:sys} and \ref{tab:sysRSD} for BAO and FS, respectively. 
\item {\it Reference cosmology systematics}. We test the arbitrary choice of the reference cosmology (both to convert redshift into distances and to choose the modelling template), for both BAO  and FS analysis. We test the relative differences between $\Delta x=x-x^{0}$ for four different reference cosmologies, and take the highest observed deviation (noted as `limit' in Table~\ref{tab:systot}),  where the super-index `0' corresponds to the parameters inferred using its own true cosmology as reference cosmology. In particular we test the differences between $\boldsymbol{\Theta}_{\rm Nseries}$ and $\boldsymbol{\Theta}_{\rm fid}$, $\boldsymbol{\Theta}_{X}$, $\boldsymbol{\Theta}_{Y}$, $\boldsymbol{\Theta}_{Z}$. These results can be found in Tables \ref{tab:cosmodependence} and  \ref{tab:cosmodependenceRSD} for BAO and FS analysis, respectively. 
\item {\it Observational systematics}, such as the effect of redshift failures, collisions and completeness, and systematics derived from the radial integral constraint. In order to test these types of systematics we take the difference between the fit on the \textsc{EZmocks}, when these effects are applied, with respect to the fit to those raw \textsc{EZmocks} previous to the appliance of the effect. We consider two separate cases, {\it i)} collisions, failures and completeness effects (FCC); and {\it ii)} the radial integral constrain effect (RIC).\footnote{In order to remove the RIC we generate a common random catalogue from all the individual 1000 random catalogues, taking a random $0.1\%$ fraction of the objects.} We only compute the contribution of the FCC and RIC systematics for the FS and consensus FS+BAO cases, as the BAO peak position is very insensitive to such effects. These results have been presented in Table~\ref{tab:sysRSD}.
\end{itemize}

For simplicity we consider only the results from the fit to the mean of the mocks, as it is less sensitive to noise effects compared to the mean of individual fits. We also consider that a systematic is detected if the deviation between the expected and measured variable is higher than $2\sigma$. In case of no detection we assign as a systematic contribution the corresponding $2\sigma$ value, which sets a limit in sensitivity. Note that $2\sigma$ corresponds to the $95\%$ confidence level of the mean of the mocks, whose effective volume is 113 times larger, for the \textsc{Nseries} mocks, and 1000 times larger for the \textsc{EZmocks}, than the DR16 CMASS+eBOSS LRG dataset.  Table~\ref{tab:systot} displays the full systematic contribution on the variables of interest. 

For the post-reconstruction BAO analysis we detect a $\sim0.5\%$ systematic shift induced by the modelling systematic on $\alpha_\parallel$, and none for $\alpha_\perp$, with a resolution limit of $\sim0.2\%$. The choice of reference cosmology places an error of about $\sim1\%$ and $\sim0.9\%$ on $\alpha_\parallel$ and $\alpha_\perp$. Such shifts are observed to be higher for the $\boldsymbol{\Theta}_Z$ cosmology. When both effects are taken into account we find that the total systematic contribution increases by $\sim10\%$ for both $\alpha_\parallel$ and $\alpha_\perp$. The reader might think that these error-bars are unrealistically inflated as the $\boldsymbol{\Theta}_Z$ cosmology represents a cosmology strongly disfavoured by state-of-the-art CMB measurements. If we only consider $\boldsymbol{\Theta}_{\rm fid}$ and $\boldsymbol{\Theta}_{X}$ as acceptable  templates instead, the systematic shifts are reduced to $0.5\%$ and $0.3\%$ for $\alpha_{\parallel}$ and $\alpha_\perp$, respectively (similar to what was found in \citealt{gil-marin_clustering_2016-1} for the DR12 LRG sample). In this less conservative case the error-bars would increase by $7\%$, instead. Therefore, the total systematic error budget is not strongly modified by these `priors' on the selection of reference cosmologies. As a conservative choice, we keep the total error budget as the most conservative one, where all four studied templates are considered. 

When we look at the variables of the FS analysis we find that, for $\alpha_\parallel$ the dominant source of systematics are FCC and the choice of reference cosmology, both contributing to about $\sim0.7\%$. The RIC contributes $0.3\%$ and we do not resolve any modelling systematic contribution ($<0.6\%$). The total systematic contribution enlarges the error budget by $7\%$. 
For $\alpha_\perp$ the dominant source of systematics is the modelling, with about $1\%$ systematic contribution. The other sources of systematics correspond to $0.2\%$ and $0.3\%$ for FCC and RIC, respectively. We do not detect any systematic related to the choice of the reference cosmology below the resolution limit ($<0.5\%$). The total error contribution of $\alpha_\perp$ increases by $8\%$ due to systematics. 
For $f\sigma_8$ the dominant source of systematic is the reference cosmology, which represents a shift of $\Delta f\sigma_8\simeq 0.009$ (about $2\%$), which corresponds to the $\boldsymbol{\Theta}_X$ reference cosmology. The FCC and RIC generate shifts of around $0.007$, and we observe no significant shift caused by modelling systematics. In total, the errors are increased by 8\% due to the systematic contribution.

When we analyse BAO and FS simultaneously we obtain systematic shifts which are comparable to those obtained by considering these analyses individually. For the scaling parameters the most important source of systematics is the choice of reference cosmology, which produces systematic shifts of about $0.7\%$, which corresponds to the $\boldsymbol{\Theta}_Y$ and $\boldsymbol{\Theta}_Z$ cosmologies. If only the $\boldsymbol{\Theta}_{\rm fid}$ and $\boldsymbol{\Theta}_{X}$ reference cosmologies were considered, these shifts would be reduced to $0.6\%$ and $0.4\%$ for $\alpha_\parallel$ and $\alpha_\perp$, respectively. As before, we take the conservative choice where all the reference cosmologies are considered, which does not modify significantly the final errors. We find that the total errors on $\alpha_\parallel$ and $\alpha_\perp$ are increased by 8\%, and 10\%, respectively. For $f\sigma_8$ we find that the dominant source of systematic is the modelling, with a shift of $0.018$ (about $4\%$). The total error budget increases by $15\%$ due to systematics. 

We have not included any BAO-type systematic from the \textsc{OuterRim-HOD} mocks in this section. The reason is that in any of the cases studied no such systematic shift was detected. However, the resolution limit of these mocks is poor given their low effective volume ($\sim27\, {\rm Gpc}^3$ for Threshold2 HOD types). Consequently, according to our detection criterion, only shifts of order $2-6\%$ would be detected. These figures would set limits for potential systematic which are above a reasonable value, as these models (both for BAO and FS) have been tested in the past using different sets of tracers, and such large systematics would have been already identified. Also, adding a $2\sigma$ resolution effect would have artificially inflated our systematic errors, simply due to poor statistical power in these mocks rather than a reasonable limitation of our modelling.

We modify the error covariance of the different data-vectors by replacing the statistical contribution only in the diagonal elements, by the total systematic plus statistical contribution. In other words, the total covariance elements, $c^{\rm tot}_{ij}$, become, $c^{\rm tot}_{ij}=c^{\rm sta}_{ij}+c^{\rm sys}_{i}\delta^{\rm Kr.}_{ij}$, where $c^{\rm sta}_{ij}$ are the elements only accounting for the statistical contribution, and the $c^{\rm sys}_{i}$ terms correspond to $\sigma_X^2$ according to Table~\ref{tab:systot}. We note that by doing this we assume that the systematic errors are un-correlated among them, as they are only added on the diagonal of the covariance. This is probably not accurate, but we take this as a conservative choice, as correlation among systematics would reduce their effect in the final covariance matrix.

\begin{table*}
\caption{Systematic error budget summary for cosmological parameters of interest: $\alpha_\parallel^{\rm post}$ and $\alpha_\perp^{\rm post}$ from a BAO analysis on the post-reconstructed catalogues; $\alpha_\parallel^{\rm FS},\,\alpha_\perp^{\rm FS}$ and $f{\sigma_8}^{\rm FS}$ for a FS analysis and $\alpha_\parallel^{\rm sim},\,\alpha_\perp^{\rm sim}$ and $f{\sigma_8}^{\rm sim}$ for the simultaneous BAO+FS fit, in all cases using the standard pipelines described in \S\ref{sec:BAO} and \S\ref{sec:RSD}.  The results show the observed relative systematic shift, along with 2 times the statistical precision inferred from the mean of the mocks. We consider a detection of systematic error when the deviation with respect to the expected value is larger than $2\sigma$. If no systematic shift is found within this limit, we adopt as a systematic contribution the $2\sigma$ value, as a conservative resolution limit. The potential sources for systematic studied are: modelling, the arbitrary choice of reference cosmology, and the observational weights: redshift failures, collisions and completeness (FCC) and radial integral constrain (RIC). $\sigma_X$ represents the total systematic contribution: $\sigma_X^2=\sum_i \sigma_i^2$, where $\sigma_i$ are the individual systematic contributions. The total error budget includes the statistical contribution, $\sigma_{\rm sta}$, added in quadrature with $\sigma_X$. }
\begin{center}
\begin{tabular}{c|c|c|c|c|c|c}
case & Modelling & Ref. Cosmology (limit) & FCC & RIC & $\sigma_{\rm X}/\sigma_{\rm sta}$ & $\sqrt{\sigma_{\rm X}^2+\sigma_{\rm est}^2}$ \\
\hline
\hline

$\Delta\alpha_\parallel^{\rm post}\pm2\sigma$ &  $-0.0048 \pm 0.0038$& $0.0106 \pm 0.0057\, (\boldsymbol{\Theta}_{\rm Z})$ & $-$ & $-$  & $0.012/0.024$ & $0.027$ \\
$\Delta\alpha_\perp^{\rm post}\pm2\sigma$ & $0.0005 \pm 0.0021$& $-0.0089 \pm 0.0032\, (\boldsymbol{\Theta}_{\rm Z})$ & $-$ & $-$  &$0.0091/0.020$ & $0.021$ \\
\hline
$\Delta\alpha_\parallel^{\rm FS}\pm2\sigma$ & $0.0016 \pm 0.0064$& $-0.0063 \pm 0.0089\, (\boldsymbol{\Theta}_{\rm Z})$ & $0.0068 \pm 0.0031$ &  $0.0033 \pm 0.0030$  & $0.013/0.036$ & $0.039$ \\
$\Delta\alpha_\perp^{\rm FS}\pm2\sigma$ & $-0.0095 \pm 0.0040$& $-0.0041 \pm 0.0056\, (\boldsymbol{\Theta}_{\rm Z})$ & $-0.0020 \pm 0.0023$ & $-0.0031 \pm 0.0023$  & $0.012/0.027$ & $0.030$ \\
$\Delta f{\sigma_8}_{\rm res}^{\rm FS}\pm2\sigma$ & $-0.0038 \pm 0.0082$& $0.009 \pm 0.012\, (\boldsymbol{\Theta}_{\rm X})$ & $-0.0065 \pm 0.0042$ & $-0.0074 \pm 0.0043$  & $0.017/0.042$ & $0.046$ \\
\hline
$\Delta\alpha_\parallel^{\rm sim}\pm2\sigma$ & $-0.0008 \pm 0.0034$& $0.0078 \pm 0.0048\, (\boldsymbol{\Theta}_{\rm Y})$ & $0.0033 \pm 0.0021$ & $0.0024 \pm 0.0022$  & $0.0094/0.023$ & $0.025$ \\
$\Delta\alpha_\perp^{\rm sim}\pm2\sigma$ & $-0.0017 \pm 0.0022$& $-0.0071 \pm 0.0030\, (\boldsymbol{\Theta}_{\rm Z})$ & $-0.0024 \pm 0.0017$ & $-0.0018 \pm 0.0017$  & $0.0080/0.017$ & $0.019$ \\
$\Delta f{\sigma_8}^{\rm sim}\pm2\sigma$ & $0.0181 \pm 0.0074$& $-0.011 \pm 0.011\, (\boldsymbol{\Theta}_{\rm Y})$ & $-0.0050 \pm 0.0037$ & $-0.0076 \pm 0.0039$  & $0.023/0.037$ & $0.043$ \\

\end{tabular}
\end{center}
\label{tab:systot}
\end{table*}

Table \ref{tab:finalresults} presents the final cosmology results with and without the full error budget for the different DR16 CMASS+eBOSS LRG type of analysis performed, both in this work and in \cite{LRG_corr}.

\section{Consensus results}\label{sec:consensus}
The DR16 CMASS+eBOSS LRG data has been analysed performing four different types of analyses: {\it i)} FS in Fourier space, {\it ii)} BAO post-reconstruction in Fourier space, {\it iii)} FS in configuration space, and {\it iv)} BAO post-reconstruction in configuration space. Analyses {\it i)} and {\it ii)} are fully described in this paper, whereas analyses {\it iii)} and {\it iv)} are presented in \cite{LRG_corr}. Although all analyses rely on the same underlying catalogues (pre-recon for FS and post-recon for BAO) their information content is not the same. This happens because 1) each space data-vector is computed directly from the catalogue and not as a Fourier Transform of the complementary-space data-vector; and 2) because data-vectors do not cover an infinite range of scales, and therefore the DFT of a finite set of elements in configuration space will never match the elements in Fourier space, and vice-versa. Since we intend to produce a single inferred set of cosmological parameters per catalogue, we aim to combine Fourier and configuration space measurements into a single consensus set of parameters. 

In order to do so, we use a similar approach to the one described in \cite{sanchez_clustering_2017}, which was used to produce the consensus results of BOSS \citep{alam_clustering_2017}. This approach, known as the `best linear unbiased estimator', consists of building a linear estimator of the consensus parameters ($\alpha_\parallel^{\rm cons},\, \alpha_\perp^{\rm cons}\, f\sigma_8^{\rm cons}$) as a function of the individual parameters estimated in Fourier and configuration space with certain coefficients. These coefficients are determined by imposing a minimum variance on the resulting consensus parameters. Thus, we require a covariance that describes the full correlation among all parameters. This matrix is given by the individual covariances among parameters of the same space. Note that the individual covariances are effectively different for each realisation of the mocks, as the noise content of each realisation is a stochastic process.  However, we still need to determine those elements of the full covariance corresponding to the blocks describing the correlation between different spaces. Such coefficients can be estimated from the mocks, by inferring the data-vector in each mock realisation. 
There are some choices to be made in the details of building the final covariance. One can take only the diagonal elements from the actual data catalogues and the remaining elements from the mocks, or take the diagonal blocks corresponding to the same space from the data, and only the off-diagonal blocks across spaces from the mocks, just as two examples. We have tested that the impact of these choices is minimal. In this work we take the approach described in sec. 3.4 of \cite{LRG_corr}. 

\begin{figure*}
\includegraphics[scale=0.33, trim={0 150 0 0}, clip=false]{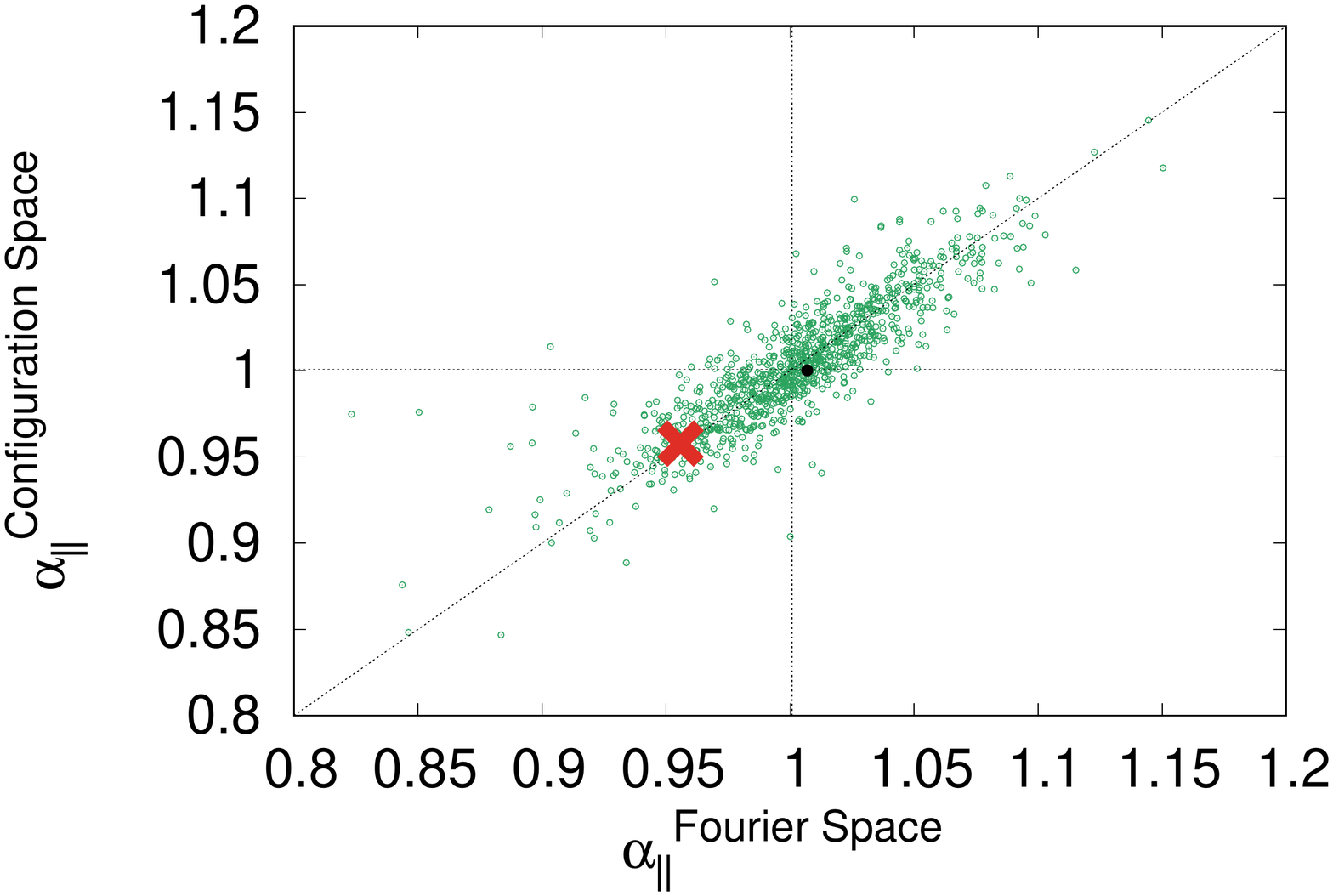}
\includegraphics[scale=0.33, trim={110 150 0 0}, clip=false]{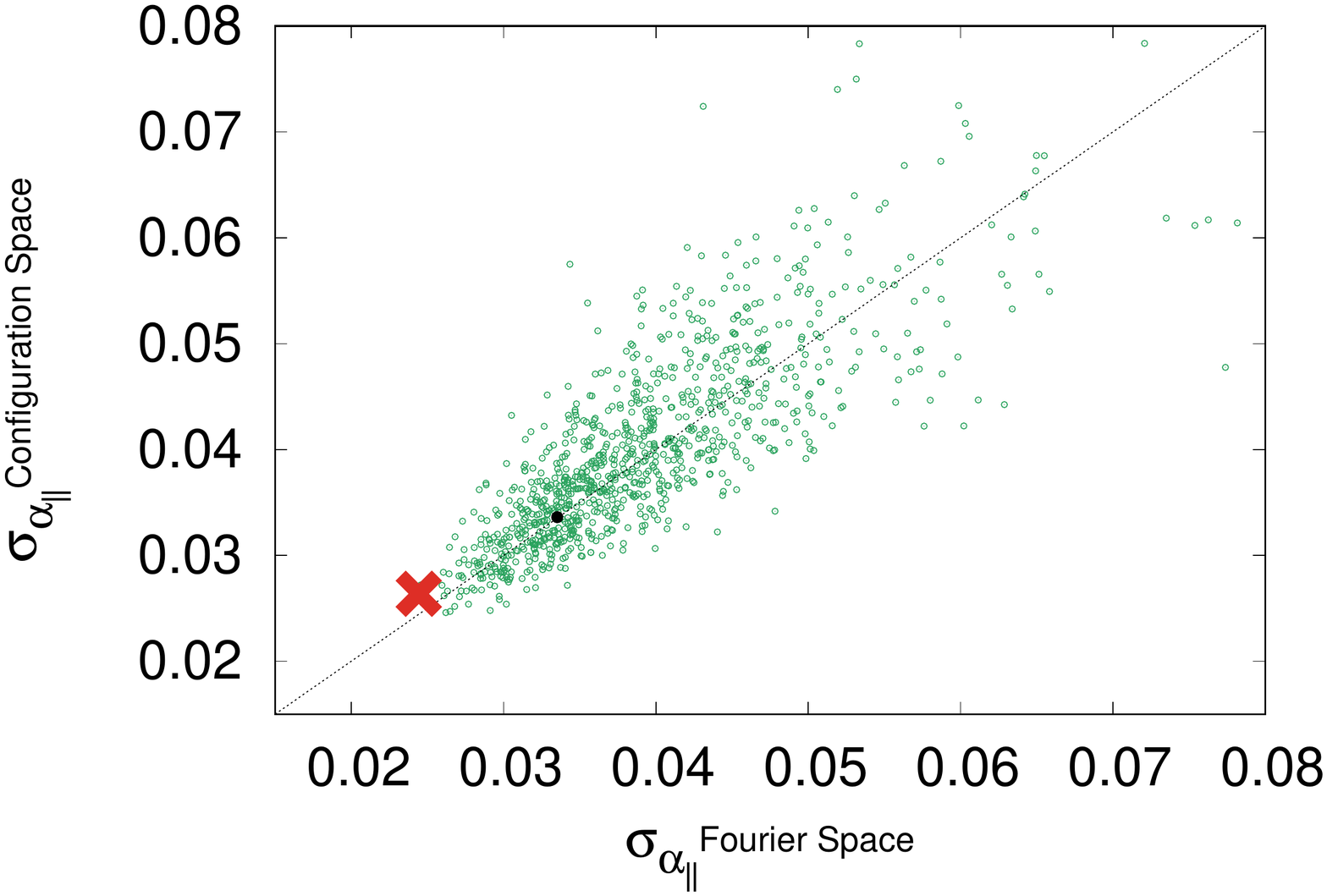}

\includegraphics[scale=0.33, trim={0 0 0 0}, clip=false]{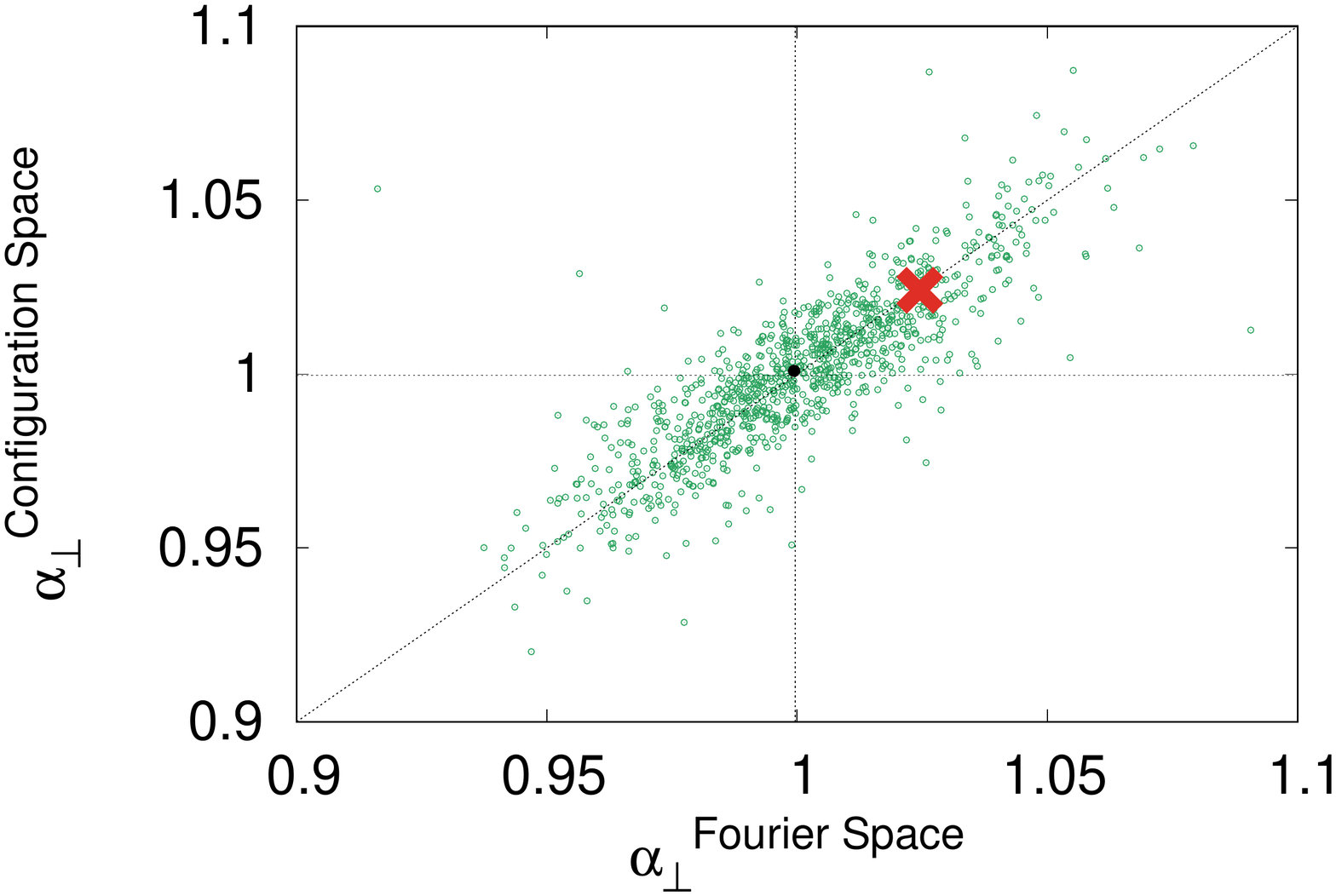}
\includegraphics[scale=0.33, trim={110 0 0 0}, clip=false]{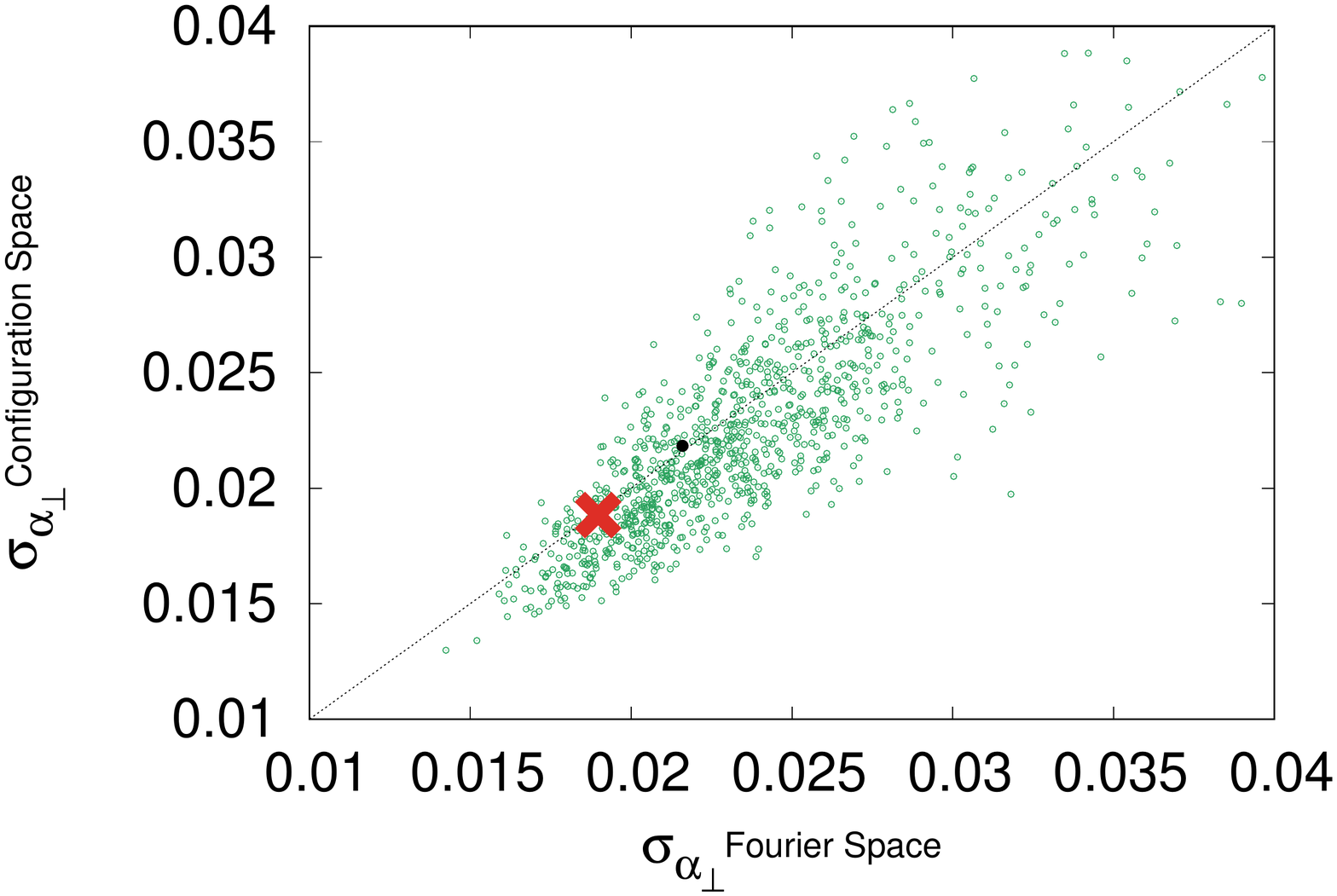}

\caption{Comparison of the of the BAO measurement on the post-reconstructed catalogues in Fourier space (this work) and in Configuration space \citep{LRG_corr}.  The $x$-axes represent the Fourier space quantities and the $y$-axes configuration space quantities. The left sub-panels display the performance on $\alpha_\parallel$ (top-left panel) and $\alpha_\perp$ (bottom-left panel), whereas the right panels display the performance on the $1-\sigma$ error of the corresponding quantities. The green symbols display the performance on the individual 1000 mocks, the black dot the performance on the mean power spectra of the 1000 mocks, and the red cross the performance on the DR16 CMASS+eBOSS LRG data. The errors correspond to $1\sigma$ and only represent the statistical contribution.}
\label{fig:baocomp}
\end{figure*}

Fig.~\ref{fig:baocomp} displays the comparison between the Fourier space results presented in this paper and the configuration space results presented in \cite{LRG_corr} for BAO analysis using the post-recon catalogues from the \textsc{EZmocks} (green points) and the DR16 CMASS+eBOSS LRG data catalogue (red cross). The panels display the comparison between the two analyses for $\alpha_\parallel$ and $\alpha_\perp$ and for their corresponding errors, as indicated. The black dot displays the result on the mean of the 1000 \textsc{EZmocks}, and for the error case, this quantity has been re-scaled by the factor $\sqrt{N_{\rm EZ}}$ to match the typical error of an individual mock. 
We find an excellent agreement, both for measurements and errors, for both mocks and data. This agreement motivates the combination of both results as they are fully consistent. The consensus results among BAO-Fourier space and BAO-configuration space are displayed in  Table~\ref{tab:finalresults}.
We note that when both spaces are combined there is a slight reduction of errors on both $D_H/r_{\rm drag}$ and $D_M/r_{\rm drag}$ parameters. The extra information driving this improvement in precision is related to the fact that Fourier and configuration space data-vectors do not contain the exact same information, although the amount of correlation is very high, with cross-correlation parameters between the $\alpha$s of the different spaces of $\rho=0.88$.

\begin{figure*}

\includegraphics[scale=0.33, trim={0 150 0 0}, clip=false]{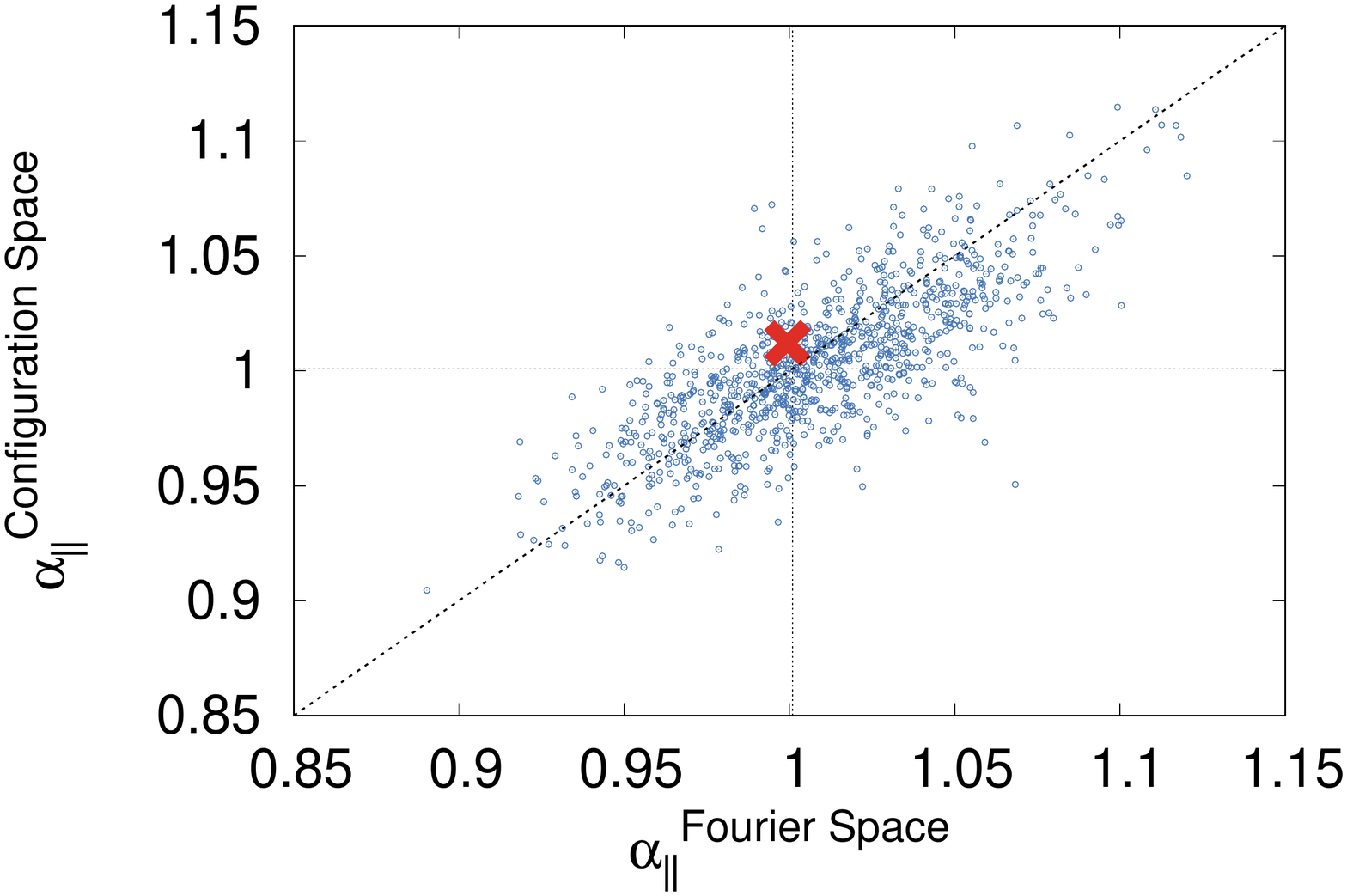}
\includegraphics[scale=0.33, trim={110 150 0 0}, clip=false]{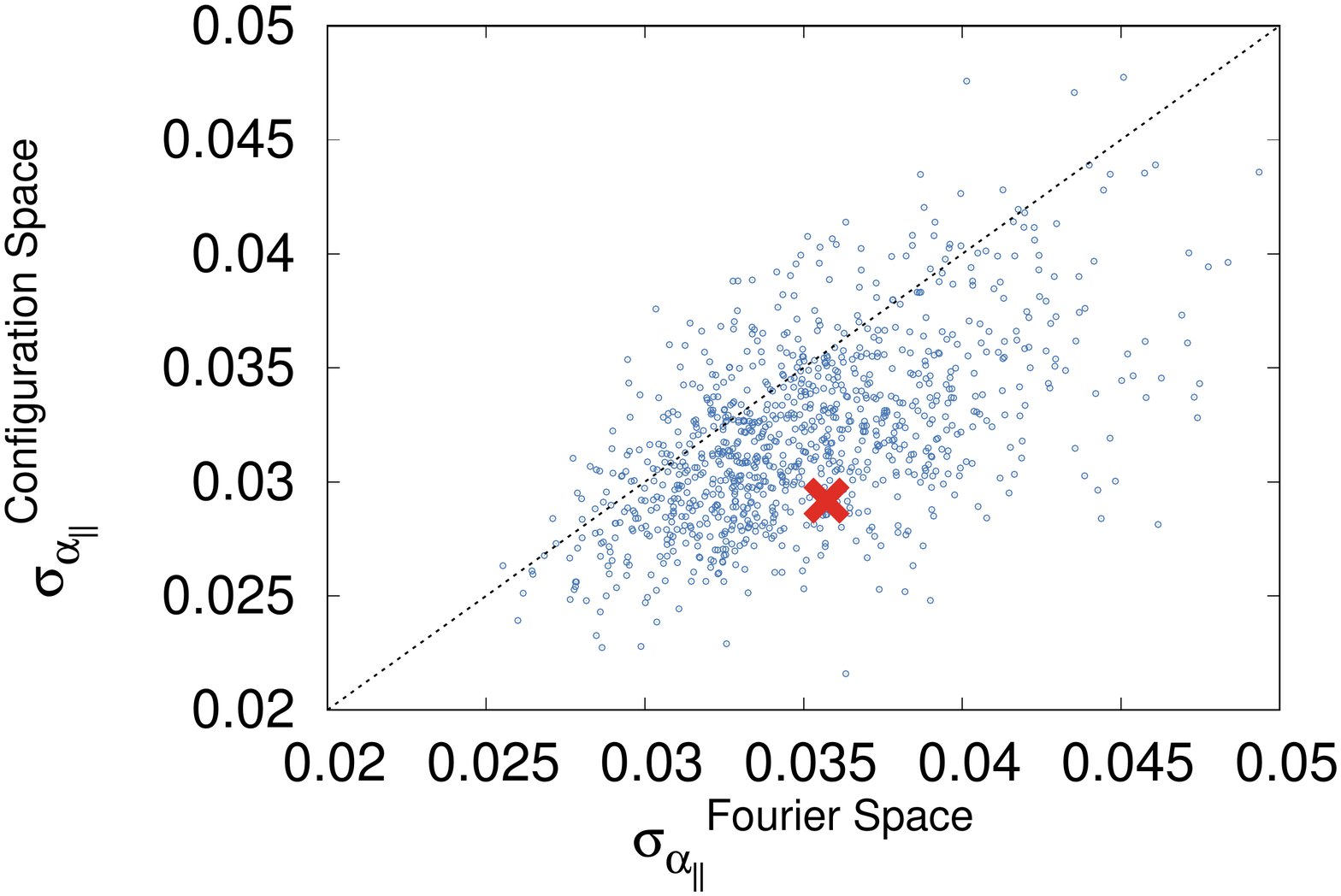}

\includegraphics[scale=0.33, trim={0 150 0 0 0}, clip=false]{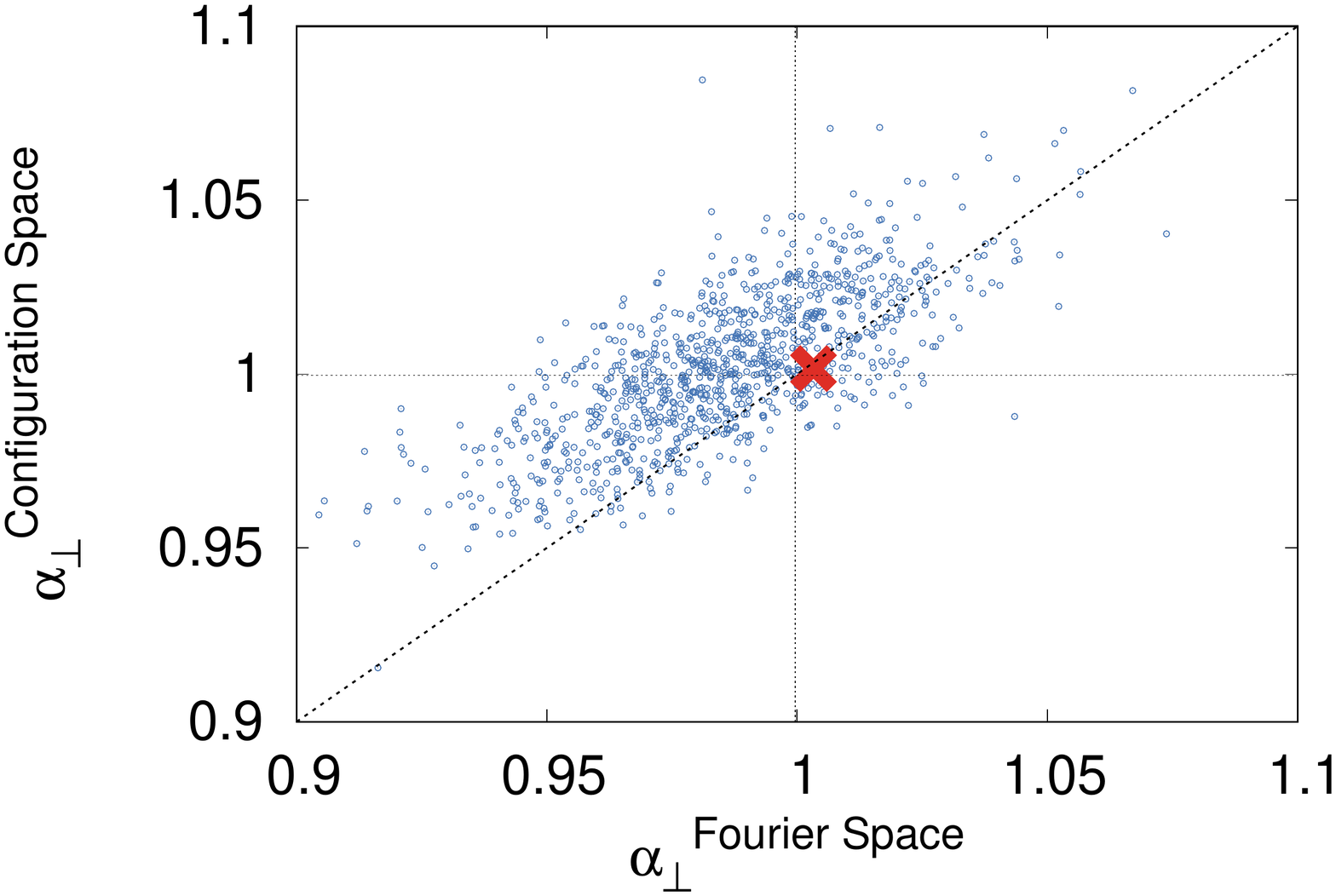}
\includegraphics[scale=0.33, trim={110 150 0 0 0}, clip=false]{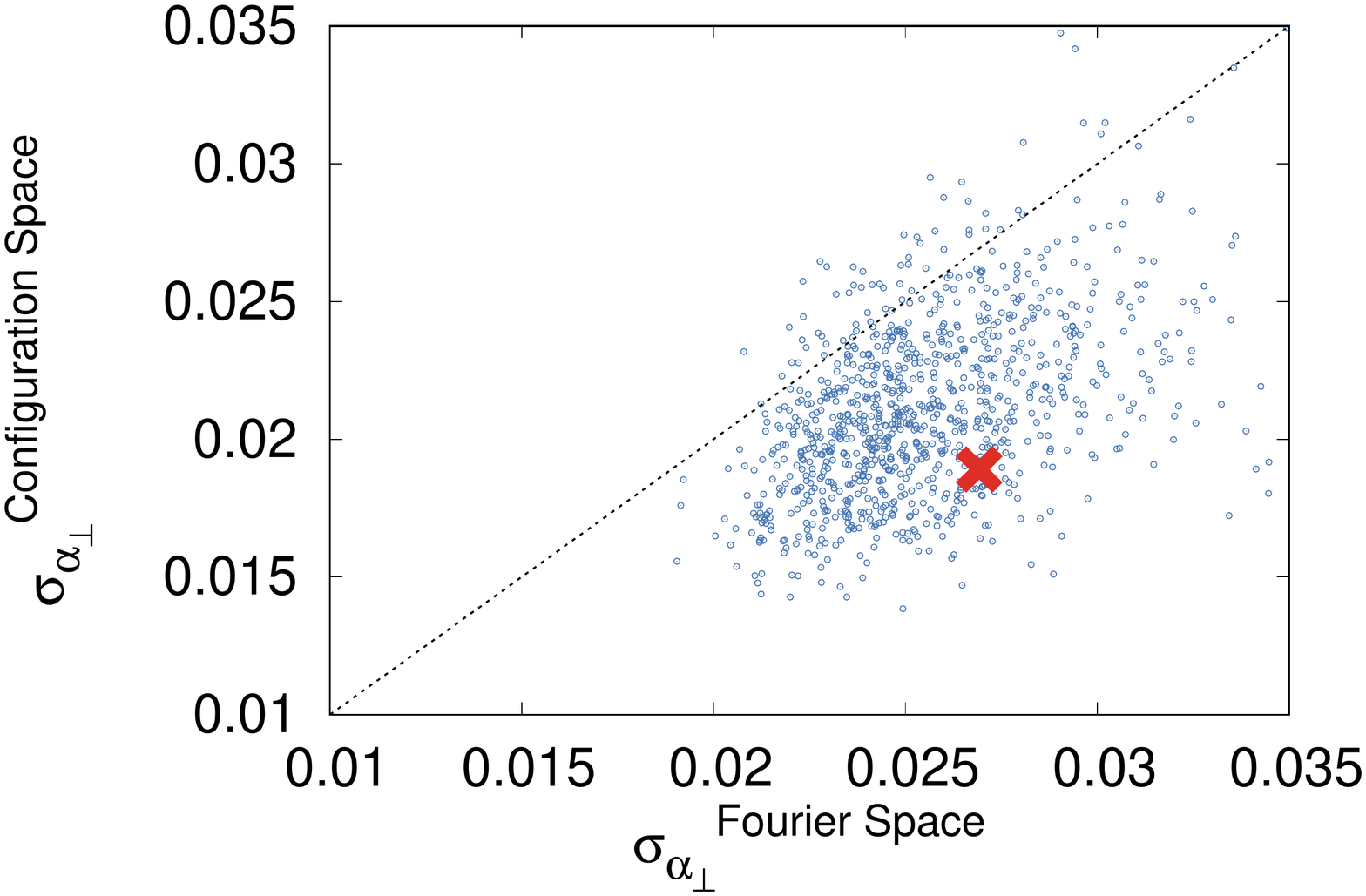}

\includegraphics[scale=0.33, trim={0 80 0 0}, clip=false]{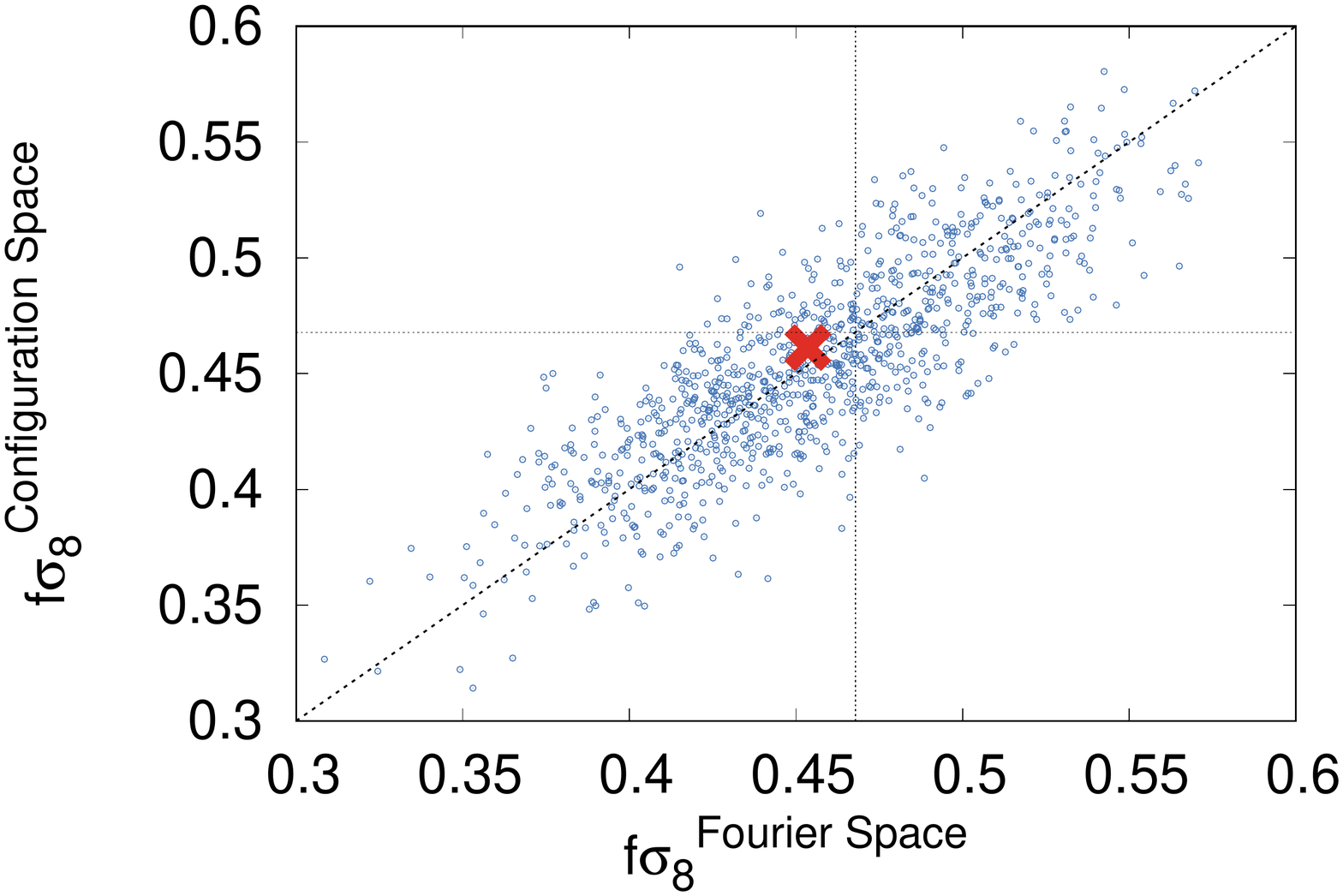}
\includegraphics[scale=0.33, trim={110 80 0 0}, clip=false]{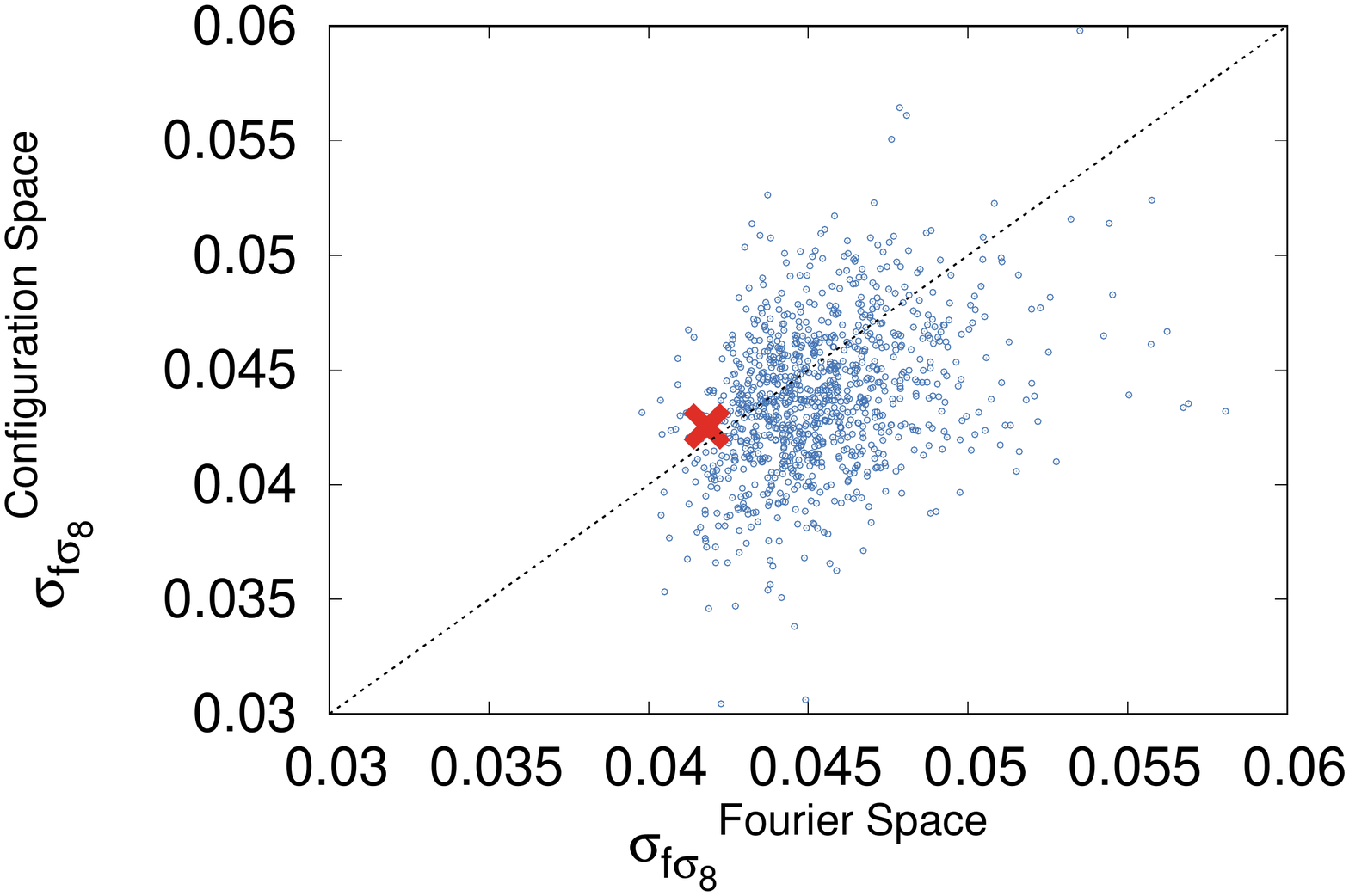}

\caption{Comparison of the of the RSD measurement in Fourier space (this work) and in configuration space \citep{LRG_corr}.  The $x$-axes represent the Fourier space quantities and the $y$-axes configuration space quantities. The left sub-panels display the performance on $\alpha_\parallel$ (top-left panel), $\alpha_\perp$ (middle-left panel) and $f\sigma_8$ (bottom-left panel), whereas the right panels display the performance on the $1-\sigma$ error of the corresponding quantities. The blue symbols display the performance on the individual 1000 mocks and the red cross the performance on the DR16 CMASS+eBOSS LRG data. }
\label{fig:rsdcomp}

\end{figure*}

Fig.~\ref{fig:rsdcomp} displays an analogous set of panels corresponding to FS-type of analysis. In this case a third pair of panels is added to account for the $f\sigma_8$ variable. As for the BAO type of analysis both Fourier and configuration show a strong correlation, for both errors and measurements. We also observe that the DR16 CMASS+eBOSS LRG catalogue behaves as expected given the performance of the mocks.  
We note that for $\alpha_\perp$ there is an offset between the Fourier and configuration space inferred values. This is caused by the $1\%$ systematic shift identified already in \S\ref{sec:RSDmodelling}. Also, as a general trend we see that configuration space errors on the scaling parameters tend to be smaller than the Fourier space one. This trend is not present in the BAO type of analysis (Fig.~\ref{fig:baocomp}). We think that this difference is caused by the shortening of the $k$-range of analysis, $0.02\leq k\,[\hompc]\leq 0.15$, with respect to the BAO, which reaches $k_{\rm max}=0.30\,{\hompc}$, which we believe adds extra BAO information. In configuration space this effect is not present as the BAO feature is very localised at scales of $\sim100\,\mpcoh$. 

\begin{table*}
\caption{Summary of the cosmology parameters inferred from the DR16 CMASS+eBOSS LRG catalogue using BAO and FS analyses, in Fourier space (this paper) and in configuration space \citep{LRG_corr}.  Fourier space, configuration space, BAO and FS results can be combined (using the parameter-level covariance inferred from \textsc{EZmocks}), which we denote as `$+$' . For the Fourier space we additionally display the result of the simultaneous BAO and FS fit (using the covariance at the $k$-bin level inferred from  \textsc{EZmocks}), which we denote as `$\times$'. The reported error-bars correspond to $1\sigma$ and contain only the statistical error budget (first half of the table) and the full error budget (second half of the table). Full resolution data-vectors and covariances can be found \href{https://www.sdss.org/dr16/}{ on-line}.}
\begin{center}
\begin{tabular}{|c|c|c|c|}
Probe  & $D_M/r_{\rm drag}$ & $D_H/r_{\rm drag}$ & $f\sigma_8$  \\
\hline
\hline
Without systematic error budget  \\
\hline
BAO $P_k$ & $17.86\pm0.34$ & $19.30\pm 0.50$ & $-$ \\
BAO $\xi_s$ & $17.86\pm0.33$ & $19.34\pm0.54$ & $-$ \\
BAO ($P_k+\xi_s$)  & $17.86\pm0.32$ & $19.31\pm0.49$ & $-$ \\
FS $P_k$ & $17.49\pm0.48$ & $20.18\pm0.73$& $0.454\pm0.042$   \\
FS  $\xi_s$ & $17.42\pm0.34$ & $ 20.46\pm0.60$ & $0.460\pm 0.044$ \\
FS  ($P_k+\xi_s$)   & $17.37\pm0.32$ & $20.39\pm0.59$ &$0.448 \pm 0.040$  \\
(BAO $+$ FS) $P_k$  & $17.72 \pm 0.31$ & $19.58 \pm 0.45$ &$0.476 \pm 0.038$  \\
(BAO $\times$ FS) $P_k$ & $17.58\pm0.30$ & $19.96\pm0.47$ &$0.466\pm0.037$  \\
(BAO $+$ FS) $\xi_s$ & $17.57\pm0.29$ & $19.95\pm0.44$ & $0.491\pm0.040$ \\
 (BAO $+$ FS) $\xi_s$ $+$ (BAO$+$FS) $P_k$ &  $17.39\pm0.27$ & $19.88\pm0.43$ &$0.475 \pm 0.037$ \\ 
 BAO ($P_k+\xi_s$) $+$ FS ($P_k+\xi_s$) & $17.55\pm0.28$ & $19.88\pm0.42$ & $0.481 \pm 0.037$ \\ 
\hline\hline
With systematic  error budget \\
\hline
BAO $P_k$ & $17.86\pm0.37$ & $19.30\pm 0.56$ & $-$ \\
BAO $\xi_s$ & $17.86\pm0.33$ & $19.34\pm0.54$ & $-$ \\
BAO ($P_k+\xi_s$)  & $17.86\pm0.33$ & $19.33\pm0.53$ & $-$ \\
FS $P_k$ & $17.49\pm0.52$ & $20.18\pm0.78$& $0.454\pm0.046$   \\
FS  $\xi_s$ & $17.42\pm0.40$ & $ 20.46\pm0.70$ & $0.460\pm 0.050$ \\
FS  ($P_k+\xi_s$)   & $17.40\pm0.39$ & $20.37\pm0.68$ &$0.449 \pm 0.044$  \\
(BAO $+$ FS) $P_k$  & $17.72 \pm 0.34$ & $19.58 \pm 0.50$ &$0.474 \pm 0.042$  \\
(BAO $\times$ FS) $P_k$ & $17.58\pm0.33$ & $19.96\pm0.50$ &$0.466\pm0.043$  \\
(BAO $+$ FS) $\xi_s$ & $17.65\pm0.31$ & $19.81\pm0.47$ & $0.483\pm0.047$ \\
 (BAO $+$ FS) $\xi_s$ $+$ (BAO$+$FS) $P_k$ &  $17.64\pm0.30$ & $19.78\pm0.46$ &$0.470 \pm 0.044$ \\ 
 BAO ($P_k+\xi_s$) $+$ FS ($P_k+\xi_s$) & $17.65\pm0.30$ & $19.77\pm0.47$ & $0.473 \pm 0.044$ \\ 
\end{tabular}
\end{center}
\label{tab:finalresults}
\end{table*}%

\subsection{BAO - FS simultaneous fit}

In this section we aim to perform a simultaneous fit using the BAO type of analysis on reconstructed catalogues and FS type  on pre-reconstructed catalogues. From the point of view of information content we are allowed to do so, because the pre- and post-reconstructed catalogues are essentially different, and there is an actual gain of information in the process of using them when extracting cosmological information.  

\begin{figure}
\centering
\includegraphics[scale=0.5]{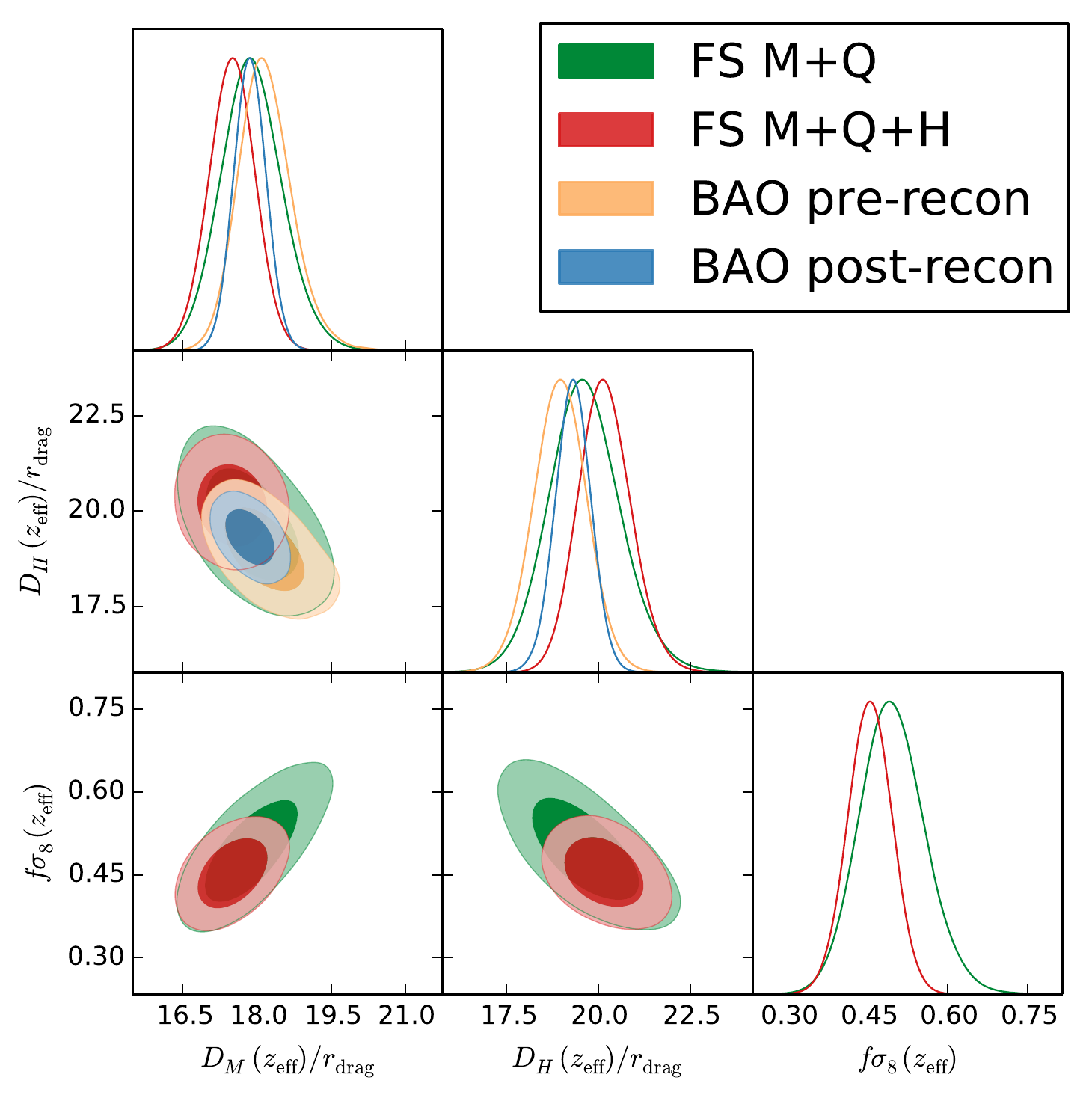}
\caption{Likelihood posterior for $1-$ and $2-\sigma$,  from the BAO and FS type of analysis on the DR16 CMASS+eBOSS LRG: BAO type of analysis on the pre-reconstructed catalogues (yellow contours) and on the post-reconstructed catalogues (blue contours), FS type of analysis when the monopole and quadrupole are the only multipoles being used (green contours), and when the hexadecapole is also included (red contours). For BAO type of analysis we use $0.02\leq k\,[\hompc]\leq 0.30$, whereas for FS analysis $0.02\leq k\,[\hompc]\leq 0.15$. For all cases the contours only account for the statistical error budget. These contours correspond to the results presented in the first half (without systematics) of Table~\ref{tab:finalresults}.}
\label{fig:contoursRSDBAO}
\end{figure}

We start by comparing all Fourier-space BAO analyses (pre- and post-reconstruction) with the FS analysis with and without the hexadecapole, and using the 50\%-prior on $A_{\rm noise}$.  Fig.~\ref{fig:contoursRSDBAO} displays the posterior likelihoods for all the Fourier space analyses presented in this paper: FS using MQ (green), FS using MQH (red), BAO on the pre-reconstructed catalogues (orange), and BAO on post-reconstructed catalogues (blue). Note that the BAO pre-reconstruction and FS M+Q data-vectors are the same, except that the BAO pre-recon data vector contains $k$-elements up to smaller scales than the FS. 
The agreement between all analyses is very good for all variables of interest. In particular we see that performing a FS analysis adding the hexadecapole helps to remove the strong correlation between the scaling factors and $f\sigma_8$.

\begin{figure}
\centering
\includegraphics[scale=0.5]{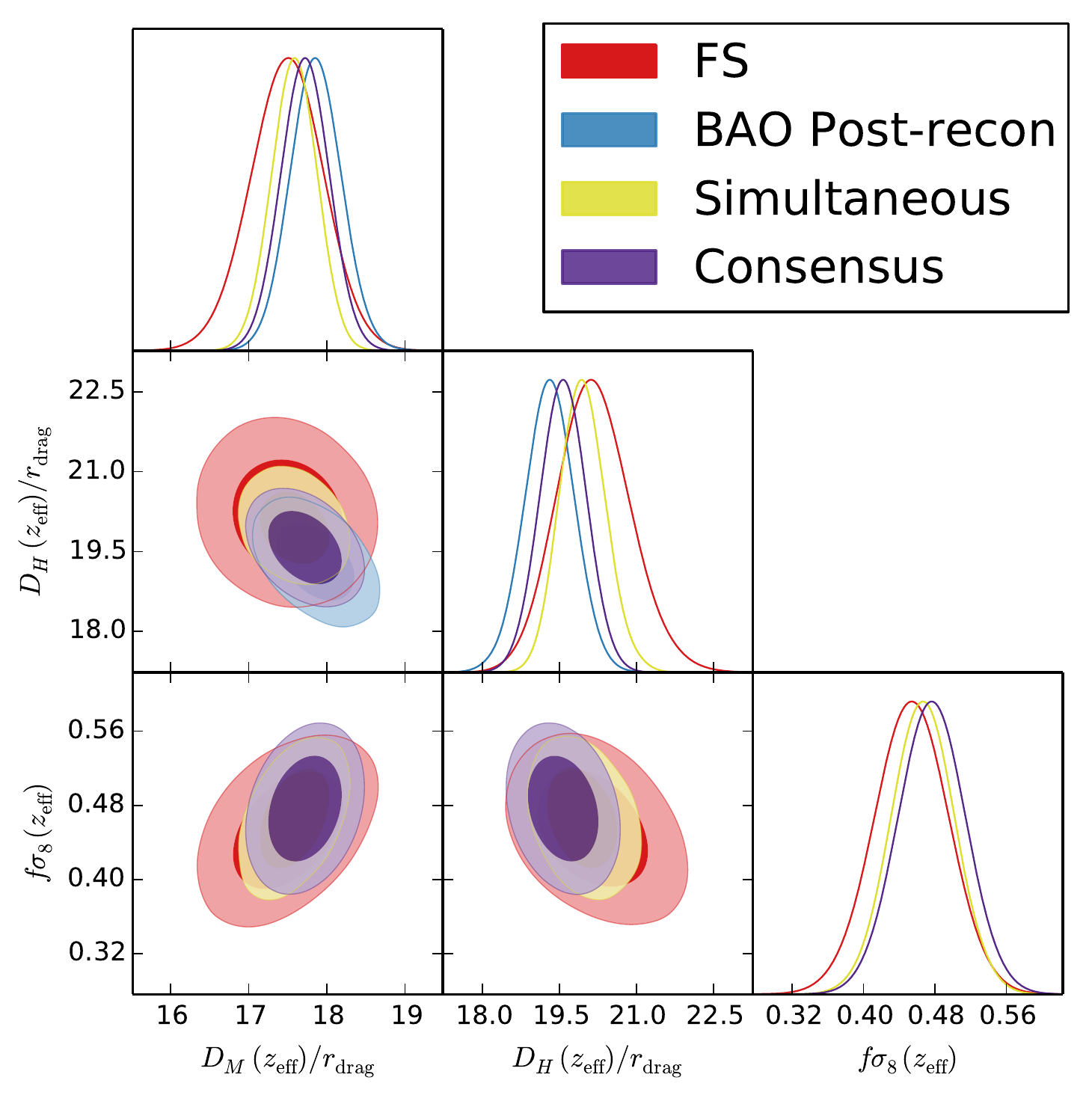}
\caption{Likelihood posterior for $1$ and $2\sigma$ contour for the Fourier space BAO and FS consensus (purple contours) and simultaneous fit (yellow contours). For reference the individual BAO post-recon (blue contours) and FS pre-recon (red contours) have also been included. In all cases the contours only account for the statistical error budget. These contours correspond to the results presented in the first half (without systematics) of Table~\ref{tab:finalresults}.}
\label{fig:consensusFourier}
\end{figure}

Fig.~\ref{fig:consensusFourier} presents in purple contours the result of combining BAO post-recon (blue contours) and FS (red contours) Fourier space analyses. This result has been obtained by applying the same technique used to the corresponding Fourier-space and configuration-space results. As before this approach suffers from having to estimate cross-method coefficients from the mocks. This may have an impact on the final contours, as it could fail to accurately describe the exact correlation that variables among the two spaces have for a specific realisation. The numerical results of this Fourier consensus are presented in Table~\ref{tab:finalresults}, as `(BAO $+$ FS) $P_k$'.

We follow an alternative analysis of extracting the combined BAO post-recon and FS pre-recon information without relying on the cross-coefficients of parameters estimated from the mocks. We do so by performing a simultaneous fit on both reconstructed and pre-reconstructed data-vectors using the BAO and FS analysis respectively, simultaneously fitting $\alpha_\parallel$ and $\alpha_\perp$. As for the individual BAO and FS analyses, we estimate the full covariance matrix using the \textsc{EZmocks}. In Appendix \ref{appenddix:cov} we show what the off-diagonal elements of this matrix look like. Since the data-vectors differ, this matrix is not singular and can be safely inverted. However, the off-diagonal cross-correlation coefficients describing the pre- and post- data-vector elements with the same $k$-bin and $\ell$-multipole, can be as high as $\sim0.9$, which inevitably will introduce some noise when inverting the matrix. 
We validate this approach by applying this pipeline to the \textsc{Nseries} mocks. Table~\ref{tab:consensusmocks} displays the performance of the simultaneous BAO and FS fit, along with the individual BAO and FS analyses, for both a fit on the mean of the mocks and the mean of the 84 individual fits. In both cases the result is very similar. We see how this combined analysis can actually recover well the expected cosmological parameters with better precision than the individual BAO and FS analyses. Also the {\it rms} and error of the mean for the combined fit is smaller than any of the individual fits, confirming the gain of information. In particular we note that by performing the consensus fits we obtain lower systematic shifts in the scaling factor variables than by performing the FS analysis alone. In fact we observe a $\sim 1\%$ systematic shift on $\alpha_\perp$ on the FS-alone, whereas for the simultaneous fit this shift is smaller than $0.5\%$. Conversely, for $f\sigma_8$, FS-alone reported a shift of $\sim -0.004$ and for the FS $\times$ BAO this has been increased up to  $\sim0.018$.

Employing the \textsc{Nseries} and \textsc{EZmocks} mocks we find that using a 3rd-order polynomial to perform the BAO part of the combined fit is not sufficient to achieve a sufficiently high accuracy on $f\sigma_8$. This is caused by small inaccuracies in reproducing the BAO post-reconstruction broadband which are severely leaked into the FS analysis through the cross-covariance terms of the matrix. As a consequence, these BAO broadband inaccuracies produced biased results on $f\sigma_8$ as well as a bad-$\chi^2$ fits. This behaviour was also reported in the companion paper by \cite{demattia20a} when performing a similar combined fit. We increase the BAO polynomial order up to 5 and find that such behaviour vanishes and we are able to recover the expected $f\sigma_8$ in mocks. 

The results of applying this combined fit methodology to the data is shown by the yellow contours in Fig. \ref{fig:consensusFourier} with only the statistical error contribution, as well reported in Table~\ref{tab:finalresults}, under the notation `(BAO $\times$ FS) $P_k$'. We observe some differences between the two approaches of using both FS and BAO data, but the overall result is very similar, especially for $f\sigma_8$. We observe that for the consensus result $D_M/r_{\rm drag}$ and $D_H/r_{\rm drag}$ are closer to the BAO-only results, opposite to the behaviour observed when the final results are produced by using the simultaneous fit. Also, bear in mind that when combining BAO and FS analysis, either by doing a consensus or a simultaneous fit, we do improve the $f\sigma_8$ measurement. This might seem paradoxical as BAO analysis do not constrain $f\sigma_8$ information. However we obtain an indirect gain on this variable through a better measurement of the BAO scaling parameters, $\alpha_\parallel$ and $\alpha_\perp$, which are significantly correlated with $f\sigma_8$. In terms of information content, the reconstructed catalogue is produced under the assumption of GR in order to undo the non-linear physics that degrade the BAO-peak significance. In this sense, the results on $f\sigma_8$ coming from either combined or simultaneous fit, have stronger priors on gravity than those derived from the FS analysis on its own. 

The general agreement between the simultaneous and consensus fit in Fourier space serves as a validation of the consensus methodology applied to combine result from both spaces.

\subsection{Consensus final LRG results from BOSS and eBOSS}

In this section we present the most relevant results and corresponding covariance matrices of this paper.\footnote{The results of all the cases can be found \href{web}{online}.} For reference, all the results correspond to those including the full systematic budget and represented by the second half of Table \ref{tab:finalresults}.

For the DR16 CMASS+eBOSS LRG BAO-only analysis in Fourier space the data vector and covariance matrix are given by, 
  \begin{equation}
   D^{P_k}_{{\rm BAO}}=
 \begin{pmatrix}
D_M/r_{\rm drag} \\
D_H/r_{\rm drag} \\
 \end{pmatrix}=
  \begin{pmatrix}
17.8637  \\
19.3033
 \end{pmatrix},
 \end{equation}
and, 
 \begin{equation} 
C^{P_k}_{{\rm BAO}} = 10^{-2}
 \begin{pmatrix}
13.9254 & -7.35600 \\
 & 30.8339 \\
 \end{pmatrix}.
\end{equation}
For the FS-only analysis in Fourier space we find that, 
  \begin{equation}
   D^{P_k}_{{\rm FS}}=
 \begin{pmatrix}
D_M/r_{\rm drag} \\
D_H/r_{\rm drag} \\
f\sigma_8 \\
 \end{pmatrix}=
  \begin{pmatrix}
17.4929  \\
20.1817\\
0.453576
 \end{pmatrix},
 \end{equation}
 and, 
 \begin{equation} 
C^{P_k}_{{\rm FS}} = 10^{-3}
 \begin{pmatrix}
267.860 & -39.8061 & 8.53160 \\
 &  607.292 & -10.5863 \\
 & & 2.10103
 \end{pmatrix}.
\end{equation}
By simultaneously fitting BAO and FS in Fourier space (the BAO $\times$ FS case in Table~\ref{tab:finalresults}), the data vector and covariance matrix are, 
  \begin{equation}
   D^{P_k}_{{\rm BAO\times FS}}=
 \begin{pmatrix}
D_M/r_{\rm drag} \\
D_H/r_{\rm drag} \\
f\sigma_8 \\
 \end{pmatrix}=
  \begin{pmatrix}
17.5840  \\
19.9603\\
0.466130
 \end{pmatrix},
 \end{equation}
and, 
 \begin{equation} 
C^{P_k}_{{\rm BAO \times FS}} = 10^{-3}
 \begin{pmatrix}
109.7713 & -32.1161 & 4.70509 \\
& 252.282 & -4.95629 \\
 & &  1.87876
 \end{pmatrix}.
\end{equation}
The full consensus results between BAO and FS, and between Fourier and configuration space, the  BAO ($P_k+\xi_s$) $+$ FS ($P_k+\xi_s$) case in Table~\ref{tab:finalresults}, are presented in Table~\ref{tab:BOSScov}, along with the lower redshift bins of the DR12 BOSS LRG measurements from  \citealt{alam_clustering_2017}. These results cover the full redshift range $0.2<z<1.0$ using LRG spectroscopic clustering measurements and are divided in a total of three redshift bins. The first two lowest redshift bins measured by BOSS overlap: $0.2<z<0.5$ with $z_{\rm eff}=0.38$, and $0.4<z<0.6$ with $z_{\rm eff}=0.51$. The third non-overlapping redshift bin, consisting of a combination of BOSS CMASS and eBOSS LRG observations and spanning $0.6<z<1.0$  with $z_{\rm eff}=0.698$, is used for the main results of this paper. Table~\ref{tab:BOSScov} presents all of them consistently in the same units, $D_M/r_{\rm drag}$, $D_H/r_{\rm drag}$ and $f\sigma_8$, and making explicit the correlation coefficients that need to be used when the 3 of them are simultaneously used.  Additionally, we rescale the original $f\sigma_8$ measurements by Eq. \ref{eq:sigma8res} to be fully consistent with our approach. These corrections are extremely sub-dominant and represent shifts of less than $1\%$, which is less than 1/10 of the total error budget.  This covariance is used in the cosmological interpretation of the eBOSS results in \cite{eBOSS_Cosmology}. 

\begin{table*}
\caption{Legacy BOSS+eBOSS LRG cosmological measurements and covariance matrix within the redshift range $0.2<z<1.0$. The table presents the results of the low- ($0.2<z<0.5,\, z_{\rm eff}=0.38$) and middle-redshift bin ($0.4<z<0.6,\,z_{\rm eff}=0.51$) of the DR12 BOSS galaxies. The new high-redshift bin  ($0.6<z<1.0,\, z_{\rm eff}=0.698$) is inferred from the DR16 CMASS + eBOSS LRG galaxies. Note that the low- and middle-redshift bins are overlapping in $z$, and therefore correlated, whereas the highest redshift bin does not overlap with any of the other two, and therefore is considered uncorrelated. The results are drawn from the combination of BAO post-reconstruction and Full Shape analyses, both in Fourier and configuration spaces. The covariance matrix elements include the full systematic budget. The low- and middle-redshift bin figures are inferred from those presented in table 8 of \citealt{alam_clustering_2017}. $f\sigma_8$ values of BOSS DR12 redshift bins have been rescaled by Eq.~\ref{eq:sigma8res} to use the same methodology as for the result of the high redshift bin.}
\begin{center}
\begin{tabular}{c|c|c|c|c|c|c|c|c|c|c|}
& Mean & &  & &  $c_{ij}\times10^4$ & & & & & \\
\hline
\hline
$D_M(0.38)/r_{\rm drag}$ & $10.274$  & $228.97$ & $-200.70$ & $26.481$ & $134.87$ & $-81.402$ & $10.292$ &0 &0 & 0 \\
$D_H(0.38)/r_{\rm drag}$ & $24.888$ &$-$ & $3384.9$ & $-85.213$ & $-160.24$ & $1365.2$ & $-38.002$ & 0& 0& 0\\
$f\sigma_8(0.38)$ & $0.49729$  &$-$ &$-$ & $20.319$ & $13.250$ & $-23.012$ & $8.14158$ &0 & 0& 0\\
\hline
$D_M(0.51)/r_{\rm drag}$ & $13.381$ & $-$ &$-$  &$-$ & $321.58$ & $-200.91$ & $26.409$ & 0 & 0 & 0\\ 
$D_H(0.51)/r_{\rm drag}$ & $22.429$  & $-$ & $-$ & $-$ & $-$ & $2319.2$ & $-55.377$ & 0 & 0 & 0 \\
$f\sigma_8(0.51)$ & $0.45902$  & $-$ & $-$ & $-$ & $-$ & $-$ & $14.322$ & 0 & 0 & 0 \\
\hline
$D_M(0.698)/r_{\rm drag}$ & $17.646$  & $-$ & $-$  & $-$ & $-$ & $-$ &  $-$& $911.40$ & $-337.89$ & $24.686$ \\
$D_H(0.698)/r_{\rm drag}$ & $19.770$   & $-$ & $-$  & $-$ & $-$ & $-$ & $-$ & $-$ & $2200.9$ & $-36.088$ \\
$f\sigma_8(0.698)$ & $0.47300$ & $-$ & $-$  & $-$ & $-$ & $-$ & $-$ & $-$ & $-$ & $19.616$ \\

\end{tabular}
\end{center}
\label{tab:BOSScov}
\end{table*}

\section{Discussion}\label{sec:cosmo}

In this section we present a brief interpretation of the results inferred from the DR16 CMASS+eBOSS LRG samples presented in this paper in combination with the configuration space counterpart presented in \cite{LRG_corr}. A full and consistent cosmological analysis is discussed in \cite{eBOSS_Cosmology}. 

\begin{figure}
\centering
\includegraphics[scale=0.8]{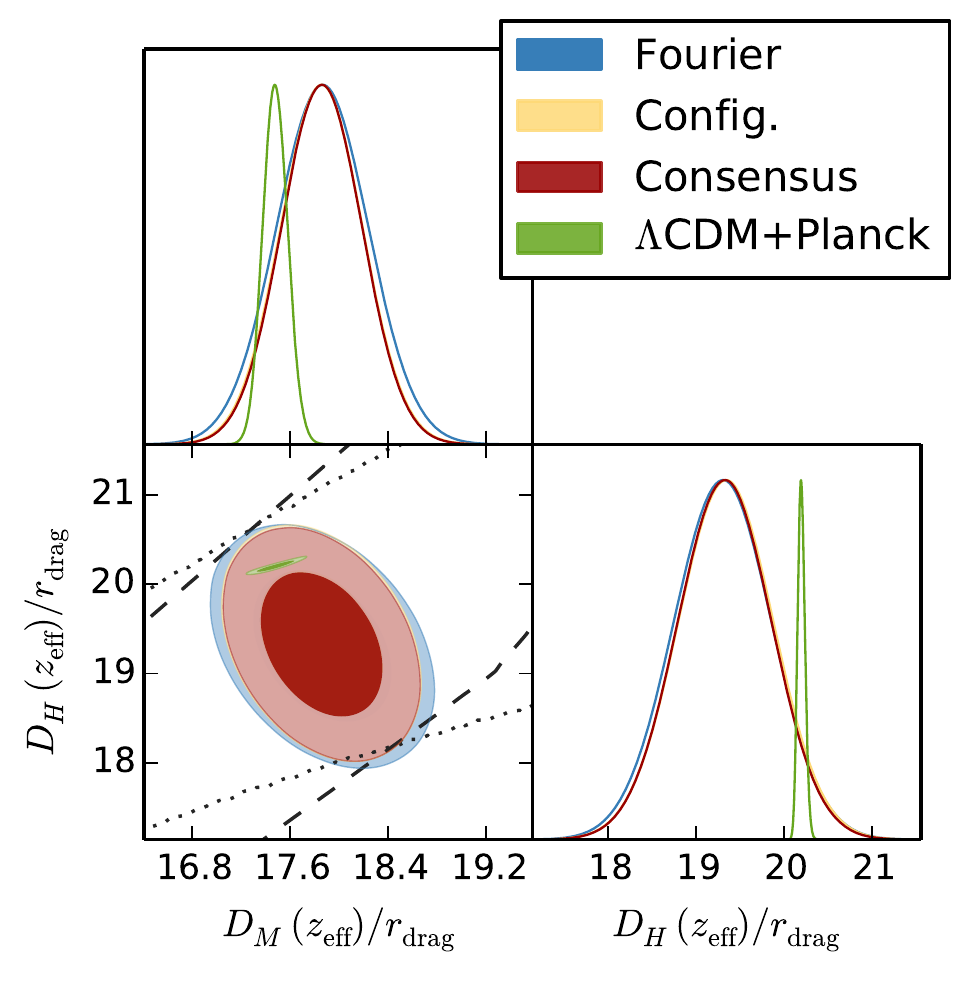}
\includegraphics[scale=0.53]{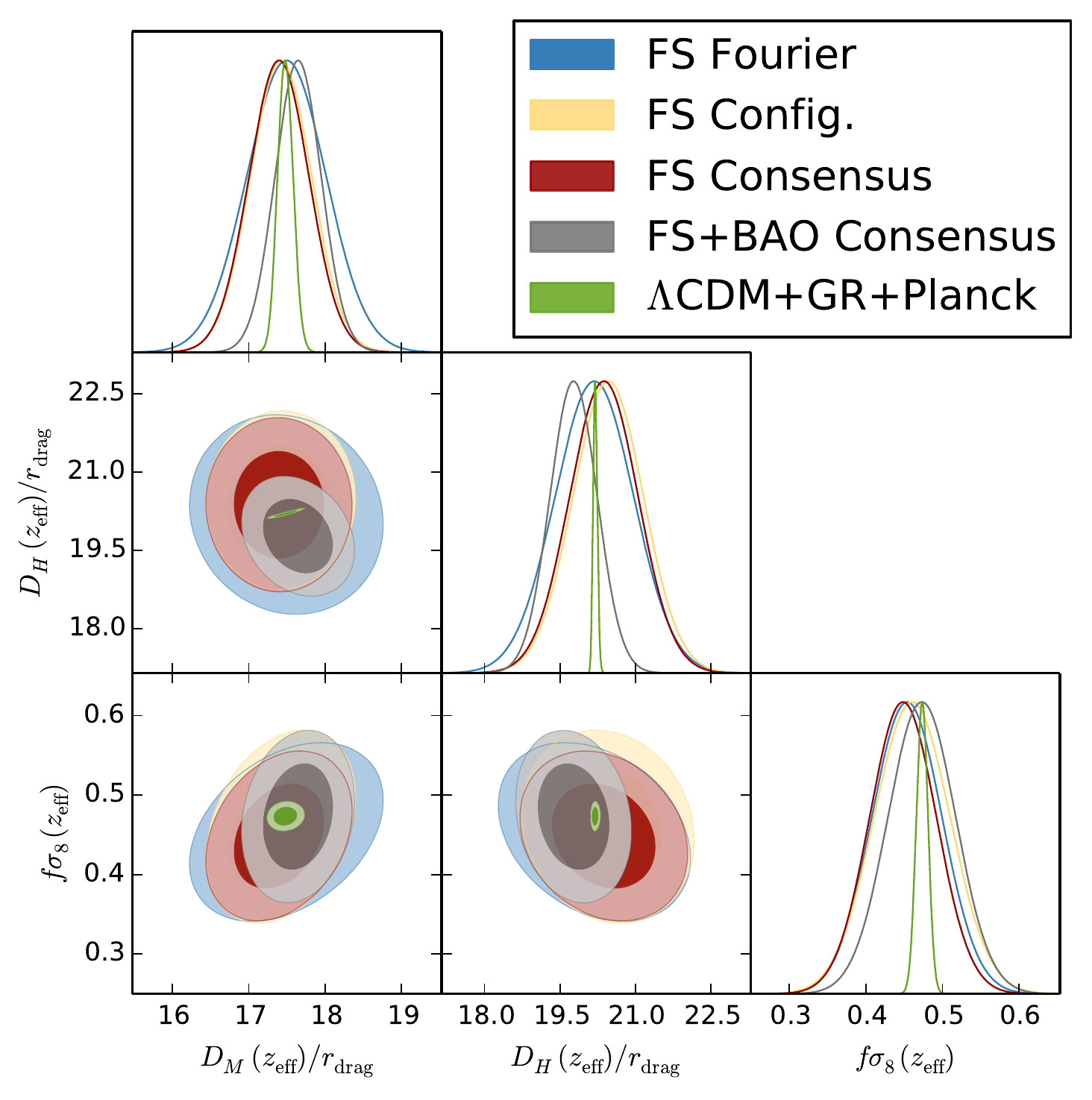}
\caption{{\it Top panel}: Likelihood posteriors from BAO reconstruction analysis , inferred from Fourier space (this work), configuration space \citep{LRG_corr}, and its consensus. The black dashed and dotted lines show the limits imposed by a flat-$\Lambda$CDM model with two sets of wide priors (see text). {\it Bottom panel}: Same as the top panel but from FS analysis, additionally a full consensus between Fourier space, configuration space, BAO and FS analysis is added. In all cases the contours do include the systematic error budget. For reference we include the prediction of flat-$\Lambda$CDM and GR using Planck measurements.}
\label{fig:cosmo}
\end{figure}

Fig.~\ref{fig:cosmo} displays the comparison between the DR16 CMASS+eBOSS LRG analyses on Fourier space (blue contours), configuration space (yellow contours) and its combination (red contours), for the BAO-only analysis on reconstructed catalogues (top panel) and FS analyses (bottom panel). Additionally, in the bottom panel we display the full BAO and FS consensus (gray contours). In green we display the prediction of a flat-$\Lambda$CDM model using the values reported by Planck \citep{planck_collaboration_planck_2018}. For the $f\sigma_8$ panel, the additional relation set by GR is used to infer $f(z)=\Omega_m^{6/11}(z)$ and the linear growth factor $D(z)$, which propagates $\sigma_8(z)=D(z)\sigma_8(z=0)$ to the redshift of interest, $z=0.698$. 

The agreement between Fourier and configuration space is very good, as we already reported in Table~\ref{tab:finalresults} and \S\ref{sec:consensus}. When BAO reconstructed information is used in combination with FS-only analyses we obtain the tighter constrains of this paper. In all cases the agreement with the flat-$\Lambda$CDM+GR model prediction is excellent. 

We remark that the methodology used in this paper to infer $D_M(z_{\rm eff})/r_{\rm drag}$ and $D_H(z_{\rm eff})/r_{\rm drag}(z)$ does not assume the internal $\Lambda$CDM prior:
$D_M(z)=\int_0^z dz'\, D_H(z')$. This relation sets additional limits on the $D_H/r_{\rm drag}-D_M/r_{\rm drag}$ parameter space, which in the top panel of Fig.~\ref{fig:cosmo} is shown as black lines when the following hard priors are used: $\{135<r_{\rm drag}\,[{\rm Mpc}]<165;\, 0.25<\Omega_m<0.90;\,  0.55<h<0.80\}$ in dashed lines; and $\{135<r_{\rm drag}\,[{\rm Mpc}]<165;\, 0.145<\Omega_mh^2<0.200;\,\}$\footnote{The hard low prior on $\Omega_mh^2$ seems to exclude the value preferred by Planck+$\Lambda$CDM. This effect is caused by the wide prior on $r_{\rm drag}$. When the prior on $r_{\rm drag}$ is tightened, the prior on $\Omega_mh^2$ needs to be relaxed to maintain the same limits on the $D_H/r_{\rm drag}-D_M/r_{\rm drag}$ plane, which would make the $\Omega_mh^2$ prior consistent with Planck best-fit.} in dotted lines. Using the wide $\Omega_x$-type of $\Lambda$CDM  priors (dashed lines) is not the optimal approach, as it easily hits the LSS contours even in this wide-prior scenario. This situation can be partially solved by imposing $\Omega_xh^2$-type of priors instead (dotted lines). However, one has to control the effect that priors on $\Omega_bh^2$ and $\Omega_mh^2$ has on $r_{\rm drag}$, which we do not study here. Therefore, those LSS analyses that iteratively change the shape of the power spectrum according to $\Lambda$CDM templates (see for e.g. \cite{Damicoetal2020,Ivanovetal:2020,Trosteretal:2020}) have to carefully asses the impact of these type priors on their analyses.
\begin{figure}
\centering
\includegraphics[scale=0.3]{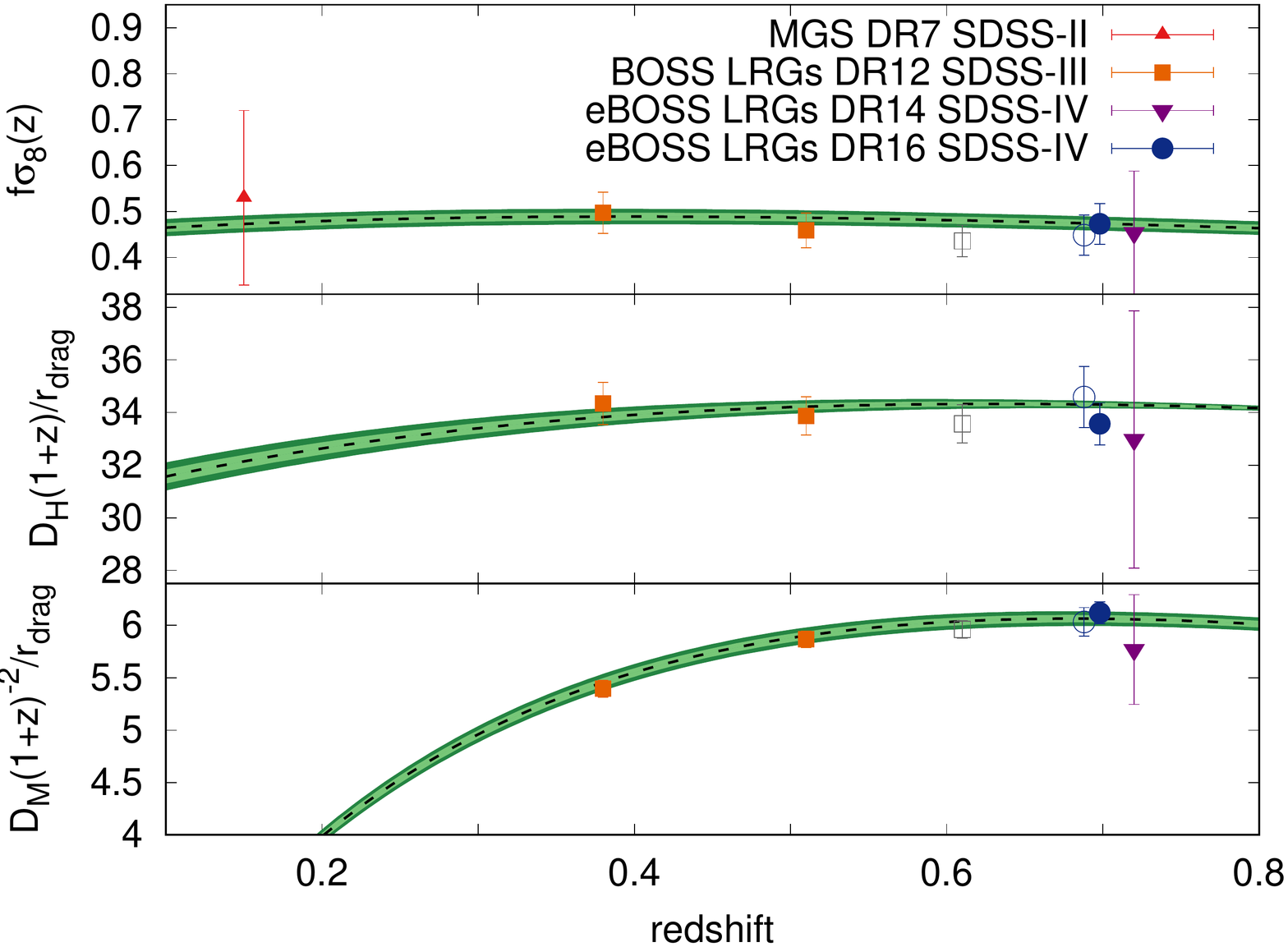}
\caption{Cosmology measurements based on low-redshift galaxies, for the DR7 Main Galaxy Sample \citep{Howlettetal:2015}  at $z=0.15$, DR12 BOSS LRG sample \citep{alam_clustering_2017} at $z=\{0.38,\, 0.51,\,0.61\}$, DR14 eBOSS LRG sample \citep{icaza-lizaola_clustering_2019} at $z=0.72$, and DR16 CMASS+eBOSS LRG sample (this work in combination with \citealt{LRG_corr}) at $z=0.698$. The DR16 CMASS+eBOSS LRG empty symbol correspond to the values inferred from the FS-only analysis, whereas the filled symbol to the full consensus of FS $+$ BAO reconstruction. Note that, {\it i)} the low and middle DR12 BOSS LRG sample measurements are correlated; {\it ii)} the high redshift bin of DR12 BOSS LRG sample (in gray) is fully contained by the DR16 CMASS+eBOSS LRG sample and the middle redshift bin of  DR12 BOSS LRG sample, and therefore, does not add any extra information. For reference in green bands the constraints inferred by flat-$\Lambda$CDM and GR using Planck measurements \citep{planck_collaboration_planck_2018} is also shown.  }
\label{fig:zevo}
\end{figure}
Fig.~\ref{fig:zevo} displays the predicted evolution with redshift of the parameters $D_M(z)/r_{\rm drag}$, $D_H(z)/r_{\rm drag}$ and $f\sigma_8(z)$ predicted by the $\Lambda$CDM model and GR using the Planck measurements (green contours) for the $1$ and $2\sigma$ confidence levels.  The symbols show the measurements by the main galaxy sample (MGS, \citealt{Howlettetal:2015}), DR12 BOSS LRG sample \citep{alam_clustering_2017}, DR14 eBOSS LRG sample \citep{icaza-lizaola_clustering_2019}, and DR16 CMASS+eBOSS LRG sample (this work in combination with \citealt{LRG_corr}) at $z=0.698$, for the FS analysis (empty symbol), and for the FS $+$ BAO analysis (filled symbol). 

We see the great improvement in the constraining power between the former DR14 eBOSS LRG analysis and the current work. Part of this gain is explained by the larger volume of the DR16 sample, a factor of $\sim 3$ larger, which explains a reduction of about a factor of 2 in the error-bars. Additional reduction is provided by the use of the hexadecapole signal in the DR16 analysis, which helps to break degeneracies between parameters, and can explain the further observed gain. 

The results from the DR12 BOSS LRG sample are shown in orange for the two lowest redshift bins, and in gray for the highest redshift bin. We remind the reader that with the current DR16 CMASS+eBOSS LRG sample in play, all the LRG galaxies contained in the BOSS high redshift bin are also contained either by either the BOSS middle redshift bin catalogue, or by the DR16 CMASS+eBOSS LRG catalogue. As a consequence, the high redshift bin of BOSS is highly correlated with the adjacent redshift bins, and therefore it barely contains extra information. Therefore, we reorganise the whole set of LRG galaxies observed by BOSS+eBOSS galaxies in 3 redshift bins: the low- ($0.2<z<0.5,\, z_{\rm eff}=0.38$) and middle-redshift bin ($0.4<z<0.6,\,z_{\rm eff}=0.51$), both from the DR12 BOSS analysis, and a new high redshift bin ($0.6<z<1.0,\,z_{\rm eff}=0.698$) containing BOSS and eBOSS LRG galaxies. Table~\ref{tab:BOSScov} summarises these measurements and correlations. 

Fig.~\ref{fig:zevo} displays a very good agreement between the measured quantities and model predictions in the redshift range $0.2<z<1.0$, showing no significant discrepancy in any of the variables. 

\section{Conclusions}\label{sec:conclusions}

We have performed BAO and full shape analyses in Fourier space of the final DR16 CMASS+eBOSS LRG catalogue, consisting of 377,458 galaxies in the redshift range $0.6<z<1.0$. In order to increase the BAO signal we have applied the density-field reconstruction technique in order to remove the non-linear gravitational physics, and enhance  BAO peak detection.  We have extracted the monopole, quadrupole and hexadecapole signal of the pre- and post-reconstructed galaxy catalogues and employed them to measure the comoving angular diameter distance over the horizon scale at drag epoch, $D_M(z_{\rm eff})/r_{\rm drag}$, the Hubble distance over the horizon scale at drag epoch, $D_H(z_{\rm eff})/r_{\rm drag}$ and the logarithmic growth factor times the amplitude of dark matter fluctuations at  scales of $8\, \mpcoh$, $f\sigma_8(z_{\rm eff})$, at the effective redshift of the sample, $z_{\rm eff}=0.698$. These analyses are complementary to those performed in configuration space and presented in \cite{LRG_corr}. We have found an excellent agreement between the Fourier space and configuration space inferred parameters, both for BAO and FS-type of analysis. 

We have combined the cosmological results produced in both spaces to generate a set of consensus parameters which represents the most precise and accurate cosmological measurements at this epoch: $D_M(z_{\rm eff})/r_{\rm drag}=17.65\pm0.30$, $D_H(z_{\rm eff})/r_{\rm drag}=19.77\pm0.47$, $f\sigma_8(z_{\rm eff})=0.473 \pm 0.044$. 

We have tested the validity of the approaches used in this paper employing realistic {\it N}-body simulation catalogues. We have quantified 4 types of sources of potential systematic errors: {\it i)} systematic errors arising from the inaccuracy of the modelling; {\it ii)}  systematic errors produced by the arbitrary choice of reference cosmology, and systematic errors produced by {\it iii)} observational effects such has redshift failures, collisions, completeness effects, and {\it iv)} the radial integral constraint. The total systematic error budget that results is sub-dominant compared to the statistical errors. After propagating the systematic error into the total error budget we have observed that the error bars of the cosmological parameters have increased by about 10\%. 

We have also tested the BAO and FS models with galaxy catalogues for different types of using different types of HOD models. We have observed no significant effect on the cosmological parameters of interest, although the precision on these catalogues does not allow to resolve changes of more than few percent in the cosmological parameters of interest.  

The inferred cosmological parameters from the DR16 CMASS+eBOSS LRG sample show an excellent agreement with the predictions by the standard cosmological model, flat-$\Lambda$CDM+GR, using the cosmological parameters inferred by Planck. These observations complement those based on ELGs \citep{tamone20a,demattia20a}, quasars \citep{hou20a, neveux20a} and Ly-$\alpha$ \citep{2020duMasdesBourbouxH}. A full cosmology interpretation using these and previous BOSS analyses \citep{alam_clustering_2017} is presented in \cite{eBOSS_Cosmology}.

Next generation galaxy surveys, such as the Dark Energy Spectroscopic Instrument (DESI, \citealt{desi1,desi2}), EUCLID \citep{euclid}, the  Large Synoptic Survey Telescope (LSST, \citealt{lsst}) or the Square Kilometer Array (SKA,\citealt{ska}), will extensively re-probe the redshift range $0<z<4$ with an unprecedented level of precision never reached before, and in some cases will extend this range up to $z\simeq6$. 

The SDSS-I and -II (2004-2009), the BOSS (2009-2014), and finally the eBOSS observations (2014-2019) have probed for first time the physics of the late-time Universe using galaxies as dark matter tracers. These experiments have demonstrated that the BAO and RSD techniques can effectively be used to measure expansion and logarithmic growth rate, opening a new window for the next generation of experiments which potentially will reveal hints of new physics phenomena occurring in our Universe.

\section*{Data Availability}
The power spectra, covariance matrices, and resulting likelihoods for cosmological parameters are available via the SDSS Science Archive Server (\href{https://data.sdss.org/sas/dr16/eboss/}{https://data.sdss.org/sas/dr16/eboss/}), as well as in \href{https://icc.ub.edu/~hector/page13.html}{here}.
\section*{Acknowledgements} 
HGM acknowledges the support from la Caixa Foundation (ID 100010434) which code LCF/BQ/PI18/11630024.  RP, SdlT, and SE acknowledge support from the ANR eBOSS project (ANR-16-CE31-0021) of the French National Research Agency. SdlT and SE acknowledge the support of the OCEVU Labex (ANR-11-LABX-0060) and the A*MIDEX project (ANR-11-IDEX-0001-02) funded by the "Investissements d’Avenir" French government program managed by the ANR. MVM and SF are partially supported by Programa de Apoyo a Proyectos de Investigaci\'on e Inovaci\'on Teconol\'ogica (PAPITT)  no. IA101518, no. IA101619  and Proyecto LANCAD-UNAM-DGTIC-136. G.R. acknowledges support from the National Research Foundation of Korea (NRF) through Grants No. 2017R1E1A1A01077508 and No. 2020R1A2C1005655 funded by the Korean Ministry of Education, Science and Technology (MoEST), and from the faculty research fund of Sejong University. SA is supported by the European Research Council through the COSFORM Research Grant (\#670193). EMM is supported by the  European Research Council (ERC) under the European Union’s Horizon 2020 research and innovation programme (grant agreement No 693024).

The numerical computations presented in this work were done on \textsc{Hipatia} and \textsc{Aganice} ICC-UB BULLx High Performance ComputingCluster at the University of Barcelona. This research used resources of the National Energy Research Scientific Computing Center, a DOE Office of Science User Facility supported by the Office of Science of the U.S. Department of Energy under Contract No. DE-AC02-05CH11231. This research also uses resources of the HPC cluster ATOCATL-IA-UNAM M\'exico.

In addition, this research relied on resources provided to the eBOSS Collaboration by the National Energy Research Scientific Computing Center (NERSC).  NERSC is a U.S. Department of Energy Office of Science User Facility operated under Contract No. DE-AC02-05CH11231.

Funding for the Sloan Digital Sky Survey IV has been provided by the Alfred P. Sloan Foundation, the U.S. Department of Energy Office of Science, and the Participating Institutions. SDSS-IV acknowledges
support and resources from the Center for High-Performance Computing at
the University of Utah. The SDSS web site is www.sdss.org.

SDSS-IV is managed by the Astrophysical Research Consortium for the 
Participating Institutions of the SDSS Collaboration including the 
Brazilian Participation Group, the Carnegie Institution for Science, 
Carnegie Mellon University, the Chilean Participation Group, the French Participation Group, Harvard-Smithsonian Center for Astrophysics, 
Instituto de Astrof\'isica de Canarias, The Johns Hopkins University, Kavli Institute for the Physics and Mathematics of the Universe (IPMU) / 
University of Tokyo, the Korean Participation Group, Lawrence Berkeley National Laboratory, 
Leibniz Institut f\"ur Astrophysik Potsdam (AIP),  
Max-Planck-Institut f\"ur Astronomie (MPIA Heidelberg), 
Max-Planck-Institut f\"ur Astrophysik (MPA Garching), 
Max-Planck-Institut f\"ur Extraterrestrische Physik (MPE), 
National Astronomical Observatories of China, New Mexico State University, 
New York University, University of Notre Dame, 
Observat\'ario Nacional / MCTI, The Ohio State University, 
Pennsylvania State University, Shanghai Astronomical Observatory, 
United Kingdom Participation Group,
Universidad Nacional Aut\'onoma de M\'exico, University of Arizona, 
University of Colorado Boulder, University of Oxford, University of Portsmouth, 
University of Utah, University of Virginia, University of Washington, University of Wisconsin, 
Vanderbilt University, and Yale University.

\input{variables.tex}

\bibliographystyle{mnras}
\bibliography{DR16_LRGBAO_FourierSpace}

\appendix

\section{Effect of isotropic template on BAO determination}\label{sec:APtest}

\begin{figure}
\centering
\includegraphics[scale=0.7]{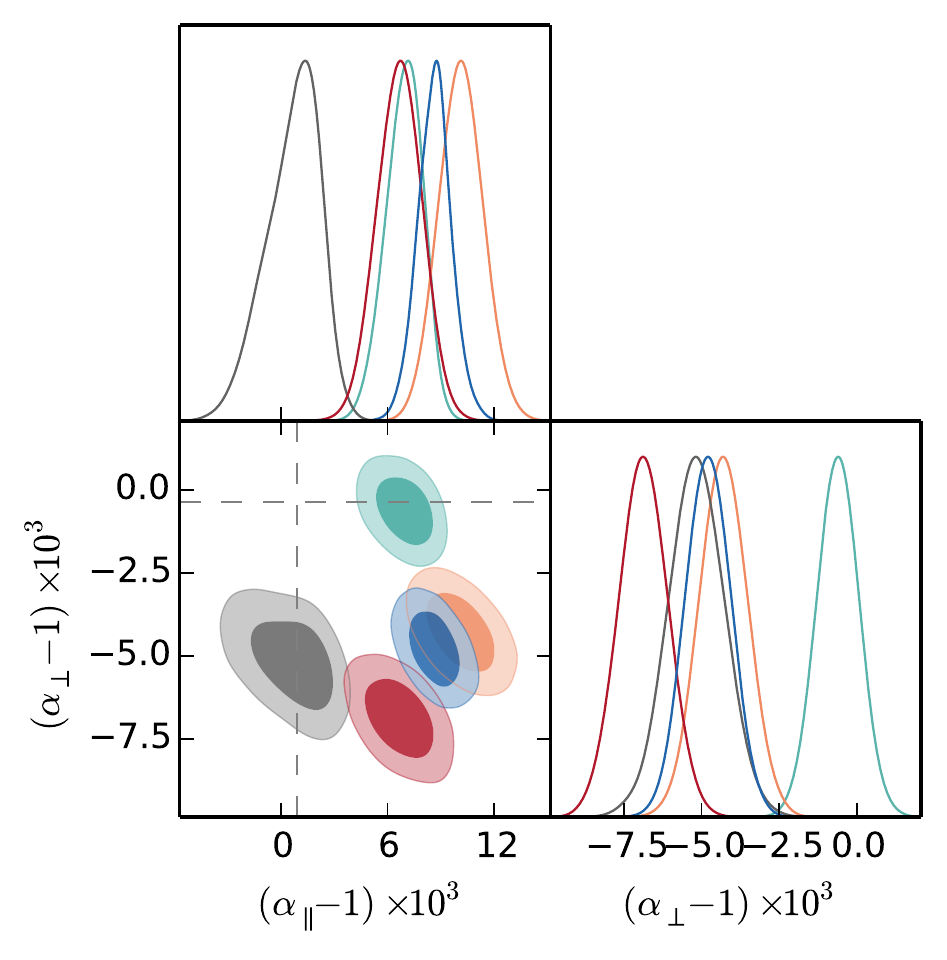}
\caption{BAO template comparison:  pre-recon anisotropic 3rd order (gray), post-recon isotropic of 3rd order, (orange), post-recon isotropic of 5th order (red), post-recon anisotropic of 3rd order (turquoise), post-recon anisotropic of 5th order (blue).  }
\label{plot:template}
\end{figure}

In this appendix we aim to show the performance of the isotropic BAO template described by Eq. \ref{eq:iso} and \ref{eq:Psm} when is applied on the analysis the full anisotropic signal, compared to the standard approach, described by Eq. \ref{eq:baolin} and \ref{eq:baoaniso}. In principle the standard approach is more complete as it describe better the damping of the BAO signal in an explicit $\mu$-dependent way, whereas the isotropic template takes only the average of this dependence for each multipole into account. However, for the post-reconstructed data, the BAO damping is weak (as most of it is removed by the reconstruction process) and both approaches converge: in the limit $\Sigma_{\perp,\,\parallel}\rightarrow0$ both approaches are equivalent. Fig. \ref{plot:template} displays the performance on the mean of the \textsc{EZmocks} for the reconstructed catalogue when the isotropic template is used using 3 (orange contours) and 5 (red contours) polynomial terms in the broadband; and when the anisotropic is also used in the same catalogue for 3 (turquoise contours which corresponds to the standard approach used in the paper) and 5 (blue contours) polynomial terms in the broadband. In all the cases, $\Sigma_0,\,\Sigma_2$ (in the isotropic template), as well as $\Sigma_\parallel,\,\Sigma_\perp$ (in the anisotropic template) have been freely fit to the mean of the mocks. For reference we also show the performance on the pre-recon catalogue of the anisotropic template with 3 terms in the polynomial broadband function. The horizontal and vertical dashed lines show the expected values given the difference between the reference (in this case the fiducial) and underlying true cosmology of the mocks. 

We see that all the post-reconstructed analyses perform very similarly on the determination of the $\alpha_\parallel$ variable, which shows an about $0.6\%$ shift with respect to the expected value, for all the studied cases. On the other hand, the different templates display a different performance when determining the $\alpha_\perp$ variable. The isotropic template tends to consistently underestimate $\alpha_\perp$ by about $0.5-0.75\%$, regardless of the polynomial order of the broadband. On the other hand, when the anisotropic template is used, having 3 broadband parameters shows unbiased results, whereas when we add two extra parameters we bias the results in about $0.5\%$. Finally, we recall that these studies are performed on fast \textsc{EZmocks}, and therefore, such conclusions should be validated with full {\it N}-body simulations.

\section{Gaussian test}\label{sec:gauss}
\begin{figure}
\centering

\includegraphics[scale=0.75]{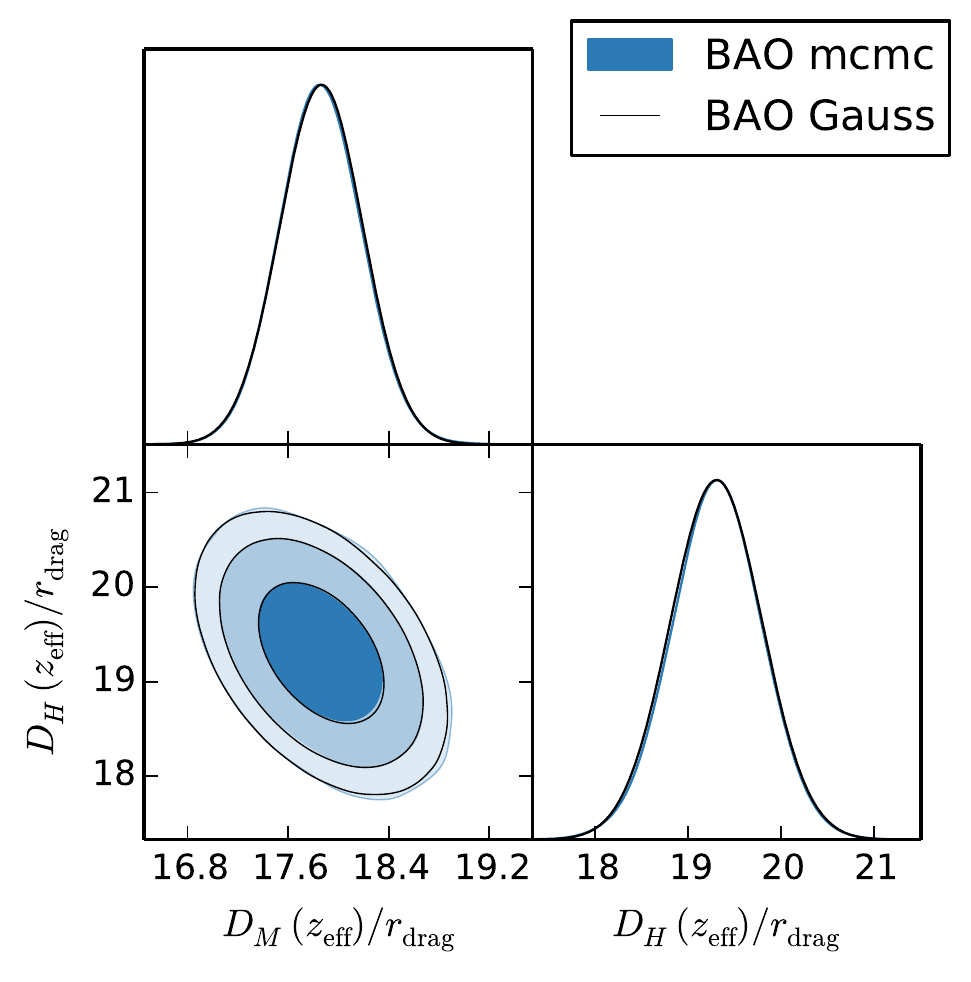}

\includegraphics[scale=0.5]{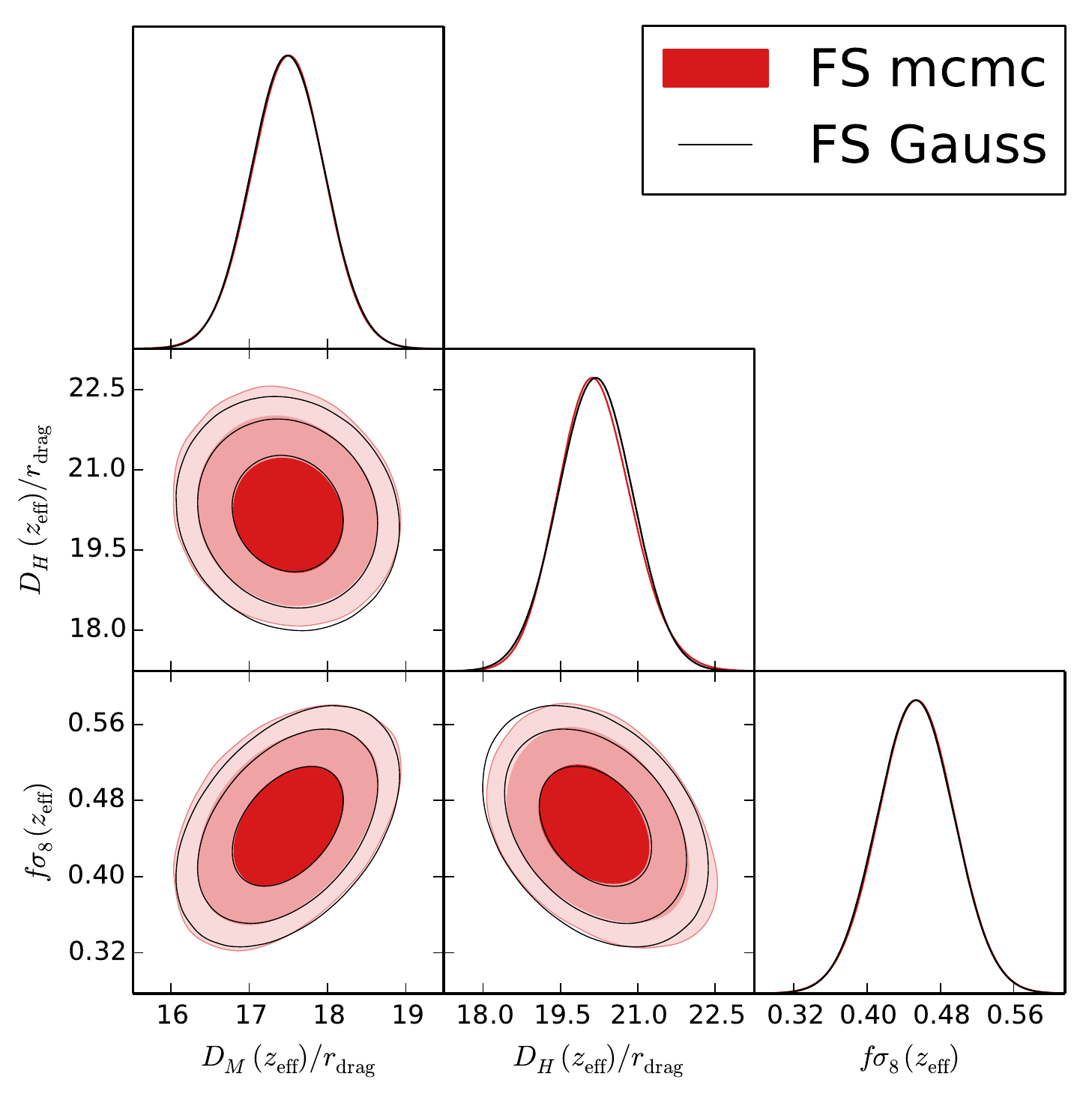}
\label{fig:gausstest} 
\caption{FS and BAO Gaussian test. The top and bottom panel displays the posterior of the  BAO post-reconstruction and FS analysis, respectively. The black curves display the Gaussian prediction, for $1$, $2$ and $3\sigma$ confidence levels. In all cases only the statistical error contribution is shown. }
\end{figure}

We perform a comparison between the outcome of the Monte Carlo Markov Chains on the DR16 CMASS+eBOSS LRG dataset and its Gaussian approximation, given by the parameters of Table~\ref{table:BAOresults} for BAO and Table~\ref{table:FSresults} for FS analysis. The comparison is displayed by Fig. \ref{fig:gausstest}, for BAO reconstructed chains (blue contours), FS chains (red contours); and their corresponding gaussian contours (black lines), where no systematic error budget has been taken into account for simplicity.  

We conclude that the Gaussian approximation is very good for BAO type of analysis up to $3\sigma$ confidence levels. The FS analysis displays some degree of non-Gaussianity at $3\sigma$, specially for $D_H/r_{\rm drag}$. This kind of behaviour is expected as the modes along the LOS present a higher level of noise, which typically induces non-Gaussian tails. However, such features only appears at the edges of the likelihood shape and have a very small impact in the cosmological constrains.  

\section{covariance}\label{appenddix:cov}
\begin{figure}
\centering
\includegraphics[scale=0.3, trim={0 100 0 0}, clip=false]{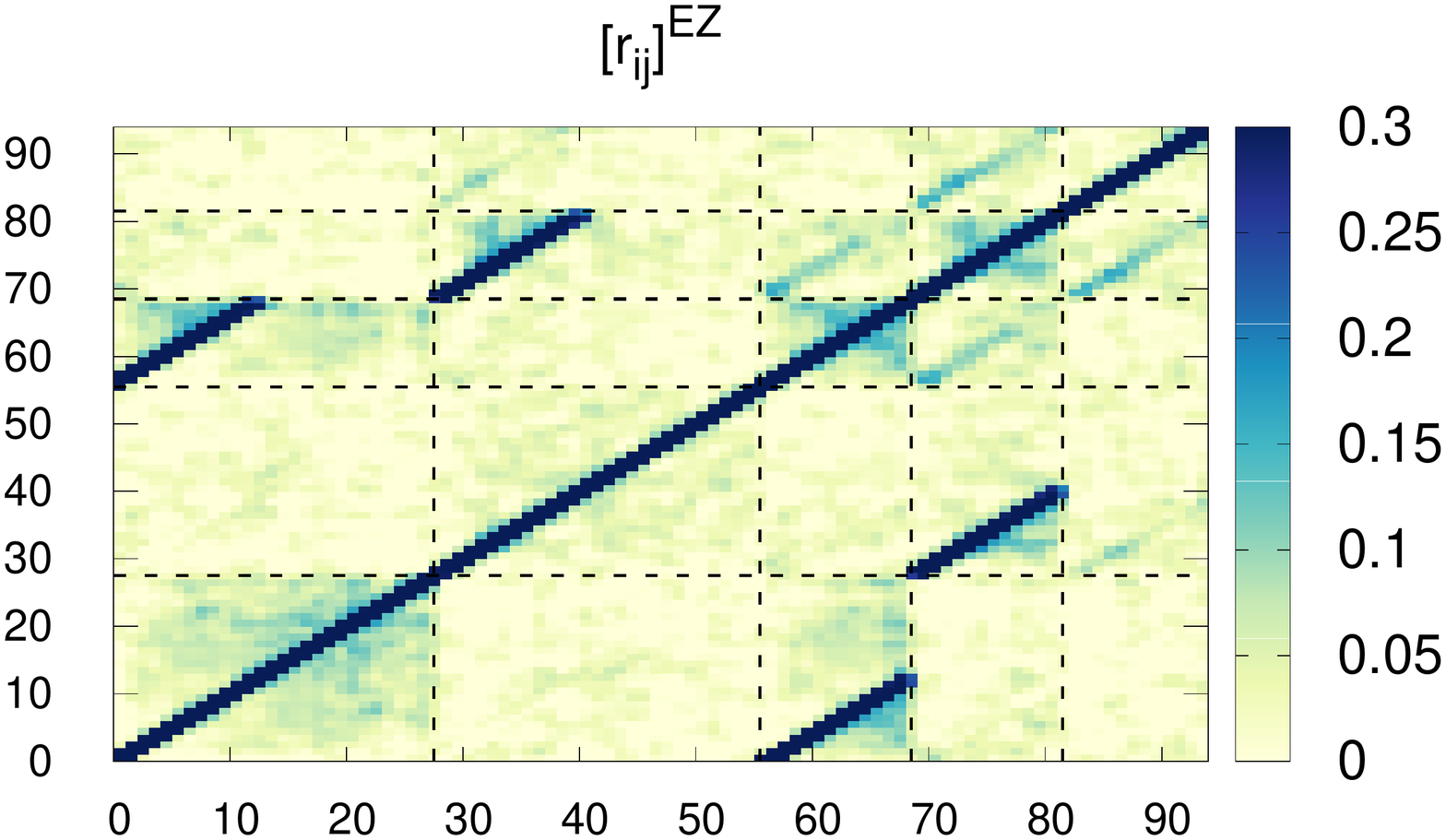}
\includegraphics[scale=0.3, trim={0 70 0 0}, clip=false]{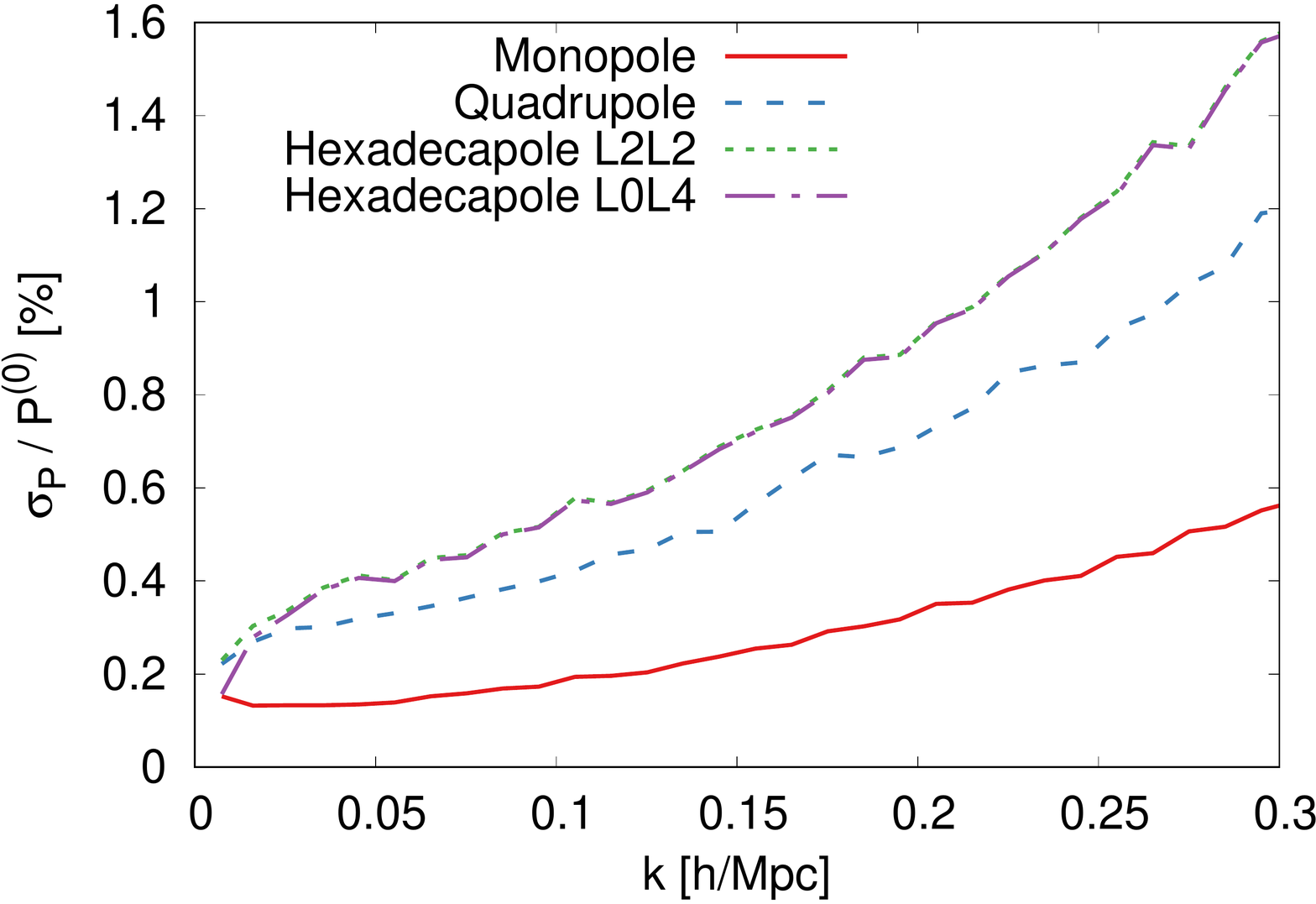}
\caption{The top panel shows the cross-correlation elements of the covariance matrix corresponding to the full power spectrum data-vector. The dashed lines mark the different blocks of the covariance: $P^{(0)}$, $P^{(2)}$ (for reconstructed catalogues), $P^{(0)}$, $P^{(2)}$, $P^{(4)}$ (for pre-reconstructed catalogues). Note that reconstructed elements are 2 times larger than the pre-reconstructed ones due to the difference in $k$-range. The bottom panel shows the statistical error estimated from 1000 realisations of the \textsc{EZmocks} of $P^{(\ell)}(k)$ relative to $P^{(0)}(k)$ for $\ell=0,\,2,\,4$, as a function of $k$. For the hexadecapole two errors are displayed, depending on the LOS treatment (see \S\ref{sec:psestimator}). In all cases only the NGC contribution is shown.}
\label{fig:cov}
\end{figure}

The top panel of Fig.~\ref{fig:cov} display the cross-correlation coefficients of the covariance matrix inferred from the 1000 realisations of the \textsc{EZmocks}.  The matrix is divided in 5 blocks corresponding to the post-reconstructed monopole and quadrupole, between $k=0.02\,$ and $k=0.30\,\hompc$; and  the pre-reconstructed monopole, quadrupole and hexadecapole, between $k=0.02\,$ and $k=0.15\,\hompc$. The high values of the off-diagonal terms corresponding to those elements cross-correlating elements with equal $k$ and $\ell$, but corresponding to pre- and post-catalogues. In order to perform the BAO type of analysis we only invert the 2 first blocks, whereas for the M+Q+H FS analysis we invert the 3 last blocks. Only when the simultaneous fit is performed we invert the 5 blocks all-togheter. 

The bottom panel of Fig.~\ref{fig:cov} display the error of $P^{(\ell)}$ relative to the value of $P^{(0)}$ as a function of $k$. For the hexadecapole, $\ell=4$ we report the error of the two estimators according to the expansions  $\mathcal{L}_4(\hat{\bf k}\cdot\hat{\bf r}_h)\rightarrow \mathcal{L}_4(\hat{\bf k}\cdot\hat{\bf r}_1)$ (L0L4) and $\mathcal{L}_4(\hat{\bf k}\cdot\hat{\bf r}_h)\rightarrow \mathcal{L}_2(\hat{\bf k}\cdot\hat{\bf r}_1) \mathcal{L}_2(\hat{\bf k}\cdot\hat{\bf r}_2)$ (L2L2), in Eq.~\ref{eq:Pyama}. We see that the variance of these two estimators of the hexadecapole is very close for the $k$-range used here, $0.02\leq k\,[\hompc]$, which implies that the wide-angle effects are in effect negligible. 

\section{Window Function}\label{sec:window}

We account for the survey selection, on a `unmasked' given power spectrum, $P^{\rm pre-mask}$ through the convolution with a mask function, which results on the `masked' power spectrum which matches the measurements, $P^{\rm post-mask.}$. In this case the survey selection function is computed from the random catalogue, and therefore only depends on the geometry of the survey and not in any clustering property. 
The convolved power spectrum $\ell$-multipoles are therefore written as the Hankel transform of $\hat{\xi}_\ell$, 
\begin{eqnarray}
P_\ell^{\rm post-mask}(k)&=& 4\pi(-i)^\ell\int dr\, r^2 \hat{\xi}_\ell(r) j_\ell(kr)
\end{eqnarray}
where $j_\ell(x)$ are the spherical Bessel functions of  $\ell$-order, and $\hat{\xi}_\ell(r)$ is given by 
\begin{eqnarray}
\nonumber\hat{\xi}_{0}(r)&=& \xi_{0}(r)W_0^2(r)+\frac{1}{5}\xi_{2}(r)W_2^2(r)+\frac{1}{9}\xi_{4}(r)W_4^2(r)\\
\\
\nonumber\hat{\xi}_{2}(r)&=& \xi_{0}(r)W_2^2(r) \\
\nonumber&+&\xi_{2}(r)\left [ W_0^2(r)+\frac{2}{7}W_2^2(r)+\frac{2}{7}W_4^2(r) \right]\\
\nonumber&+&\xi_{4}(r)\left [\frac{2}{7}W_2^2(r)+\frac{100}{693}W_4^2(r)+\frac{25}{143}W_6^2(r)\right]\\
\\
\nonumber\hat{\xi}_{4}(r)&=& \xi_{0}(r)W_4^2(r)\\
\nonumber&+&\xi_{2}(r)\left [\frac{18}{35} W_2^2(r)+\frac{20}{77}W_4^2(r)+\frac{45}{143}W_6^2(r) \right]\\
\nonumber&+&\xi_{4}(r)\left [W_0^2(r)+\frac{20}{77}W_2^2(r)\frac{162}{1001}W_4^2(r)\right.\\
&+&\left.\frac{20}{143}W_6^2(r)+\frac{490}{2431}W_8^2(r)\right]
\end{eqnarray}
where $\xi_\ell$ is the inverse Hankel Transform of  $P^{\rm pre-mask}$,
\begin{eqnarray}
\xi_\ell(r)&=& \frac{4\pi i^\ell}{(2\pi)^3}\int dk\, k^2 P^{\rm pre-mask}_\ell(k) j_\ell(kr).
\end{eqnarray}
Note that $W_0^2(r)=1$ and $W_{\ell>0}^2(r)=0$ corresponds to the case of no-selection function, where  $P^{\rm pre-mask}=\hat{P}^{\rm post-mask}$, as it happens within a cubic box with uniform mean density and periodic boundary conditions. 

\begin{figure}
\centering
\includegraphics[scale=0.30]{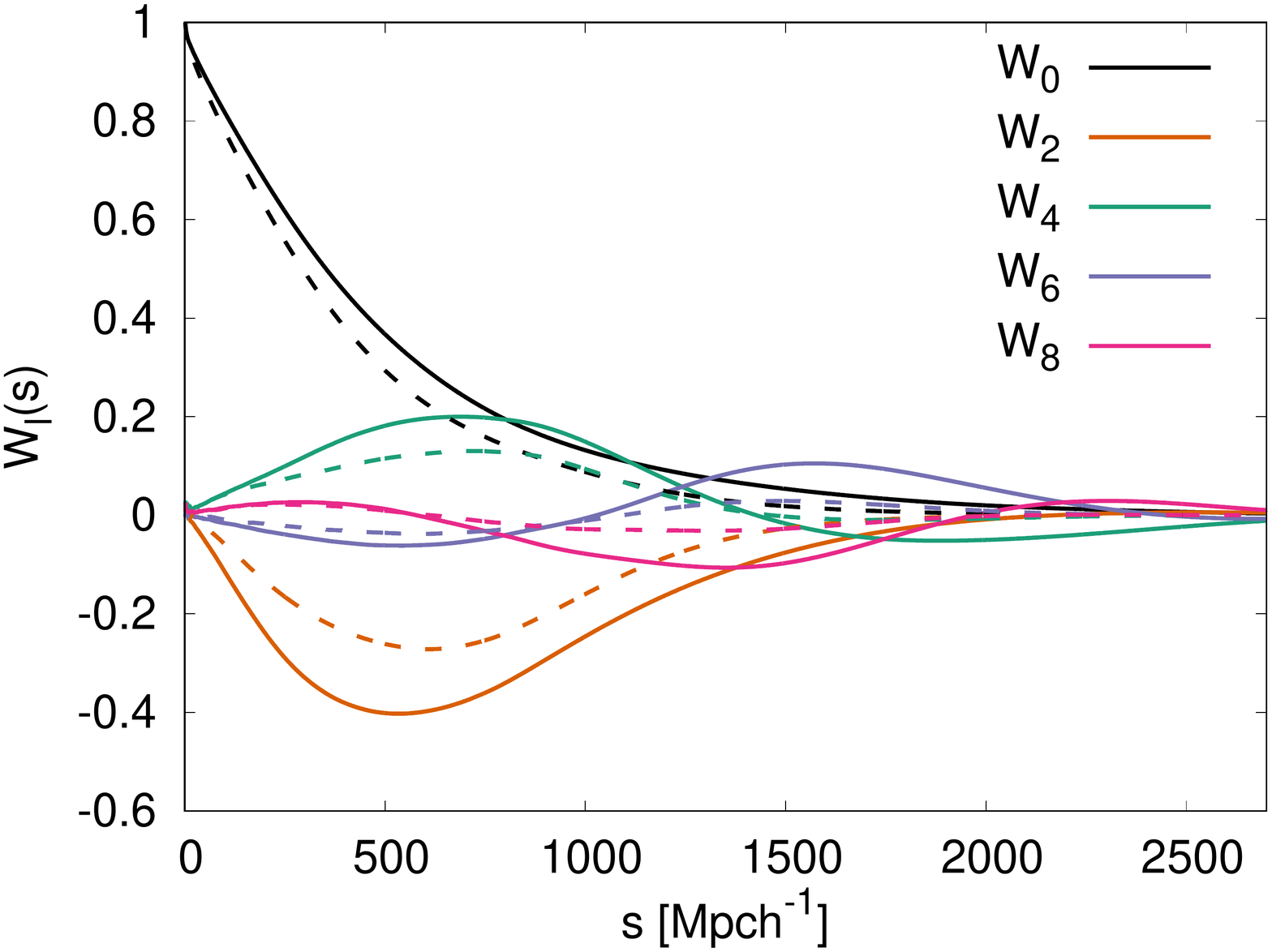}
\caption{Selection function multipoles according to Eq. \ref{eq:wl} for the DR16 CMASS+eBOSS LRG catalogue within $0.6\leq z \leq 1.0$. The solid lines represent the NGC and the dashed lines the SGC. The different colours display the performance for the even $\ell$-multipoles as indicated in the key.  }
\label{Fig:window}
\end{figure}

Fig. \ref{Fig:window} display the $W_\ell^2(r)$ functions for the NGC/SGC DR16 CMASS+eBOSS LRG sample (solid lines/dashed lines).

\section{Effect of the priors in the amplitude of shot noise}\label{sec:Anoise}

\begin{figure}
\centering
\includegraphics[scale=0.5]{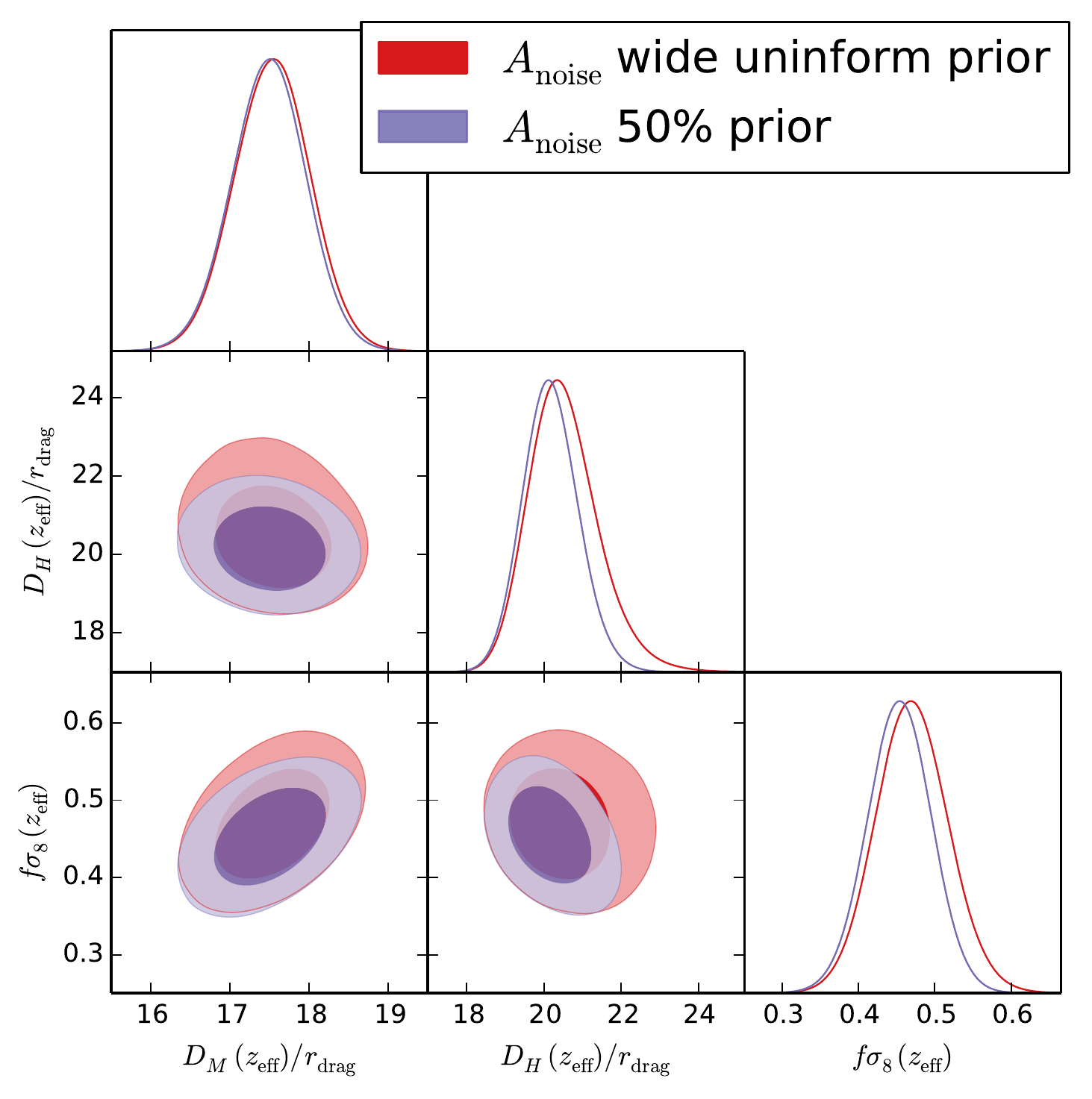}
\caption{Effect of the noise prior on the likelihood posteriors of the variables of cosmological interest for the FS analysis. The red contours show the posteriors with an uninformative prior on the amplitude of shot noise; the purple contours show the same when a hard prior has been applied to this amplitude to not differ more than 50\% from the Poissonian prediction. The effect on the rest of the model parameters is displayed in Table~\ref{plot:fullvector_Anoise}.}
\label{fig:contoursRSDBAO_Anoise}
\end{figure}

Fig.~\ref{fig:contoursRSDBAO_Anoise} displays the effect of the 50\%-noise prior, $0.5<A_{\rm noise}<1.5$ (purple contours) on the likelihood posterior of the cosmological parameters of interest, $D_H/r_{\rm drag}$, $D_M/r_{\rm drag}$ and $f\sigma_8$, for the DR16 CMASS+eBOSS LRG sample. For reference we show in red contours the posterior corresponding to a wider and uninformative prior on $A_{\rm noise}$.  The effect of the 50\%-noise prior prior is almost uninformative for $D_M/r_{\rm drag}$ and remove the non-Gaussian tail on the higher side of the likelihood and posteriors of $D_H/r_{\rm drag}$ and $f\sigma_8$, which are highly correlated. 

\begin{figure*}
\centering
\includegraphics[scale=0.45, trim={400 0 0 0}, clip=false]{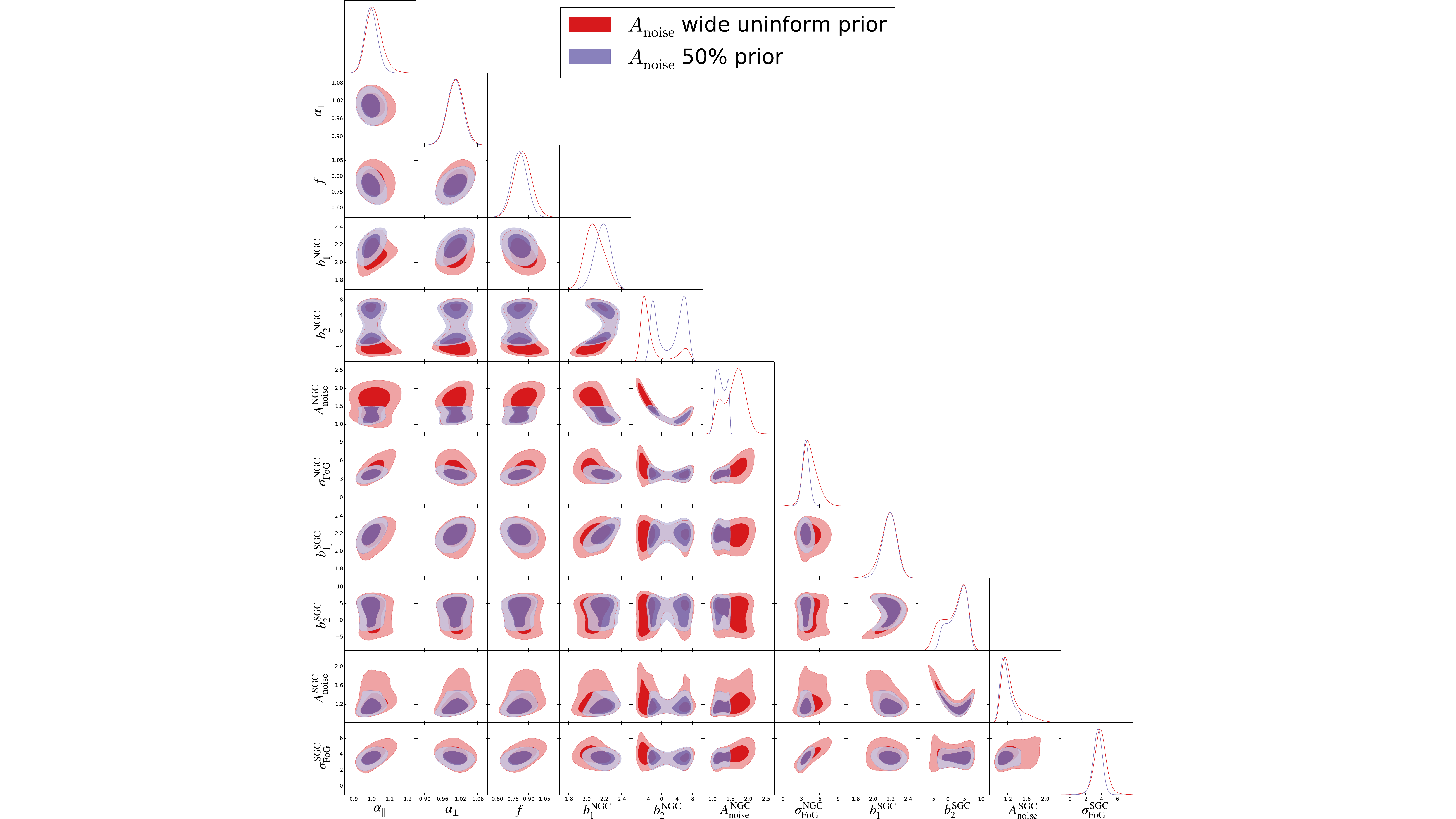}
\caption{Posterior likelihood for all the 11 parameters of the FS fit with M+Q+H and $k_{\rm max}=0.15$ to the DR16 LRG CMASS+eBOSS dataset, with an uninformative prior on $A_{\rm noise}$ (red contours) and with the prior $0.5<A_{\rm noise}<1.5$ (the main result for FS in this paper) in purple. Note at the strong correlation between $b_2$ and $A_{\rm noise}$, which drive $A_{\rm noise}$ to take unphysical values $A_{\rm noise}\sim2$. }
\label{plot:fullvector_Anoise}
\end{figure*}

Fig. \ref{plot:fullvector_Anoise} displays the same effect but extended to the full parameter-vector of the FS type of analysis. Some of the variables show a highly non-Gaussian behaviour. This is the case of $b_2$, which is poorly constrained by the power spectrum. In this case, we observe a strong banana-shape type of correlation between  $b_2$ and $A_{\rm noise}$. This effect is leaked through correlations to the rest of parameters, in particular into $\alpha_\parallel$ and $f$, which causes the non-Gaussian tails showed in Fig.~\ref{plot:fullvector_Anoise}. We can partly solves this spurious behaviour by imposing the  50\%-noise prior on $A_{\rm noise}$ (purple contours). Alternatively, including bispectrum data would also help to constrain $b_2$ and naturally help to keep $A_{\rm noise}$ posterior around values of 1, which is the Poisson prediction.

\section{Tables}

In this section we include a series of tables which can be helpful for the reader to cross-check some values of the plots displayed in the main text. We list them below briefly.

\begin{table}
\caption{Impact of non-periodicity in \textsc{OuterRim-HOD}-mocks for pre-recon catalogues. The cubic box catalogues consist of non-periodic cubic boxes of $1\,h^{-1}{\rm Gpc}$. The sky-cuts catalogues mimic a sky-mock geometry (i.e. non-uniform $n(z)$), where galaxies and randoms are placed in cubic boxes of $3\,h^{-1}{\rm Gpc}$, where $2/3$ of the volume is empty). The periodicity of the box is implicitly assumed under any discrete Fourier space calculation. We expect that the non-periodic effects are negligible for the sky-cuts mocks, but not for the cubic boxes. For simplicity only the fit to the mean is provided. }
\begin{center}
\begin{tabular}{|c|c|c|c|}
HOD-type & Catalogue & $ \alpha_\parallel  -\alpha_\parallel^{\rm exp}$ & $ \alpha_\perp  -\alpha_\perp^{\rm exp}$ \\
\hline
\hline
Hearin & Sky-cut &  $-0.022 \pm 0.014$ & $0.0108 \pm 0.0099$  \\
Hearin & Cubic box & $0.032 \pm 0.018$ & $-0.0002 \pm 0.0097$ \\
\hline
Leauthaud & Sky-cut  & $-0.011 \pm 0.018$ & $0.003 \pm 0.011$ \\
Leauthaud & Cubic box & $0.023 \pm 0.018$ & $0.0030 \pm 0.0097$ \\
\hline
Tinker & Sky-cut & $0.002 \pm 0.018$ & $-0.005 \pm 0.012$ \\
Tinker & Cubic box & $0.017 \pm 0.024$ & $0.002 \pm 0.012$ \\
\end{tabular}
\end{center}
\label{table:periodicity}
\end{table}%

Table~\ref{table:periodicity} accounts for the impact of non-periodic boundary conditions of the $1\,h^{-1}{\rm Gpc}$ cubic boxes from the \textsc{OuterRim}-HOD mocks by comparing them with the  $3\,h^{-1}{\rm Gpc}$ padded sky-cut samples.

\begin{table}
\caption{Impact of different HOD types and `flavours' on pre-recon BAO fits on cubic boxes without periodic boundary conditions. For simplicity only fit on the mean is provided. }
\begin{center}
\begin{tabular}{|c|c|c|c|c|c}
HOD-type & HOD-flavour &  $\alpha_\parallel  -\alpha_\parallel^{\rm exp}$ & $ \alpha_\perp  -\alpha_\perp^{\rm exp}$ \\
\hline
\hline
Hearin & standard & $0.028 \pm 0.010$ & $-0.0005 \pm 0.0054$ \\
Hearin & Threshold 1 & $ 0.024 \pm 0.011$ & $-0.0002 \pm 0.0061$ \\
Hearin & Threshold 2 & $0.032 \pm 0.018$ & $-0.0002 \pm 0.0097$ \\
\hline
Leauthaud & standard & $0.030 \pm 0.010$ & $-0.0016 \pm 0.0052$ \\
Leauthaud & Threshold 1 & $0.029 \pm 0.010$ & $-0.0013 \pm 0.0052$ \\
Leauthaud & Threshold 2 & $0.023 \pm 0.018$ & $0.0030 \pm 0.0097$ \\
\hline
Tinker & standard & $0.038 \pm 0.010$ & $-0.0005 \pm 0.0054$ \\
Tinker & Threshold 1 & $0.025 \pm 0.010$ & $-0.0013 \pm 0.0055$ \\
Tinker & Threshold 2 & $0.017 \pm 0.024$ & $0.002 \pm 0.012$ \\
\hline
Zheng & standard & $0.024 \pm 0.010$ & $-0.0005 \pm 0.0060$ \\
Zheng & Threshold 1 & $0.027 \pm 0.011$ & $-0.0006 \pm 0.0062$ \\
Zheng & Threshold 2 & $ 0.031 \pm 0.013$ & $-0.0014 \pm 0.0075$ \\
\end{tabular}
\end{center}
\label{tab:HOD4BAO}
\end{table}

Table~\ref{tab:HOD4BAO} lists the effect of different HOD models and flavours of the $1\,h^{-1}{\rm Gpc}$ cubic sub-boxes drawn from the \textsc{OuterRim}-HOD mocks on BAO pre-recon analysis.

\begin{table}
\caption{Parameters from FS analysis corresponding to $k_{\rm max}=0.15\,\hompc$ using the power spectrum monopole and quadrupole (M+Q); and also the hexadecapole (M+Q+H) which is the main FS result of this paper. In both cases the results are obtained with with 50\% prior on $A_{\rm noise}$. $f$ should be rescaled by the fiducial $\sigma_8(\alpha_0)$ value according to Eq.~\ref{eq:sigma8res}, which for the used template is $\sigma_8(\alpha_0=1.0020)=0.55825$ (for M+Q+H fit); and $\sigma_8(\alpha_0=1.0081)=0.55590$ (for the M+Q fit). For reference the $\sigma_8(\alpha_0=1)=0.55901$.}
\begin{center}
\begin{tabular}{|c|c|c}
Parameter & Value M+Q & Value M+Q+H\\
\hline
\hline
$\alpha_\parallel$ & $0.9724 \pm 0.0496$  & $0.9994 \pm 0.0357$\\
$\alpha_\perp$ & $1.0265 \pm 0.0363$ & $1.0033 \pm 0.0269$ \\
$f$ & $0.892 \pm 0.111$ &  $0.8125 \pm 0.0749$ \\
\hline
$b_1^{\rm NGC}$ & $2.1466 \pm 0.0942$  &  $2.1851 \pm 0.0891$ \\
$b_1^{\rm SGC}$ & $2.1488 \pm 0.0881$ & $2.1896 \pm 0.0828$ \\
\hline
$b_2^{\rm NGC}$ & $3.16\pm3.69$ & $2.42\pm3.71$\\
$b_2^{\rm SGC}$ & $3.07\pm2.68$ &$3.06\pm 2.71$\\
\hline
$A_{\rm noise}^{\rm NGC}$ & $1.257 \pm 0.133$ & $ 1.254 \pm 0.140$\\
$A_{\rm noise}^{\rm SGC}$ & $1.188 \pm 0.119$  & $1.172 \pm 0.120$\\
\hline
$\sigma_{\rm FoG}^{\rm NGC}$ & $3.658 \pm 0.604$ & $3.757 \pm 0.598$\\
$\sigma_{\rm FoG}^{\rm SGC}$ & $3.563 \pm 0.605$ & $ 3.525 \pm 0.606$\\
\hline
$\chi^2/{\rm dof}$ & $38/(52-11)$ &  $77/(78-11)$
\end{tabular}
\end{center}
\label{tab:resultsPk}
\end{table}

Table~\ref{tab:resultsPk} provides the FS best-fitting parameters to the DR16 CMASS+eBOSS LRG catalogues, for M+Q and M+Q+H cases. Note that these are the raw results, performed at a given fixed template amplitude. Therefore, the $f$ and biases values need to be re-scaled the template amplitude, to be physically interpreted as cosmology-reference invariant parameters. The rescaling factors are given in the table caption. 

\begin{table*}
\caption{Impact of the reference cosmology on  \textsc{Nseries} mocks. The different cosmology models are listed in Table~\ref{tab:cosmo}. For $\boldsymbol{\Theta}_{\rm Nseries}$ we expect that both $\alpha$-parameters are 1; for $\boldsymbol{\Theta}_{\rm fid}$ we expect $\alpha_\parallel^{\rm exp}=0.9875$ and $\alpha_\perp^{\rm exp}=0.9787$; for $\boldsymbol{\Theta}_X$ we expect $\alpha_\parallel^{\rm exp} = 0.9846$ and $\alpha_\perp^{\rm exp} = 0.9620$; for $\boldsymbol{\Theta}_Y$ we expect $\alpha_\parallel^{\rm exp} = 0.9543$ and $\alpha_\perp^{\rm exp} = 0.9325$; for $\boldsymbol{\Theta}_Z$ we expect $\alpha_\parallel^{\rm exp} = 0.9557$ and $\alpha_\perp^{\rm exp} = 0.9291$. We use the same notation of  Table~\ref{tab:sys} where Mean catalogue display the fits to the mean with the error on the mean, and Individual catalogue display the mean of the 84 individual fits reporting the ${\rm \it rms}/\sqrt{N}$. The bottom panel of Fig.~\ref{fig:olin} displays the performance of the results. }
\begin{center}
\begin{tabular}{|c|c|c|c|c}
Reference cosmology & catalogue & $ \alpha_\parallel -\alpha_\parallel^{\rm exp}$ & $\alpha_\perp -\alpha_\perp^{\rm exp}$ & $N_{\rm det}/N_{\rm tot}$\\
\hline
\hline
$\boldsymbol{\Theta}_{\rm Nseries}$ & Mean post-recon  & $-0.0048\pm 0.0019$ & $0.0005\pm 0.0010$ & $1/1$\\
$\boldsymbol{\Theta}_{\rm Nseries}$ & Individual post-recon &  $-0.0016\pm 0.0033$ & $-0.0030\pm 0.0018$ & $84/84$\\
\hline
$\boldsymbol{\Theta}_{\rm fid}$ & Mean post-recon & $0.0006 \pm 0.0018$ & $-0.0026 \pm 0.0012$ & $1/1$\\  
$\boldsymbol{\Theta}_{\rm fid}$ & Individual post-recon & $ 0.0010 \pm0.0030$ &  $-0.0027\pm 0.0017$ & $84/84$\\
\hline
$\boldsymbol{\Theta}_X$ & Mean post-recon & $0.0023 \pm 0.0020$ & $-0.0065 \pm 0.0012$ & $1/1$\\  
$\boldsymbol{\Theta}_X$ & Individual post-recon & $0.0076 \pm 0.0026$ &  $-0.0068 \pm 0.0016$ & $84/84$\\
\hline
$\boldsymbol{\Theta}_Y$ & Mean post-recon & $0.0037 \pm 0.0021$ & $-0.0022 \pm 0.0015$ & $1/1$\\  
$\boldsymbol{\Theta}_Y$ & Individual post-recon & $0.0078 \pm 0.0025$ &  $-0.0024 \pm 0.0016$ & $84/84$\\
\hline
$\boldsymbol{\Theta}_Z$ & Mean post-recon & $0.0055 \pm 0.0020$ & $ -0.0078 \pm 0.0012$ & $1/1$\\  
$\boldsymbol{\Theta}_Z$ & Individual post-recon & $0.0031 \pm 0.0026$ &  $-0.0078 \pm 0.0016$ & $84/84$\\
\end{tabular}
\end{center}
\label{tab:cosmodependence}
\end{table*}%

Table~\ref{tab:cosmodependence} corresponds to the middle panel of Fig. \ref{fig:olin} and displays the impact of the arbitrary choice of cosmology on the post-reconstructed \textsc{Nseries} mocks on the BAO analysis, for both the fit to the mean of the mocks, and the mean of the 84 individual fits. 

\begin{table*}
\caption{Impact of reference cosmology on  \textsc{Nseries} mocks for FS analyses. The different cosmology models are listed in Table~\ref{tab:cosmo}. The expected $\alpha$ values are the same as those from Table~\ref{tab:cosmodependence}. In all the cases we expect to recover the same expected growth of structure, $f\sigma_8^{\rm exp}=0.4702$. We use the same notation of  Table~\ref{tab:sys} where Mean catalogue display the fits to the mean with the error on the mean, and Individual catalogue display the mean of the 84 individual fits reporting the ${\rm \it rms}/\sqrt{N}$.  The bottom panel of Fig.~\ref{fig:olin} displays the performance of the results. }
\begin{center}
\begin{tabular}{|c|c|c|c|c|c}
Reference cosmology & catalogue & $ \alpha_\parallel -\alpha_\parallel^{\rm exp}$ & $\alpha_\perp -\alpha_\perp^{\rm exp}$ & $f\sigma_8-f\sigma_8^{\rm exp}$ & $N_{\rm det}/N_{\rm tot}$\\
\hline
\hline
$\boldsymbol{\Theta}_{\rm Nseries}$ & Mean&  $0.0016\pm 0.0032$  & $-0.0095\pm 0.0020$ & $-0.0038\pm0.0041$  & $1/1$\\
$\boldsymbol{\Theta}_{\rm Nseries}$ & Individual &  $0.0082 \pm 0.0040$  & $-0.0089 \pm 0.0021$  & 0.0073$\pm0.0043$ & $84/84$\\
\hline
$\boldsymbol{\Theta}_{\rm fid}$ & Mean & $0.0003\pm0.0031$ & $-0.0087\pm  0.0019$  & $-0.0055\pm0.0040$ & $1/1$\\
$\boldsymbol{\Theta}_{\rm fid}$ & Individual & $0.0060 \pm 0.0039$ & $-0.0084 \pm 0.0020$  &$0.0107\pm0.0042$ & $84/84$\\
\hline
$\boldsymbol{\Theta}_X$ & Mean & $-0.0046 \pm 0.0031$ & $-0.0115 \pm 0.0020$  & $-0.0050 \pm 0.0042$ & $1/1$\\
$\boldsymbol{\Theta}_X$ & Individual & $0.0005 \pm 0.0039$ & $-0.0111 \pm 0.0019 $  & $0.0106 \pm 0.0049$ & $84/84$\\
\hline
$\boldsymbol{\Theta}_Y$ & Mean & $ 0.0028 \pm 0.0032$ & $ -0.0062 \pm 0.0019$  & $-0.0054 \pm 0.0041$ & $1/1$\\
$\boldsymbol{\Theta}_Y$ & Individual & $0.0102\pm 0.0038$ & $-0.0064 \pm 0.0019 $  & $0.0004 \pm 0.0046$ & $84/84$\\
\hline
$\boldsymbol{\Theta}_Z$ & Mean & $ -0.0045 \pm 0.0030$ & $-0.0126 \pm 0.0018$  & $0.0039\pm0.0041$ & $1/1$\\
$\boldsymbol{\Theta}_Z$ & Individual & $-0.0020\pm 0.0032$ & $-0.0123 \pm0.0018$  & $0.0077\pm0.0045$ & $84/84$\\

\end{tabular}
\end{center}
\label{tab:cosmodependenceRSD}
\end{table*}

Table~\ref{tab:cosmodependenceRSD} corresponds to the bottom panel of Fig. \ref{fig:olin} and lists the impact of the arbitrary choice of cosmology on the \textsc{Nseries} catalogues on the FS type of analysis, for both the fit to the mean of the mocks, and the mean of the 84 individual fits. 

\begin{table*}
\caption{Impact of non-periodicity in HOD-mocks for pre-recon catalogues or FS analyses. For simplicity only the fit to the mean is provided. }
\begin{center}
\begin{tabular}{c|c|c|c|c|c}
Multipoles & HOD-type & Catalogue & $ \alpha_\parallel  -\alpha_\parallel^{\rm exp}$ & $ \alpha_\perp  -\alpha_\perp^{\rm exp}$ & $f\sigma_8-f\sigma_8^{\rm exp}$ \\
\hline
\hline
M+Q+H& Hearin & Sky-cut &  $-0.010 \pm 0.013$ & $-0.0020 \pm 0.0089$ & $0.019\pm0.019$ \\
M+Q+H& Hearin & Cubic box & $0.010 \pm 0.013$ & $-0.0019 \pm 0.0085$ & $ 0.015\pm 0.019$ \\
M+Q& Hearin & Sky-cut & $-0.002 \pm 0.017$ & $-0.001 \pm 0.011$ & $0.009\pm0.023$ \\
M+Q& Hearin & Cubic box & $0.022 \pm 0.016$ & $-0.0010 \pm 0.0092$ & $ 0.005\pm 0.023$ \\
\hline
M+Q+H& Leauthaud & Sky-cut  & $0.003 \pm 0.014$  & $-0.0111 \pm 0.0094$ & $-0.004\pm0.020$ \\
M+Q+H& Leauthaud & Cubic box & $-0.003 \pm 0.012$ & $0.0047 \pm 0.0095$ & $0.013\pm0.019$ \\
M+Q& Leauthaud & Sky-cut  & $0.007 \pm 0.018$  & $-0.006 \pm 0.011$ & $-0.001\pm0.024$ \\
 M+Q& Leauthaud & Cubic box & $0.002\pm0.016$ & $0.010\pm0.010$ & $0.022\pm0.023$ \\
\hline
M+Q+H& Tinker & Sky-cut & $0.006 \pm  0.014$ & $-0.018 \pm 0.011$ & $-0.007\pm0.021$  \\
M+Q+H& Tinker & Cubic box & $0.004 \pm 0.014$ & $-0.007  \pm 0.010$ & $0.004\pm0.021$ \\
M+Q& Tinker & Sky-cut & $0.012 \pm  0.019$ & $-0.012 \pm 0.013$ & $-0.005\pm0.026$ \\
M+Q& Tinker & Cubic box & $-0.004 \pm  0.019$ & $0.004 \pm 0.011$ & $0.025\pm0.025$ \\
\end{tabular}
\end{center}
\label{tab:RSDperiodicity}
\end{table*}%

Table~\ref{tab:RSDperiodicity} shows an analogous result displayed by Table~\ref{table:periodicity}, but for the FS type of analysis.

\begin{table*}
\caption{Impact of different HOD types and `flavours' on pre-recon FS fits on cubic boxes without periodic boundary conditions. For simplicity only fit on the mean is provided where monopole, quadrupole and hexadecapole are used up to $k_{\rm max}=0.15\,\hompc$.}
\begin{center}
\begin{tabular}{|c|c|c|c|c|c|c}
HOD-type & HOD-flavour &  $\alpha_\parallel  -\alpha_\parallel^{\rm exp}$ & $ \alpha_\perp  -\alpha_\perp^{\rm exp}$ & $f\sigma_8-f\sigma_8^{\rm exp}$\\
\hline
\hline
Hearin & standard & $0.0121 \pm0.0069$ & $-0.0031 \pm 0.0051$ & $0.0040 \pm0.0077$ \\
Hearin & Threshold 1 & $0.0106 \pm 0.0068$ & $-0.0032 \pm0.0052$ & $-0.0112 \pm0.0066$ \\
Hearin & Threshold 2 & $0.010 \pm 0.013$ & $-0.0019 \pm 0.0085$ & $ 0.015\pm 0.019$ \\
\hline
Leauthaud & standard & $0.0133 \pm 0.0066$ & $-0.0055 \pm0.0049$ & $-0.0003\pm0.0074$\\
Leauthaud & Threshold 1 & $0.0093 \pm 0.0064$ & $-0.0035\pm 0.0046$ & $-0.0076\pm0.0069$\\
Leauthaud & Threshold 2 & $-0.003 \pm 0.012$ & $0.0047 \pm 0.0095$ & $0.013\pm0.019$ \\
\hline
Tinker & standard & $0.0178 \pm0.0068$ & $-0.0027 \pm 0.0050$ &$0.0124\pm0.0084$\\
Tinker & Threshold 1 & $0.0107 \pm 0.0066$ & $ -0.0043 \pm0.0049$ & $-0.0055\pm0.0065$\\
Tinker & Threshold 2 & $0.004 \pm 0.014$ & $-0.007  \pm 0.010$ & $0.004\pm0.021$ \\
\hline
Zheng & standard & $ 0.0108 \pm0.0067$ & $-0.0025 \pm0.0051$ & $-0.0083 \pm0.0065$\\
Zheng & Threshold 1 & $ 0.0107 \pm0.0069$ & $-0.0035 \pm0.0052$ & $-0.0085\pm0.0073$\\
Zheng & Threshold 2 & $ 0.0138 \pm 0.0095$ & $-0.0049 \pm 0.0072$ & $0.008 \pm 0.013$\\
\end{tabular}
\end{center}
\label{tab:RSDHOD}
\end{table*}%

Table~\ref{tab:RSDHOD}  displays the effect  on the FS analysis of a broad type of HOD and `flavours' using the \textsc{OuterRim}-HOD mocks, analogously to the results displayed on Table~\ref{tab:HOD4BAO} for the BAO type of analysis.

\begin{table*}
\caption{Results from the BAO post-recon only analysis, FS pre-reconstruction analysis and simultaneous  BAO+FS fit (BAO $\times$ FS); on the. \textsc{Nseries} mocks. The \textsc{Nseries} cosmology has been used as a reference cosmology. BAO stands for post-recon. For the individual fits we report the mean and the {\it rms} / $\sqrt{84 }$, whereas for the `Mean' we report the best-fit and the error of the mean. For the BAO only analysis we use 3 polynomial terms for describing the broadband. When the simultaneous  BAO+FS fit is performed we use 5 polynomial terms for describing the BAO broadband (see main text for the full motivation of this approach). }
\begin{center}
\begin{tabular}{c|c|c|c|c|c}
Mock name & type of fit &$ \alpha_\parallel -\alpha_\parallel^{\rm exp}$ & $\alpha_\perp -\alpha_\perp^{\rm exp}$ & $f\sigma_8-f\sigma_8^{\rm exp}$ & $N_{\rm det}/N_{\rm tot}$\\
\hline
\hline
Mean  \textsc{Nseries} & BAO $\times$ FS  & $-0.0008 \pm 0.0017$  & $-0.0017 \pm 0.0011$ & $0.0181 \pm 0.0037$  & $1/1$\\
Individual  \textsc{Nseries}  & BAO $\times$ FS& $0.0062\pm0.0024$  & $-0.0068\pm0.0015$  & $0.0193\pm0.0041$ & $84/84$\\
\hline
Mean  \textsc{Nseries} & FS only  & $0.0015\pm 0.0032$  & $-0.0095\pm 0.0020$ & $-0.0038\pm0.0041$  & $1/1$\\
Individual  \textsc{Nseries}  & FS only& $0.0082 \pm 0.0040$  & $-0.0089 \pm 0.0021$  & 0.0073$\pm0.0043$ & $84/84$\\
\hline
Mean  \textsc{Nseries} & BAO only &  $-0.0048 \pm 0.0019$ &  $0.0005 \pm 0.0010$   & - &  $1/1$\\
Individual  \textsc{Nseries} & BAO only &  $ -0.0016 \pm0.0033$  & $-0.0030\pm 0.0018$  & - & $84/84$\\
\end{tabular}
\end{center}
\label{tab:consensusmocks}
\end{table*}%

Table~\ref{tab:consensusmocks} displays the fits on the \textsc{Nseries} mocks of the BAO post-recon analysis, the FS analysis, and a simultaneous FS and BAO analysis. The reference cosmology chosen in this case is the \textsc{Nseries} own cosmology.

\bsp	
\label{lastpage}
\end{document}

%% file: variables.tex
%
%
%


\def\jnl@style{\it}
\def\aaref@jnl#1{{\jnl@style#1}}

\def\aaref@jnl#1{{\jnl@style#1}}

\def\aj{\aaref@jnl{AJ}}                   
\def\araa{\aaref@jnl{ARA\&A}}             
\def\apj{\aaref@jnl{ApJ}}                 
\def\apjl{\aaref@jnl{ApJ}}                
\def\apjs{\aaref@jnl{ApJS}}               
\def\ao{\aaref@jnl{Appl.~Opt.}}           
\def\apss{\aaref@jnl{Ap\&SS}}             
\def\aap{\aaref@jnl{A\&A}}                
\def\aapr{\aaref@jnl{A\&A~Rev.}}          
\def\aaps{\aaref@jnl{A\&AS}}              
\def\azh{\aaref@jnl{AZh}}                 
\def\baas{\aaref@jnl{BAAS}}               
\def\jrasc{\aaref@jnl{JRASC}}             
\def\memras{\aaref@jnl{MmRAS}}            
\def\mnras{\aaref@jnl{MNRAS}}             
\def\pra{\aaref@jnl{Phys.~Rev.~A}}        
\def\prb{\aaref@jnl{Phys.~Rev.~B}}        
\def\prc{\aaref@jnl{Phys.~Rev.~C}}        
\def\prd{\aaref@jnl{Phys.~Rev.~D}}        
\def\pre{\aaref@jnl{Phys.~Rev.~E}}        
\def\prl{\aaref@jnl{Phys.~Rev.~Lett.}}    
\def\pasp{\aaref@jnl{PASP}}               
\def\pasj{\aaref@jnl{PASJ}}               
\def\qjras{\aaref@jnl{QJRAS}}             
\def\skytel{\aaref@jnl{S\&T}}             
\def\solphys{\aaref@jnl{Sol.~Phys.}}      
\def\sovast{\aaref@jnl{Soviet~Ast.}}      
\def\ssr{\aaref@jnl{Space~Sci.~Rev.}}     
\def\zap{\aaref@jnl{ZAp}}                 
\def\nat{\aaref@jnl{Nature}}              
\def\iaucirc{\aaref@jnl{IAU~Circ.}}       
\def\aplett{\aaref@jnl{Astrophys.~Lett.}} 
\def\apspr{\aaref@jnl{Astrophys.~Space~Phys.~Res.}}
\def\bain{\aaref@jnl{Bull.~Astron.~Inst.~Netherlands}} 
\def\fcp{\aaref@jnl{Fund.~Cosmic~Phys.}}  
\def\gca{\aaref@jnl{Geochim.~Cosmochim.~Acta}}   
\def\grl{\aaref@jnl{Geophys.~Res.~Lett.}} 
\def\jcp{\aaref@jnl{J.~Chem.~Phys.}}      
\def\jgr{\aaref@jnl{J.~Geophys.~Res.}}    
\def\jqsrt{\aaref@jnl{J.~Quant.~Spec.~Radiat.~Transf.}}
\def\memsai{\aaref@jnl{Mem.~Soc.~Astron.~Italiana}}
\def\nphysa{\aaref@jnl{Nucl.~Phys.~A}}   
\def\physrep{\aaref@jnl{Phys.~Rep.}}   
\def\physscr{\aaref@jnl{Phys.~Scr}}   
\def\planss{\aaref@jnl{Planet.~Space~Sci.}}   
\def\procspie{\aaref@jnl{Proc.~SPIE}}   
\def\jcap{\aaref@jnl{J. Cosmology Astropart. Phys.}}

\let\astap=\aap
\let\apjlett=\apjl
\let\apjsupp=\apjs
\let\applopt=\ao

\newcommand{\etal}{et al.\ }

\newcommand{\mpc}{\, {\rm Mpc}}
\newcommand{\kpc}{\, {\rm kpc}}
\newcommand{\hmpc}{\, h^{-1} \mpc}
\newcommand{\ihmpc}{\, h\, {\rm Mpc}^{-1}}
\newcommand{\ikms}{\, {\rm s\, km}^{-1}}
\newcommand{\kms}{\, {\rm km\, s}^{-1}}
\newcommand{\hkpc}{\, h^{-1} \kpc}
\newcommand{\lya}{Ly$\alpha$\ }
\newcommand{\lyb}{Lyman-$\beta$\ }
\newcommand{\lyaf}{Ly$\alpha$ forest}
\newcommand{\lr}{\lambda_{{\rm rest}}}
\newcommand{\bF}{\bar{F}}
\newcommand{\bS}{\bar{S}}
\newcommand{\bC}{\bar{C}}
\newcommand{\bB}{\bar{B}}
\newcommand{\vdF}{{\mathbf \delta_F}}
\newcommand{\vdS}{{\mathbf \delta_S}}
\newcommand{\vdf}{{\mathbf \delta_f}}
\newcommand{\vdn}{{\mathbf \delta_n}}
\newcommand{\vdC}{{\mathbf \delta_C}}
\newcommand{\vdX}{{\mathbf \delta_X}}
\newcommand{\xrei}{x_{rei}}
\newcommand{\lrmin}{\lambda_{{\rm rest, min}}}
\newcommand{\lrmax}{\lambda_{{\rm rest, max}}}
\newcommand{\lmin}{\lambda_{{\rm min}}}
\newcommand{\lmax}{\lambda_{{\rm max}}}
\newcommand{\hi}{\mbox{H\,{\scriptsize I}\ }}
\newcommand{\heii}{\mbox{He\,{\scriptsize II}\ }}
\newcommand{\vp}{\mathbf{p}}
\newcommand{\vq}{\mathbf{q}}
\newcommand{\vxperp}{\mathbf{x_\perp}}
\newcommand{\vkperp}{\mathbf{k_\perp}}
\newcommand{\vrperp}{\mathbf{r_\perp}}
\newcommand{\vx}{\mathbf{x}}
\newcommand{\vy}{\mathbf{y}}
\newcommand{\vk}{\mathbf{k}}
\newcommand{\vR}{\mathbf{r}}
\newcommand{\tdtwo}{\tilde{b}_{\delta^2}}
\newcommand{\tstwo}{\tilde{b}_{s^2}}
\newcommand{\tbthree}{\tilde{b}_3}
\newcommand{\tadtwo}{\tilde{a}_{\delta^2}}
\newcommand{\tastwo}{\tilde{a}_{s^2}}
\newcommand{\tabthree}{\tilde{a}_3}
\newcommand{\vnabla}{\mathbf{\nabla}}
\newcommand{\tpsi}{\tilde{\psi}}
\newcommand{\vv}{\mathbf{v}}
\newcommand{\fnl}{{f_{\rm NL}}}
\newcommand{\tfnl}{{\tilde{f}_{\rm NL}}}
\newcommand{\gnl}{g_{\rm NL}}
\newcommand{\orderfour}{\mathcal{O}\left(\delta_1^4\right)}
\newcommand{\SDSSPF}{\cite{2006ApJS..163...80M}}
\newcommand{\PF}{$P_F^{\rm 1D}(k_\parallel,z)$}
\newcommand\ionalt[2]{#1$\;${\scriptsize \uppercase\expandafter{\romannumeral #2}}}%
\newcommand{\vxone}{\mathbf{x_1}}
\newcommand{\vxtwo}{\mathbf{x_2}}
\newcommand{\vRot}{\mathbf{r_{12}}}
\newcommand{\cm}{\, {\rm cm}}